\documentclass[aps,prd,preprint,amsmath,amssymb]{revtex4}

\usepackage{graphicx}
\usepackage{bm}
\usepackage{bbm}
\usepackage{hyperref}
\usepackage{url}
\usepackage{booktabs}

\allowdisplaybreaks

\begin{document}

\preprint{}

\title{Rescaled Leptonic Unitarity Triangles and Rephasing Invariants \vspace{3mm}}

\author{\bf Shu Luo}
\email{luoshu@xmu.edu.cn}
\affiliation{Department of Astronomy, Xiamen University, Xiamen, Fujian 361005, China\vspace{12mm}}

\begin{abstract}
\vspace{3mm}
The field of neutrino physics has made significant progress in measuring the strength and frequency of neutrino and antineutrino oscillations in the past two decades.
It is clear that the amplitudes involved in the neutrino oscillation probabilities are all rephaping invariants of the quartet forms of the elements of the PMNS mixing matrix.
We show in this paper how these quartet observables can be directly linked to the rescaled leptonic unitarity triangles within the framework of three active neutrinos.
We provide a systematic discussion of the nine CP-conserving quartets ${\cal R}^{}_{\gamma k} \equiv {\rm Re} \left [ V^{}_{\alpha i} V^{}_{\beta j} V^{*}_{\alpha j} V^{*}_{\beta i}\right ] $ along with the universal Jarlskog invariant of CP violation ${\cal J} \equiv \sum_\gamma \epsilon^{}_{\alpha\beta\gamma} \sum_k \epsilon^{}_{ijk} \; {\rm Im} \left [ V^{}_{\alpha i} V^{}_{\beta j} V^{*}_{\alpha j} V^{*}_{\beta i} \right ]$, and place particular emphasis on the matter effect on these quartets.
In addition to the well-known Naumov relation for the Jarlskog invariant ${\cal J}$, similar relations connecting ${\cal R}$ in vacuum and its effective counterparts $\widetilde{\cal R}$ in matter are introduced and examined in detail.
We find that the effective CP-conserving invariants $\widetilde{\cal R}^{}_{\alpha i}$ in matter can be regarded as linear combinations of their vacuum counterparts.
The corresponding composition matrices are primarily determined by the matter potential $A^{}_{CC}$, the neutrino mass-squared differences and the first row of the PMNS matrix, due to the fact that only the charged-current interaction affects neutrino oscillation behaviors in matter.
This finding indicates that with the already measured parameters we are able to calculate the composition matrix to a great accuracy, which valid the exploration of the yet unknown neutrino oscillation parameters in the future.
With the latest global fit data of neutrino masses and mixing elements, numerical analyses are carried out to give an intuitive understanding of how these rephasing invariants evolve as the matter density increases.
\vspace{3mm}
\end{abstract}

\maketitle

\section{CP-violating and CP-conserving Rephasing Invariants}

In the past two decades, the field of neutrino physics has made enormous progress in measuring the strength and frequency of neutrino and antineutrino oscillations~\cite{ParticleDataGroup:2022pth}.
A large set of oscillation results obtained from various experimental configurations can be interpreted within the framework of three active neutrinos, whose mass and flavor eigenstates are related by a 3 × 3 unitary mixing matrix, the Pontecorvo-Maki-Nakagawa-Sakata (PMNS) lepton flavor mixing matrix $V$~\cite{Maki:1962mu,Pontecorvo:1967fh}, conventionally parameterized by three mixing angles $\theta^{}_{12}$, $\theta^{}_{23}$, $\theta^{}_{13}$ and a CP-violating phase $\delta$ as~\cite{ParticleDataGroup:2022pth}
\begin{eqnarray}
V \; = \; \left ( \begin{matrix} c^{}_{12} c^{}_{13} & s^{}_{12} c^{}_{13} & s^{}_{13} e^{-i\delta}_{} \cr -s^{}_{12} c^{}_{23} - c^{}_{12} s^{}_{23} s^{}_{13} e^{i\delta}_{} & c^{}_{12} c^{}_{23} - s^{}_{12} s^{}_{23} s^{}_{13} e^{i\delta}_{} & s^{}_{23} c^{}_{13} \cr s^{}_{12} s^{}_{23} - c^{}_{12} c^{}_{23} s^{}_{13} e^{i\delta}_{} & - c^{}_{12} s^{}_{23} - s^{}_{12} c^{}_{23} s^{}_{13} e^{i\delta}_{} & c^{}_{23} c^{}_{13} \end{matrix} \right ) \; .
\label{1}
\end{eqnarray}
Here we do not consider the possible Majorana phases, as they are irrelevant to neutrino oscillations both in vacuum and in matter.
Given the unitarity of the $3\times 3$ mixing matrix $V$, the disappearance and appearance neutrino oscillation probabilities in vacuum can be written as
\begin{eqnarray}
P ( \stackrel{(-)}{\nu}^{}_{\alpha} \rightarrow  \stackrel{(-)}{\nu}^{}_{\alpha} ) & = & 1 - 4 \sum^{}_{j > i} | V^{}_{\alpha i} |^{2} | V^{}_{\alpha j} |^{2} \sin^2 \frac{\Delta^{}_{ji} L}{4 E} \; ,
\label{2} \\
P ( \stackrel{(-)}{\nu}^{}_{\alpha} \rightarrow  \stackrel{(-)}{\nu}^{}_{\beta} ) & = & - 4 \sum^{}_{j > i} {\rm Re} \left [ V^{}_{\alpha i} V^{}_{\beta j} V^{*}_{\alpha j} V^{*}_{\beta i} \right ] \sin^2 \frac{\Delta^{}_{ji} L}{4 E} \mp 8 {\cal J} \prod^{}_{j > i} \sin \frac{\Delta^{}_{ji} L}{4 E} \; , 
\label{3}
\end{eqnarray}
where the oscillating frequencies are determined by the baseline length $L$, the neutrino beam energy $E$, and the neutrino mass-squared differences $\Delta^{}_{ji} \equiv m^{2}_{j} - m^{2}_{i}$ (for $i, j = 1, 2, 3$ and $i < j$).
The amplitudes involved in the appearance probabilities are all rephasing invariants of the quartet forms of the mixing matrix elements.
They are all independent of any redefinition of the phases for charged lepton and neutrino fields. Of which nine are involved in the CP-conserving terms~\cite{Harrison:2006bj},
\begin{eqnarray}
{\cal R}^{}_{\gamma k} & \equiv & {\rm Re} \left [ V^{}_{\alpha i}  V^{}_{\beta j} V^{*}_{\alpha j} V^{*}_{\beta i}\right ] \; ,
\label{4}
\end{eqnarray}
where $\alpha$, $\beta$ and $\gamma$ run co-cyclically over $e$, $\mu$ and $\tau$, while $i$, $j$ and $k$ run co-cyclically over $1$, $2$ and $3$.
In other words, we simply use the two indices not included in the quartet as subscripts to identify the specific CP-conserving invariant.
These nine real components can be arranged in a matrix ${\cal R}$ as follows: 
\begin{eqnarray}
\hspace{-2mm}{\cal R} & \equiv & \left ( \begin{matrix} {\cal R}^{}_{e 1} & {\cal R}^{}_{e 2} & {\cal R}^{}_{e 3} \cr {\cal R}^{}_{\mu 1} & {\cal R}^{}_{\mu 2} & {\cal R}^{}_{\mu 3} \cr {\cal R}^{}_{\tau 1} & {\cal R}^{}_{\tau 2} & {\cal R}^{}_{\tau 3} \cr \end{matrix} \right ) 
\equiv \left ( \begin{matrix} {\rm Re} \left [ V^{}_{\mu 2}  V^{}_{\tau 3} V^{*}_{\mu 3} V^{*}_{\tau 2}\right ] & {\rm Re} \left [ V^{}_{\mu 3} V^{}_{\tau 1} V^{*}_{\mu 1} V^{*}_{\tau 3} \right ] & {\rm Re} \left [ V^{}_{\mu 1} V^{}_{\tau 2} V^{*}_{\mu 2} V^{*}_{\tau 1} \right ] \cr {\rm Re} \left [ V^{}_{\tau 2} V^{}_{e 3} V^{*}_{\tau 3} V^{*}_{e 2} \right ] & {\rm Re} \left [ V^{}_{\tau 3} V^{}_{e 1} V^{*}_{\tau 1} V^{*}_{e 3} \right ] & {\rm Re} \left [ V^{}_{\tau 1} V^{}_{e 2} V^{*}_{\tau 2} V^{*}_{e 1} \right ] \cr {\rm Re} \left [ V^{}_{e 2} V^{}_{\mu 3} V^{*}_{e 3} V^{*}_{\mu 2} \right ] & {\rm Re} \left [ V^{}_{e 3} V^{}_{\mu 1} V^{*}_{e 1} V^{*}_{\mu 3} \right ] & {\rm Re} \left [ V^{}_{e 1} V^{}_{\mu 2} V^{*}_{e 2} V^{*}_{\mu 1} \right ] \cr \end{matrix} \right ) \; .
\label{5}
\end{eqnarray}
And another is the well-known Jarlskog invariant~\cite{Jarlskog:1985ht,Jarlskog:1985cw}
\begin{eqnarray}
{\cal J} & \equiv & \sum_\gamma \epsilon^{}_{\alpha\beta\gamma} \sum_k \epsilon^{}_{ijk} \; {\rm Im} \left [ V^{}_{\alpha i} V^{}_{\beta j} V^{*}_{\alpha j} V^{*}_{\beta i} \right ] \; ,
\label{6}
\end{eqnarray}
which is a unique measure of the leptonic CP violation.
 
Apparently, we have
\begin{eqnarray}
( {\cal R}^{}_{\gamma k} )^{2}_{} + {\cal J}^{2}_{} & = & |V^{}_{\alpha i}|^{2} |V^{}_{\alpha j}|^{2} |V^{}_{\beta i}|^{2} |V^{}_{\beta j}|^{2} \; .
\label{7}
\end{eqnarray}
We can find that the sum of two rephasing invariants ${\cal R}^{}_{\alpha i}$ in the same row or column results in a combination of the quadratic parameters $|V^{}_{\alpha i}|^{2}$,
\begin{eqnarray}
{\cal R}^{}_{\gamma j} + {\cal R}^{}_{\gamma k} \; = \; - |V^{}_{\alpha i}|^{2} |V^{}_{\beta i}|^{2} \; , \nonumber\\
{\cal R}^{}_{\beta k} + {\cal R}^{}_{\gamma k} \; = \; - |V^{}_{\alpha i}|^{2} |V^{}_{\alpha j}|^{2} \; .
\label{8}
\end{eqnarray}
One may immediately see that $|V^{}_{\alpha i}|^{2} |V^{}_{\alpha j}|^{2}$ corresponds to the amplitudes in the disappearance oscillation probabilities $P ( \stackrel{(-)}{\nu}^{}_{\alpha} \rightarrow  \stackrel{(-)}{\nu}^{}_{\alpha} )$, whereas the combination $|V^{}_{\alpha i}|^{2} |V^{}_{\beta i}|^{2}$ is involved in the ultra-high-energy (UHE) neutrino transition probabilities $P ( \stackrel{(-)}{\nu}^{}_{\alpha} \rightarrow  \stackrel{(-)}{\nu}^{}_{\beta} ) = \sum^{3}_{i=1} |V^{}_{\alpha i}|^{2} |V^{}_{\beta i}|^{2}$.
Again, in Eqs.~(\ref{7}) and (\ref{8}), $\alpha$, $\beta$ and $\gamma$ run co-cyclically over $e$, $\mu$ and $\tau$, while $i$, $j$ and $k$ run co-cyclically over $1$, $2$ and $3$.

As the precision measurement era progresses, ongoing and upcoming neutrino oscillation experiments are intended to measure the neutrino energy spectrum with increasing precision, as well as explore more oscillation channels, not only to address the missing parts of the neutrino mixing picture but also to test the unitarity of the PMNS matrix and find clues for new physics beyond the SM. 
Preliminary but encouraging evidence for CP violation has recently been reported by the T2K Collaboration at the $2\sigma$ level~\cite{T2K:2019bcf,T2K:2023smv}.
We can expect that some of the leptonic unitarity triangles will be experimentally reconstructed and well constrained in the near future, similar to what has been accomplished in the quark sector.
 Compared with the mixing angles and phases that parameterized the PMNS matrix, those quartets defined in Eqs.~(\ref{4}) and (\ref{6}) are all rephasing invariants directly related to observables of various neutrino oscillation experiments.
Therefore, their phenomenological applications deserve closer examination.

\section{Rephasing Invariants in Rescaled Unitarity Triangles}

\begin{figure}[]
\begin{center}
\vspace{-7mm}
\includegraphics[width=.94\textwidth]{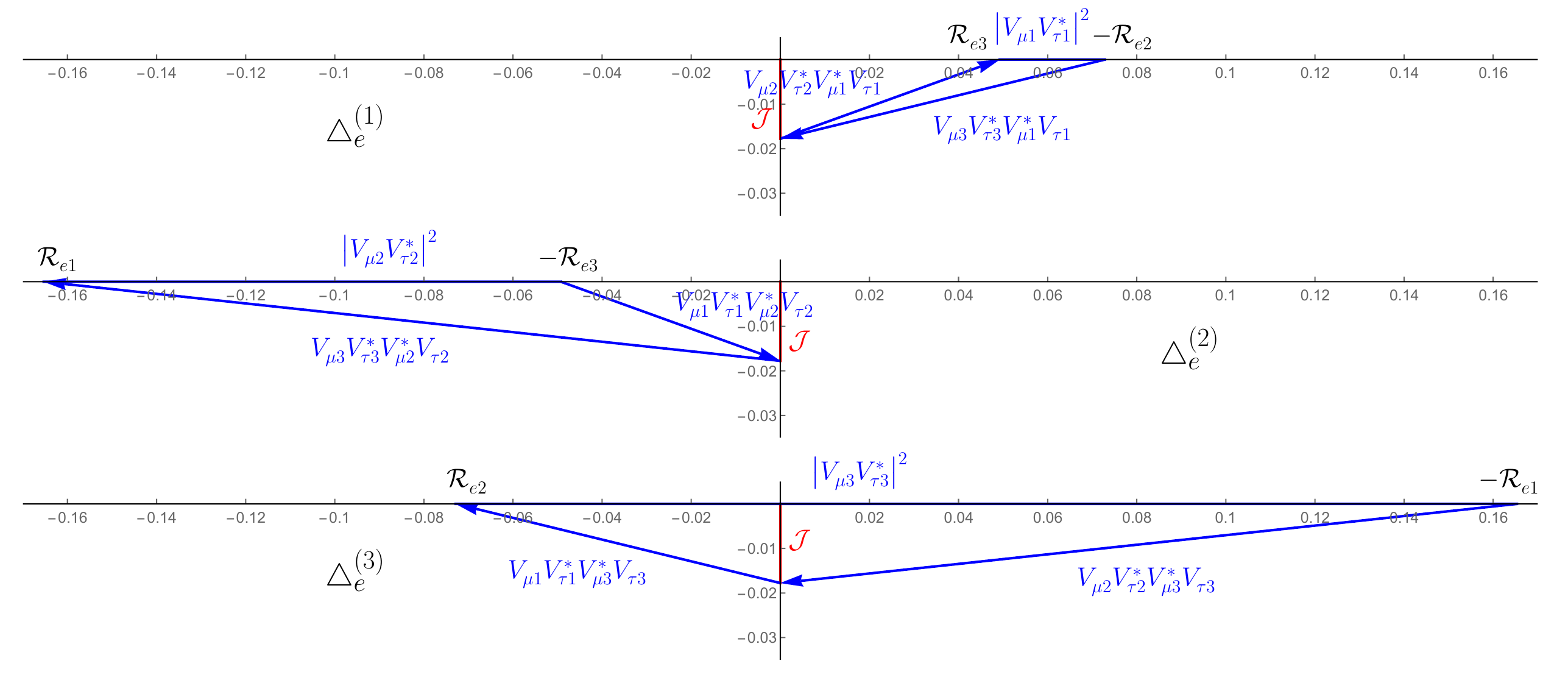}\\
\vspace{5mm}
\includegraphics[width=.94\textwidth]{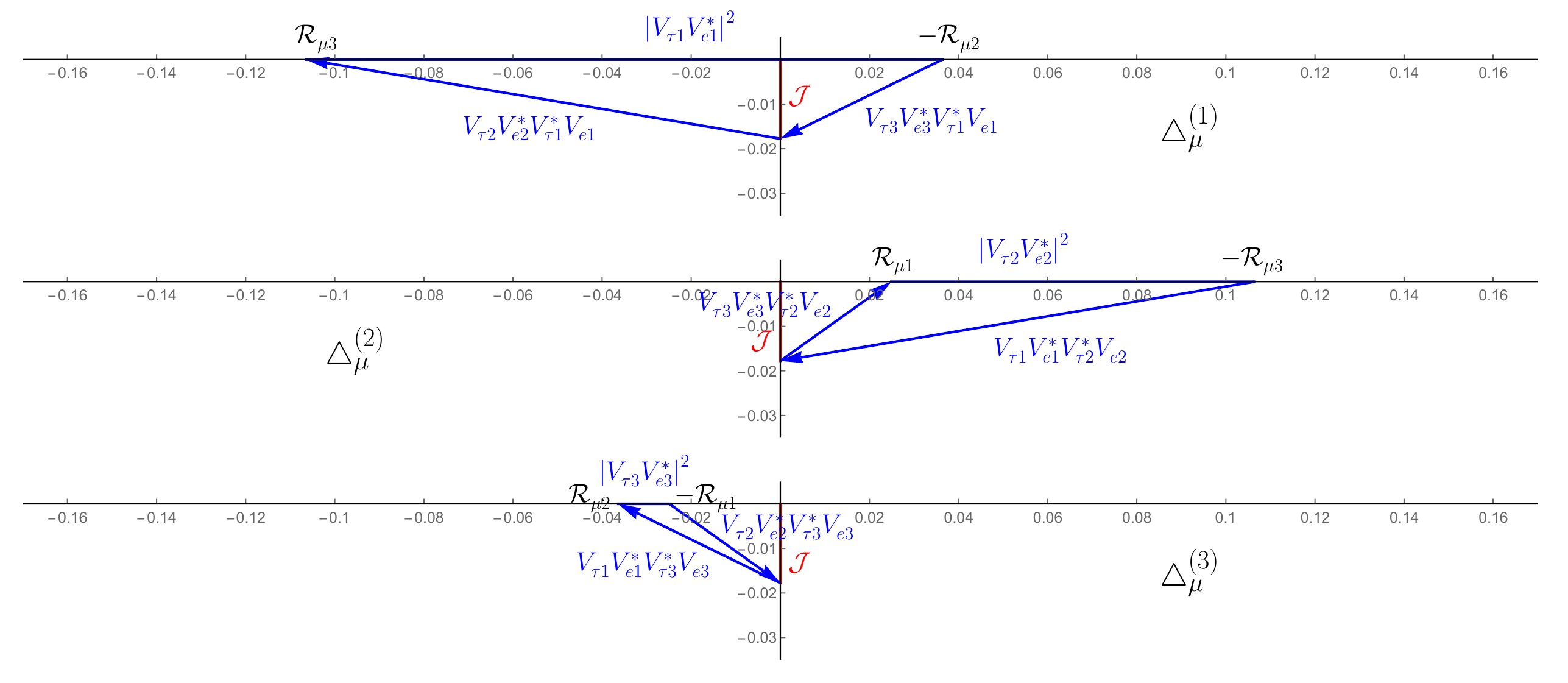}\\
\vspace{5mm}
\includegraphics[width=.94\textwidth]{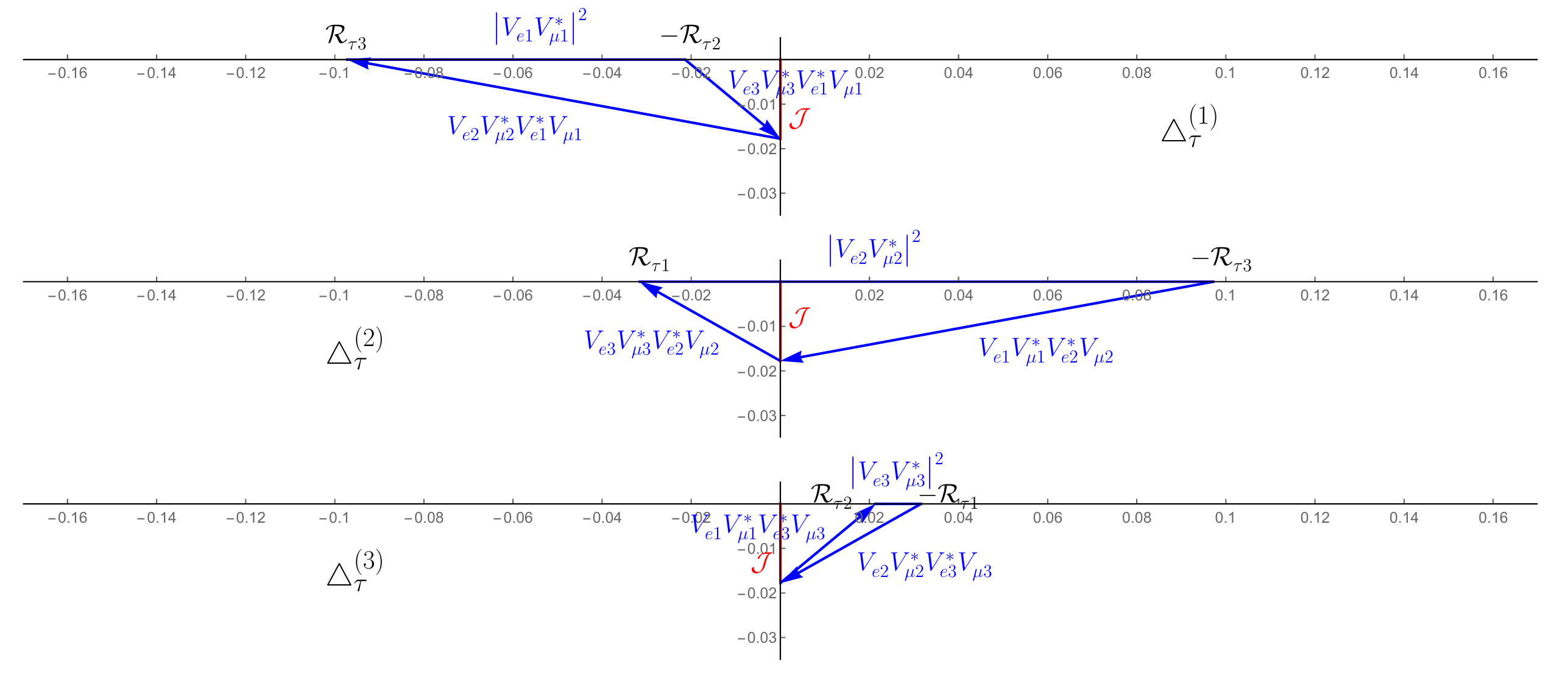}\\
\vspace{0mm}
\caption{A numerical illustration of the rescaled UTs originating from $\triangle^{}_{e}$, $\triangle^{}_{\mu}$ and $\triangle^{}_{\tau}$ respectively in the complex plane, where the best-fit values of the relevant flavor mixing and CP-violating parameters~\cite{Esteban:2024eli,Nufit} for NMO have been input.}
\label{f1}
\end{center}
\end{figure}

\begin{figure}[]
\begin{center}
\vspace{-7mm}
\includegraphics[width=.94\textwidth]{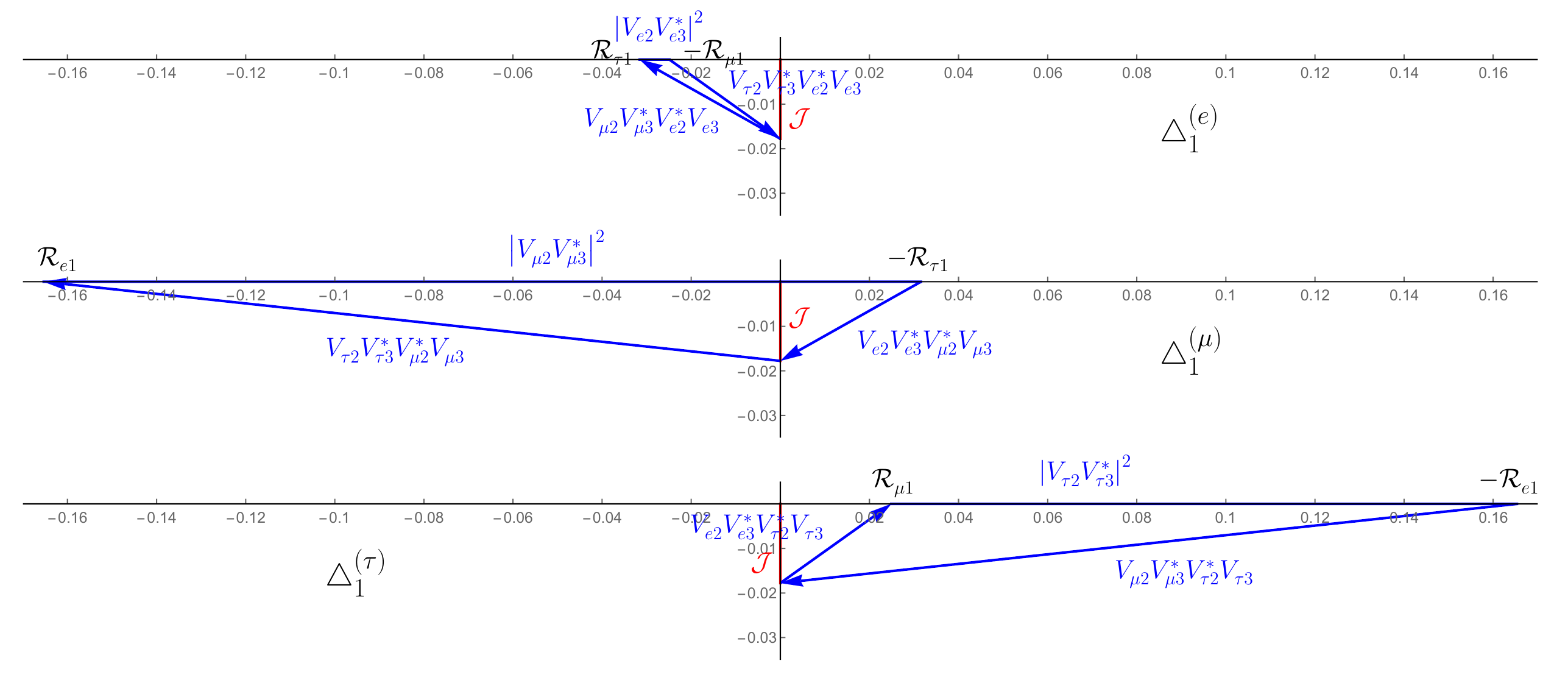}\\
\vspace{5mm}
\includegraphics[width=.94\textwidth]{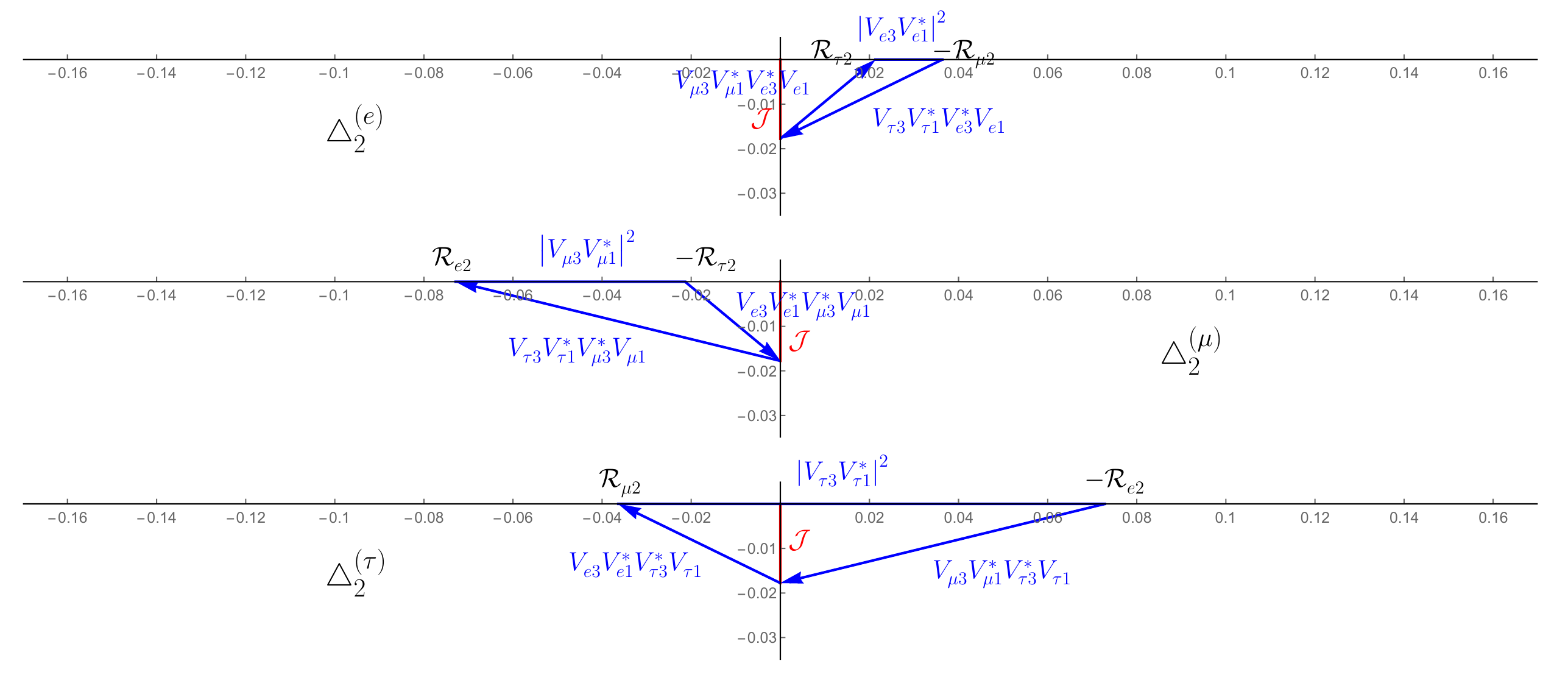}\\
\vspace{5mm}
\includegraphics[width=.94\textwidth]{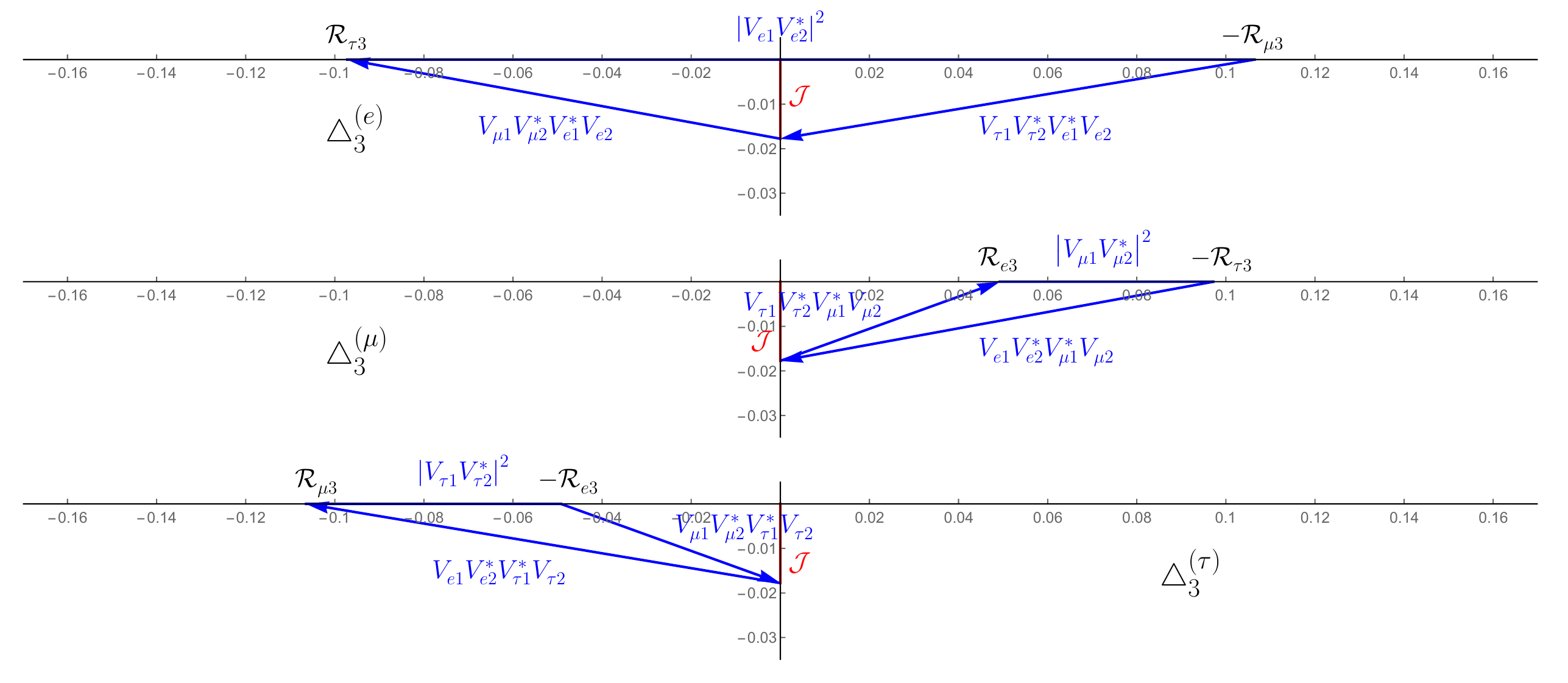}\\
\vspace{0mm}
\caption{A numerical illustration of the rescaled UTs originating from $\triangle^{}_{1}$, $\triangle^{}_{2}$ and $\triangle^{}_{3}$ respectively in the complex plane, where the best-fit values of the relevant flavor mixing and CP-violating parameters~\cite{Esteban:2024eli,Nufit} for NMO have been input.}
\label{f2}
\end{center}
\end{figure}

\begin{figure}[]
\begin{center}
\vspace{-7mm}
\includegraphics[width=.94\textwidth]{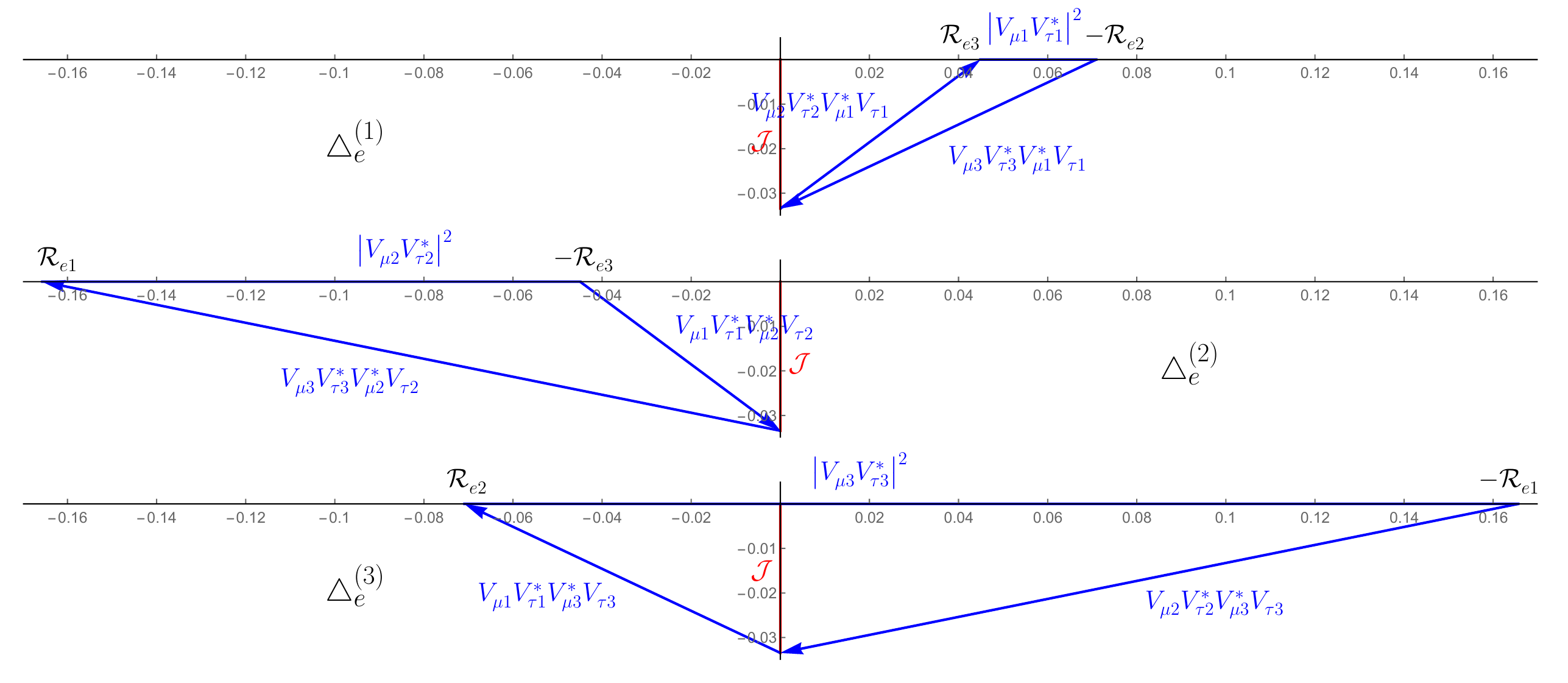}\\
\vspace{5mm}
\includegraphics[width=.94\textwidth]{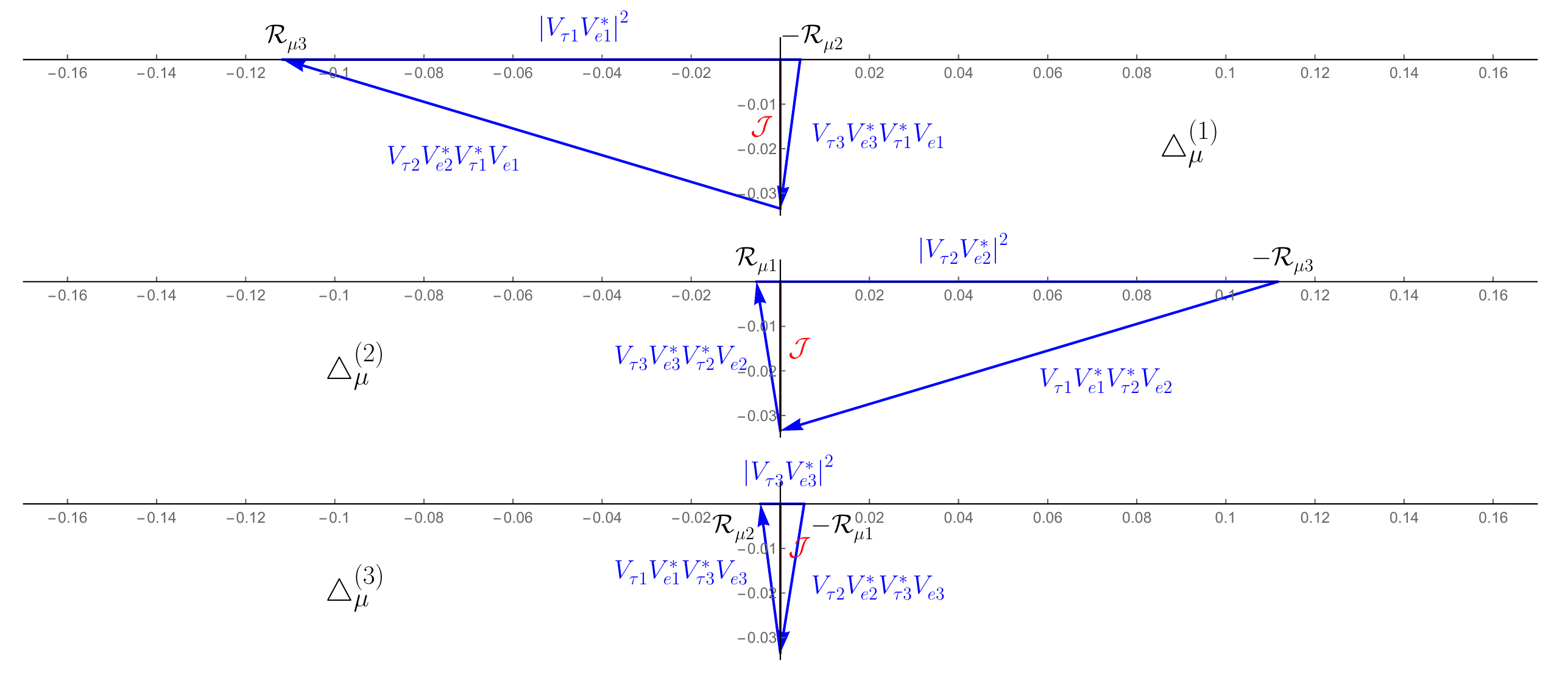}\\
\vspace{5mm}
\includegraphics[width=.94\textwidth]{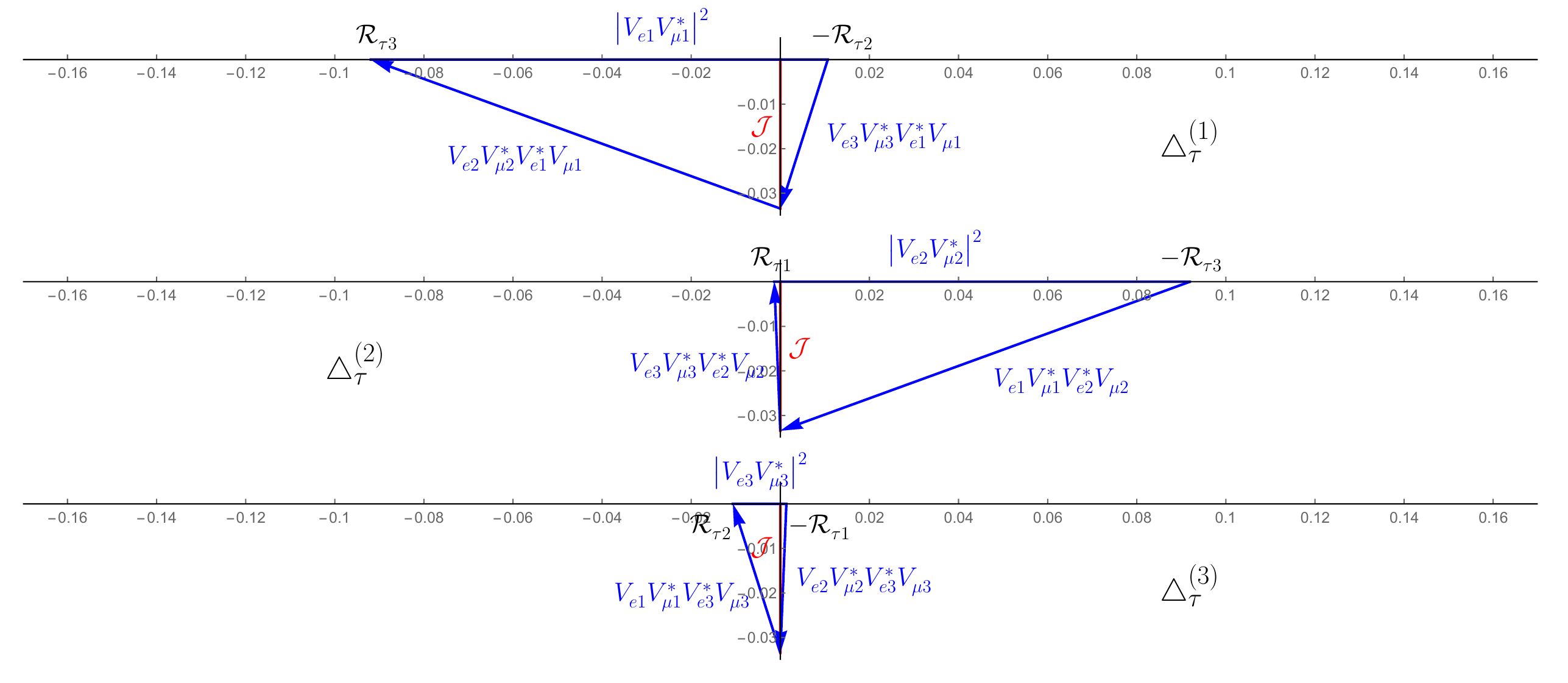}\\
\vspace{0mm}
\caption{A numerical illustration of the rescaled UTs originating from $\triangle^{}_{e}$, $\triangle^{}_{\mu}$ and $\triangle^{}_{\tau}$ respectively in the complex plane, where the best-fit values of the relevant flavor mixing and CP-violating parameters~\cite{Esteban:2024eli,Nufit} for IMO have been input.}
\label{f3}
\end{center}
\end{figure}

\begin{figure}[]
\begin{center}
\vspace{-7mm}
\includegraphics[width=.94\textwidth]{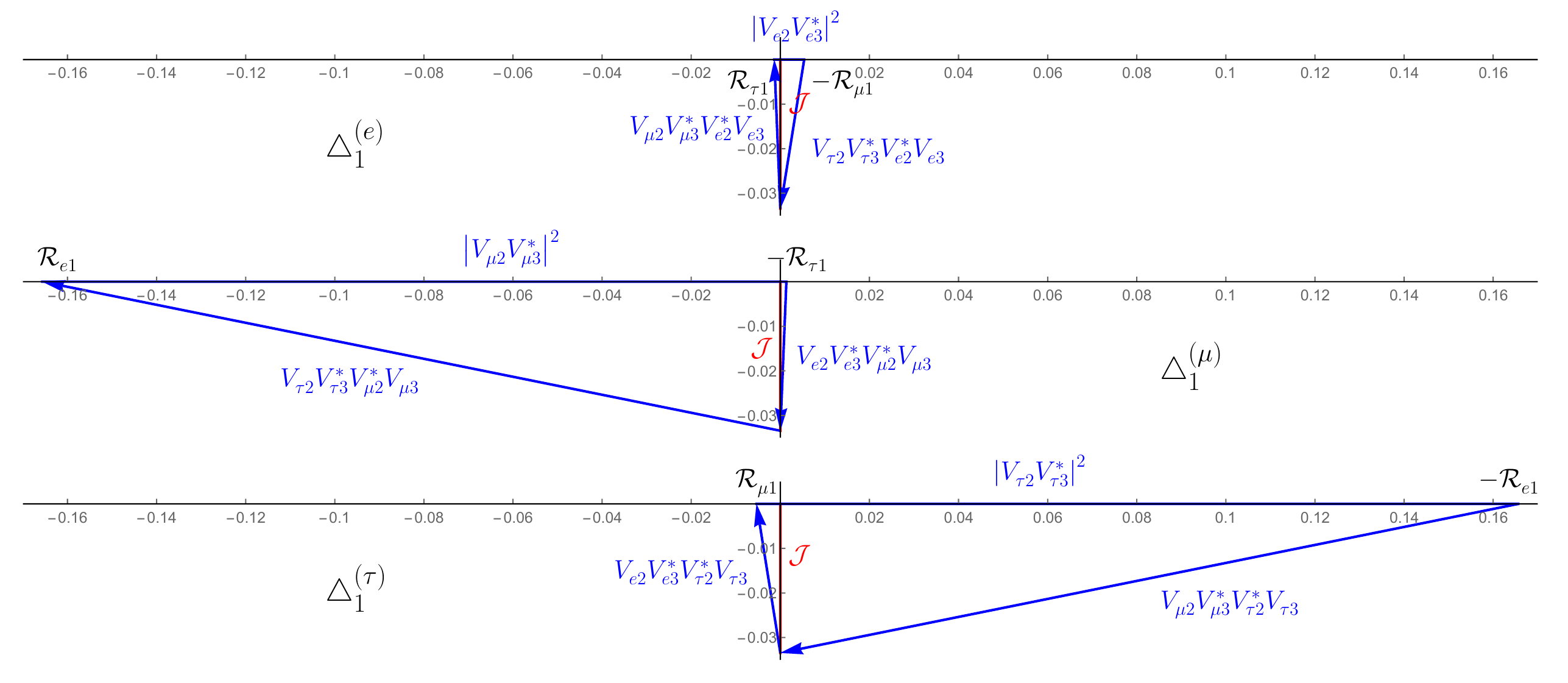}\\
\vspace{5mm}
\includegraphics[width=.94\textwidth]{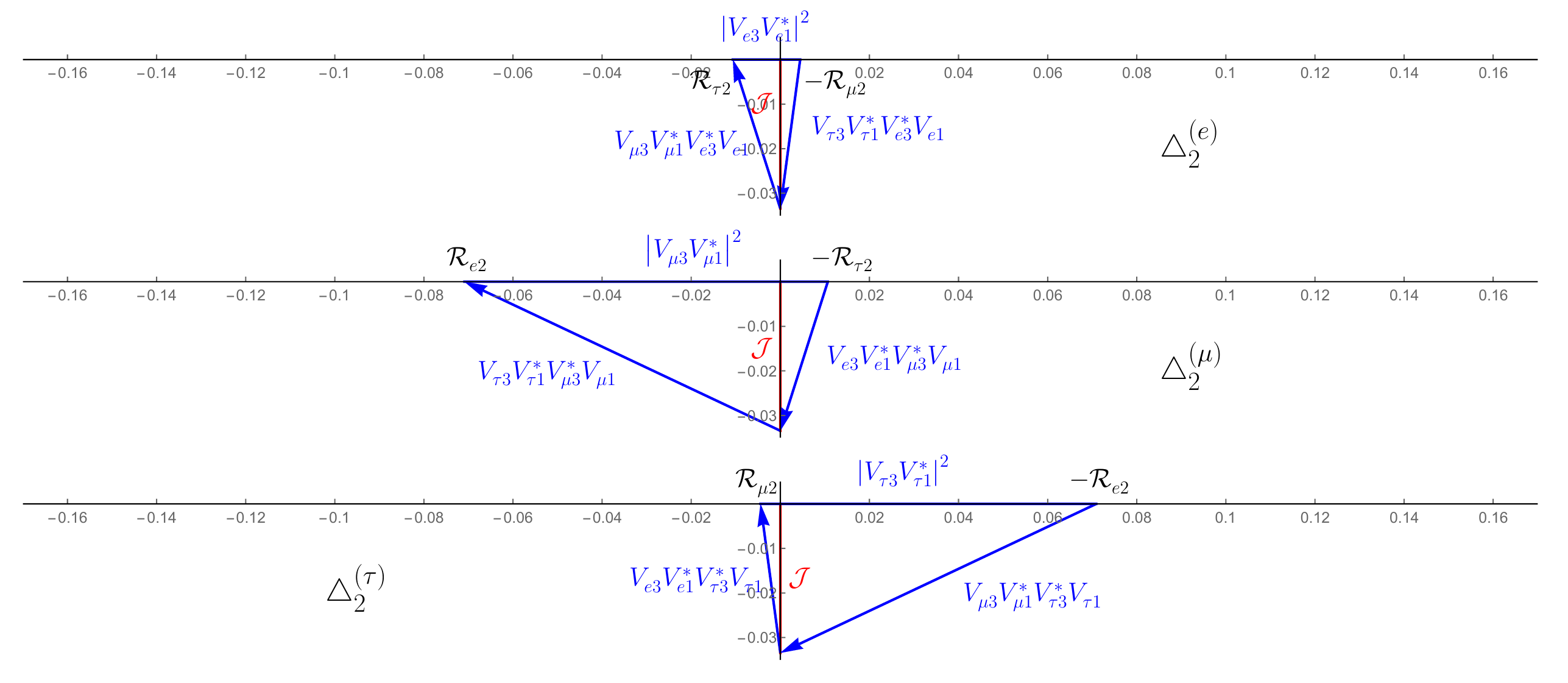}\\
\vspace{5mm}
\includegraphics[width=.94\textwidth]{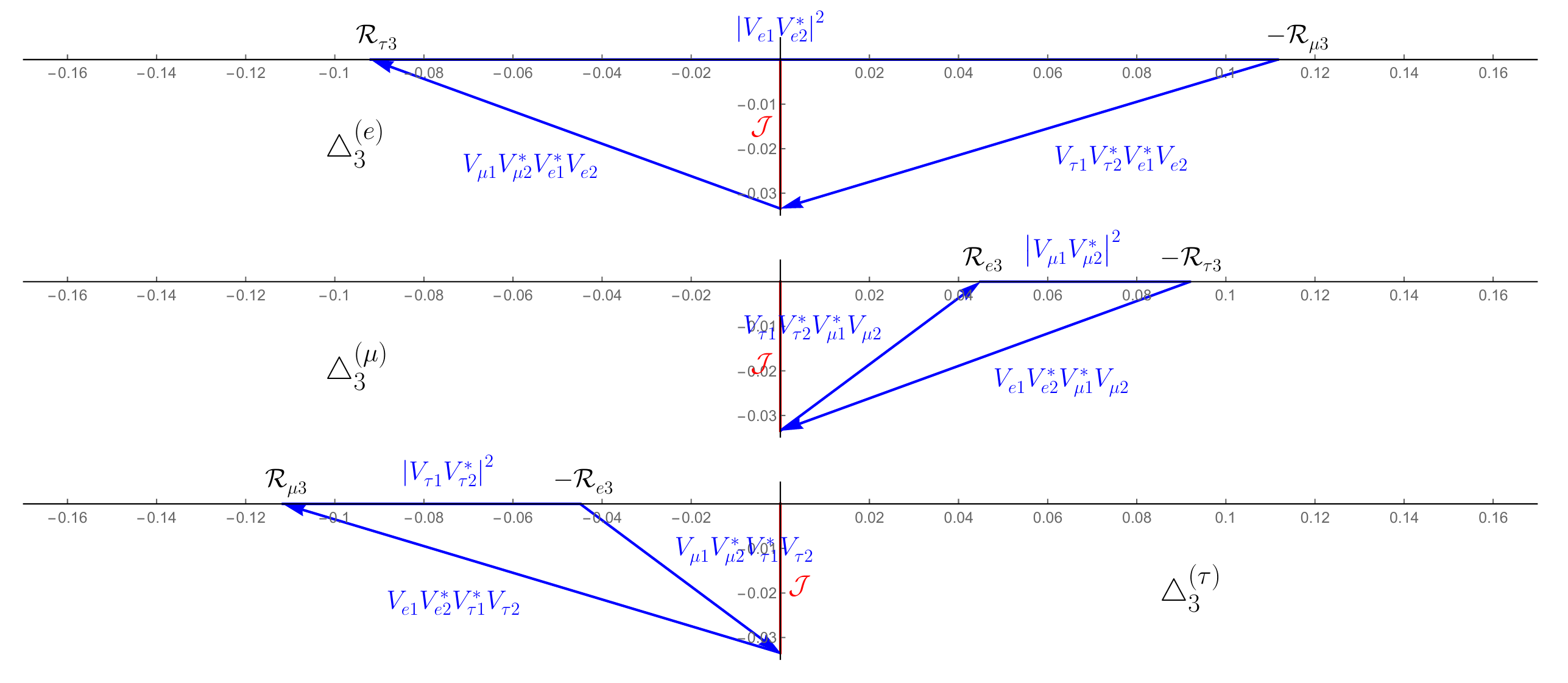}\\
\vspace{0mm}
\caption{A numerical illustration of the rescaled UTs originating from $\triangle^{}_{1}$, $\triangle^{}_{2}$ and $\triangle^{}_{3}$ respectively in the complex plane, where the best-fit values of the relevant flavor mixing and CP-violating parameters~\cite{Esteban:2024eli,Nufit} for IMO have been input.}
\label{f4}
\end{center}
\end{figure}

Unitarity triangles (UTs) are regarded as an intuitive language to geometrically describe lepton flavor mixing and CP violation~\cite{Fritzsch:1999ee,Aguilar-Saavedra:2000jom,Farzan:2002ct,Ahuja:2007cu,Dueck:2010fa,He:2013rba, He:2016dco}.
The orthogonality relations of the $3 \times 3$ unitary PMNS matrix $V$ define the six UTs in the complex plane:
\begin{eqnarray}
\triangle^{}_{e} : &~& V^{}_{\mu 1} V^{*}_{\tau 1} + V^{}_{\mu 2} V^{*}_{\tau 2} + V^{}_{\mu 3} V^{*}_{\tau 3} = 0 \; , \nonumber\\
\triangle^{}_{\mu} : &~& V^{}_{\tau 1} V^{*}_{e 1} + V^{}_{\tau 2} V^{*}_{e 2} + V^{}_{\tau 3} V^{*}_{e 3} = 0 \; , \nonumber\\
\triangle^{}_{\tau} : &~& V^{}_{e 1} V^{*}_{\mu 1} + V^{}_{e 2} V^{*}_{\mu 2} + V^{}_{e 3} V^{*}_{\mu 3} = 0 \; , \nonumber\\
\triangle^{}_{1} : &~& V^{}_{e 2} V^{*}_{e 3} + V^{}_{\mu 2} V^{*}_{\mu 3} + V^{}_{\tau 2} V^{*}_{\tau 3} = 0 \; , \nonumber\\
\triangle^{}_{2} : &~& V^{}_{e 3} V^{*}_{e 1} + V^{}_{\mu 3} V^{*}_{\mu 1} + V^{}_{\tau 3} V^{*}_{\tau 1} = 0 \; , \nonumber\\
\triangle^{}_{3} : &~& V^{}_{e 1} V^{*}_{e 2} + V^{}_{\mu 1} V^{*}_{\mu 2} + V^{}_{\tau 1} V^{*}_{\tau 2} = 0 \; .
\label{9}
\end{eqnarray}
To incorporate the quartets, we can rescale these UTs leading to three fully rephasing-invariant rescaled triangles corresponding to each UT given above~\cite{Xing:2012zv,Luo:2023xmv}. 
The sides of these rescaled triangles in the complex plane are all invariant under phase rephasing.
Note that the three rescaled UTs in each group (e.g. $\triangle^{(1)}_{e}$, $\triangle^{(2)}_{e}$, and $\triangle^{(3)}_{e})$ are similar triangles and similar to the original UT ($\triangle^{}_{e}$).
\begin{eqnarray}
\triangle^{(1)}_{e} : &~& \left | V^{}_{\mu 1} V^{*}_{\tau 1} \right |^{2} + V^{}_{\mu 2} V^{*}_{\tau 2} V^{*}_{\mu 1} V^{}_{\tau 1} + V^{}_{\mu 3} V^{*}_{\tau 3}  V^{*}_{\mu 1} V^{}_{\tau 1} = 0 \; , \nonumber\\
\triangle^{(2)}_{e} : &~& \left | V^{}_{\mu 2} V^{*}_{\tau 2} \right |^{2} + V^{}_{\mu 3} V^{*}_{\tau 3} V^{*}_{\mu 2} V^{}_{\tau 2} + V^{}_{\mu 1} V^{*}_{\tau 1}  V^{*}_{\mu 2} V^{}_{\tau 2} = 0 \; , \nonumber\\
\triangle^{(3)}_{e} : &~& \left | V^{}_{\mu 3} V^{*}_{\tau 3} \right |^{2} + V^{}_{\mu 1} V^{*}_{\tau 1} V^{*}_{\mu 3} V^{}_{\tau 3} + V^{}_{\mu 2} V^{*}_{\tau 2}  V^{*}_{\mu 3} V^{}_{\tau 3} = 0 \; , \nonumber\\[2mm]
\triangle^{(1)}_{\mu} : &~& \left | V^{}_{\tau 1} V^{*}_{e 1} \right |^{2} + V^{}_{\tau 2} V^{*}_{e 2} V^{*}_{\tau 1} V^{}_{e 1} + V^{}_{\tau 3} V^{*}_{e 3}  V^{*}_{\tau 1} V^{}_{e 1} = 0 \; , \nonumber\\
\triangle^{(2)}_{\mu} : &~& \left | V^{}_{\tau 2} V^{*}_{e 2} \right |^{2} + V^{}_{\tau 3} V^{*}_{e 3} V^{*}_{\tau 2} V^{}_{e 2} + V^{}_{\tau 1} V^{*}_{e 1}  V^{*}_{\tau 2} V^{}_{e 2} = 0 \; , \nonumber\\
\triangle^{(3)}_{\mu} : &~& \left | V^{}_{\tau 3} V^{*}_{e 3} \right |^{2} + V^{}_{\tau 1} V^{*}_{e 1} V^{*}_{\tau 3} V^{}_{e 3} + V^{}_{\tau 2} V^{*}_{e 2}  V^{*}_{\tau 3} V^{}_{e 3} = 0 \; , \nonumber\\[2mm]
\triangle^{(1)}_{\tau} : &~& \left | V^{}_{e 1} V^{*}_{\mu 1} \right |^{2} + V^{}_{e 2} V^{*}_{\mu 2} V^{*}_{e 1} V^{}_{\mu 1} + V^{}_{e 3} V^{*}_{\mu 3}  V^{*}_{e 1} V^{}_{\mu 1} = 0 \; , \nonumber\\
\triangle^{(2)}_{\tau} : &~& \left | V^{}_{e 2} V^{*}_{\mu 2} \right |^{2} + V^{}_{e 3} V^{*}_{\mu 3} V^{*}_{e 2} V^{}_{\mu 2} + V^{}_{e 1} V^{*}_{\mu 1}  V^{*}_{e 2} V^{}_{\mu 2} = 0 \; , \nonumber\\
\triangle^{(3)}_{\tau} : &~& \left | V^{}_{e 3} V^{*}_{\mu 3} \right |^{2} + V^{}_{e 1} V^{*}_{\mu 1} V^{*}_{e 3} V^{}_{\mu 3} + V^{}_{e 2} V^{*}_{\mu 2}  V^{*}_{e 3} V^{}_{\mu 3} = 0 \; , \nonumber\\[2mm]
\triangle^{(e)}_{1} : &~& \left | V^{}_{e 2} V^{*}_{e 3} \right |^{2} + V^{}_{\mu 2} V^{*}_{\mu 3}  V^{*}_{e 2} V^{}_{e 3} + V^{}_{\tau 2} V^{*}_{\tau 3}  V^{*}_{e 2} V^{}_{e 3} = 0 \; , \nonumber\\
\triangle^{(\mu)}_{1} : &~& \left | V^{}_{\mu 2} V^{*}_{\mu 3} \right |^{2} + V^{}_{\tau 2} V^{*}_{\tau 3}  V^{*}_{\mu 2} V^{}_{\mu 3} + V^{}_{e 2} V^{*}_{e 3}  V^{*}_{\mu 2} V^{}_{\mu 3} = 0 \; , \nonumber\\
\triangle^{(\tau)}_{1} : &~& \left | V^{}_{\tau 2} V^{*}_{\tau 3} \right |^{2} + V^{}_{e 2} V^{*}_{e 3}  V^{*}_{\tau 2} V^{}_{\tau 3} + V^{}_{\mu 2} V^{*}_{\mu 3}  V^{*}_{\tau 2} V^{}_{\tau 3} = 0 \; , \nonumber\\[2mm]
\triangle^{(e)}_{2} : &~& \left | V^{}_{e 3} V^{*}_{e 1} \right |^{2} + V^{}_{\mu 3} V^{*}_{\mu 1}  V^{*}_{e 3} V^{}_{e 1} + V^{}_{\tau 3} V^{*}_{\tau 1}  V^{*}_{e 3} V^{}_{e 1} = 0 \; , \nonumber\\
\triangle^{(\mu)}_{2} : &~& \left | V^{}_{\mu 3} V^{*}_{\mu 1} \right |^{2} + V^{}_{\tau 3} V^{*}_{\tau 1}  V^{*}_{\mu 3} V^{}_{\mu 1} + V^{}_{e 3} V^{*}_{e 1}  V^{*}_{\mu 3} V^{}_{\mu 1} = 0 \; , \nonumber\\
\triangle^{(\tau)}_{2} : &~& \left | V^{}_{\tau 3} V^{*}_{\tau 1} \right |^{2} + V^{}_{e 3} V^{*}_{e 1}  V^{*}_{\tau 3} V^{}_{\tau 1} + V^{}_{\mu 3} V^{*}_{\mu 1}  V^{*}_{\tau 3} V^{}_{\tau 1} = 0 \; , \nonumber\\[2mm]
\triangle^{(e)}_{3} : &~& \left | V^{}_{e 1} V^{*}_{e 2} \right |^{2} + V^{}_{\mu 1} V^{*}_{\mu 2}  V^{*}_{e 1} V^{}_{e 2} + V^{}_{\tau 1} V^{*}_{\tau 2}  V^{*}_{e 1} V^{}_{e 2} = 0 \; , \nonumber\\
\triangle^{(\mu)}_{3} : &~& \left | V^{}_{\mu 1} V^{*}_{\mu 2} \right |^{2} + V^{}_{\tau 1} V^{*}_{\tau 2}  V^{*}_{\mu 1} V^{}_{\mu 2} + V^{}_{e 1} V^{*}_{e 2}  V^{*}_{\mu 1} V^{}_{\mu 2} = 0 \; , \nonumber\\
\triangle^{(\tau)}_{3} : &~& \left | V^{}_{\tau 1} V^{*}_{\tau 2} \right |^{2} + V^{}_{e 1} V^{*}_{e 2}  V^{*}_{\tau 1} V^{}_{\tau 2} + V^{}_{\mu 1} V^{*}_{\mu 2}  V^{*}_{\tau 1} V^{}_{\tau 2} = 0 \; .
\label{10}
\end{eqnarray}
We illustrated in Figs.~\ref{f1} and \ref{f2} (or Figs.~\ref{f3} and \ref{f4}) the sizes and shapes of all the 18 rescaled UTs by using the best-fit values $\theta^{}_{12} = 33.68^\circ$ (or $33.68^\circ$), $\theta^{}_{13} = 8.56^\circ$ (or $8.59^\circ$), $\theta^{}_{23} = 43.3^\circ$ (or $47.9^\circ$) and $\delta = 212^\circ$ (or $274^\circ$) in the standard parametrization of $V$ corresponding to the normal neutrino mass ordering, NMO (or inverted mass ordering, IMO), as obtained from a latest global analysis of current neutrino oscillation data~\cite{Esteban:2024eli,Nufit}.
To give a ballpark figure of these rephasing invariants, the best-fit values, 1$\sigma$ and 3$\sigma$ ranges of ${\cal J}$ and all nine ${\cal R}^{}_{\alpha i}$ in both the NMO and IMO cases are also listed in Table~\ref{t1}.

\begin{table}[]
\footnotesize
\caption{The best-fit values, 1$\sigma$ and 3$\sigma$ ranges of the Jarlskog CP-violating invariant ${\cal J}$ and the nine CP-conserving invariants ${\cal R}^{}_{\alpha i}$, where we've adopt the latest global fit of the unitary PMNS leptonic mixing matrix given in~\cite{Esteban:2024eli,Nufit}.}
\label{t1} 
\begin{tabular}{ccccccccccccc}
\hline\noalign{\smallskip}
&& \multicolumn{5}{c}{Normal Neutrino Mass Ordering} && \multicolumn{5}{c}{Inverted Neutrino Mass Ordering}  \\
&& best-fit && 1$\sigma$ range && 3$\sigma$ range && best-fit && 1$\sigma$ range && 3$\sigma$ range \\
\noalign{\smallskip}\hline\noalign{\smallskip}
${\cal J}$ ~ & ~~~ & -0.0178 & ~ & -0.0291 $\sim$ 0.0054 & ~ & -0.0358 $\sim$ 0.0297 & ~~~ & -0.0334 & ~ & -0.0343 $\sim$ -0.0293 & ~ & -0.0360 $\sim$ -0.0111 \\
\noalign{\smallskip}
${\cal R}^{}_{e1}$ && -0.1653 && -0.1700 $\sim$ -0.1603 && -0.1776 $\sim$ -0.1473 && -0.1656 && -0.1705 $\sim$ -0.1592 && -0.1773 $\sim$ -0.1481 \\
${\cal R}^{}_{e2}$ && -0.0729 && -0.0769 $\sim$ -0.0689 && -0.0844 $\sim$ -0.0554 && -0.0709 && -0.0752 $\sim$ -0.0655 && -0.0838 $\sim$ -0.0559 \\
${\cal R}^{}_{e3}$ && 0.0493 && 0.0456 $\sim$ 0.0520 && 0.0398 $\sim$ 0.0542 && 0.0449 && 0.0429 $\sim$ 0.0476 && 0.0398 $\sim$ 0.0534 \\
${\cal R}^{}_{\mu1}$ && 0.0249 && 0.0139 $\sim$ 0.0307 && -0.0400 $\sim$ 0.0319 && -0.0054 && -0.0184 $\sim$ 0.0092 && -0.0369 $\sim$ 0.0298 \\
${\cal R}^{}_{\mu2}$ && -0.0364 && -0.0424 $\sim$ -0.0250 && -0.0441 $\sim$ 0.0289 && -0.0045 && -0.0194 $\sim$ 0.0084 && -0.0420 $\sim$ 0.0257 \\
${\cal R}^{}_{\mu3}$ && -0.1064 && -0.1135 $\sim$ -0.0977 && -0.1378 $\sim$ -0.0672 && -0.1116 && -0.1234 $\sim$ -0.1008 && -0.1367 $\sim$ -0.0691 \\
${\cal R}^{}_{\tau1}$ && -0.0316 && -0.0378 $\sim$ -0.0202 && -0.0400 $\sim$ 0.0320 && -0.0014 && -0.0163 $\sim$ 0.0112 && -0.0379 $\sim$ 0.0288 \\
${\cal R}^{}_{\tau2}$ && 0.0214 && 0.0101 $\sim$ 0.0273 && -0.0442 $\sim$ 0.0288 && -0.0107 && -0.0236 $\sim$ 0.0043 && -0.0412 $\sim$ 0.0266 \\
${\cal R}^{}_{\tau3}$ && -0.0972 && -0.1071 $\sim$ -0.0884 && -0.1332 $\sim$ -0.0631 && -0.0919 && -0.1017 $\sim$ -0.0802 && -0.1314 $\sim$ -0.0643 \\
\noalign{\smallskip}
\hline
\end{tabular}
\end{table}

Non-collapsing UTs imply the existence of appreciable effects of CP violation in neutrino oscillations.
However, unlike conventional UTs defined by Eq.~(\ref{9}) which all have areas equal to $|{\cal J}| / 2$, these rescaled UTs defined by Eq.~(\ref{10}) share a common height (with the base on the real axis) equal to the Jarlskog invariant ${\cal J}$, as clearly shown in Figs.~\ref{f1} - \ref{f4}.
Therefore, rescaled UTs lie in the upper (or lower) half-plane if ${\cal J}>0$ (or ${\cal J}<0$).

By scaling the conventional UTs defined in Eq.~(\ref{9}), we bridge the rescaled UTs directly to the observables of various neutrino oscillation experiments.
This allows us to clearly understand the experimental measurements required for determining the shape and size of a specific UT.
Note that a rescaled UT can be reproduced from either a certain oscillation channel or a variety of oscillation experiments.
For example, if ${\cal R}^{}_{\tau 2}$, ${\cal R}^{}_{\tau 3}$, and ${\cal J}$ can be well determined by long-baseline accelerator $\nu^{}_{\mu} \to \nu^{}_{e}$ such as T2K or NO$\nu$A, then $\triangle^{(3)}_{\tau}$ can be constructed, which is promising in the near future.
The JUNO reactor experiment recently reported updated results on $\left | V^{}_{e 1} V^{*}_{e 2} \right |^{2}$. If ${\cal R}^{}_{\tau 3}$ can be measured by certain short- or medium-baseline accelerator neutrino oscillation experiments, together with the measured ${\cal J}$, we may be able to directly build the triangle $\triangle^{(e)}_{3}$.

In addition, characteristics of the UTs and symmetries among these UTs can also be conveniently indicated by using the language of these quartets. 
The first, second and third rows of the ${\cal R}$ matrix are neatly related to $\triangle^{}_{e}$, $\triangle^{}_{\mu}$, and $\triangle^{}_{\tau}$, respectively, while the first, second and third columns are connected with $\triangle^{}_{1}$, $\triangle^{}_{2}$, and $\triangle^{}_{3}$, respectively. 
If the three elements in the same row or column of ${\cal R}$ have the same sign, the corresponding UT is an acute triangle; otherwise it is an obtuse triangle. If any ${\cal R}^{}_{\alpha i}$ equals zero, the corresponding two UTs $\triangle^{}_{\alpha}$ and $\triangle^{}_{i}$ are both right triangles.
We can find from Table~\ref{t1} that ${\cal R}^{}_{e 3}$ is positive, ${\cal R}^{}_{e 1}$, ${\cal R}^{}_{e 2}$, ${\cal R}^{}_{\mu 3}$, ${\cal R}^{}_{\tau 3}$, and ${\cal J}$ are all negative, while ${\cal R}^{}_{\mu 1}$, ${\cal R}^{}_{\mu 2}$, ${\cal R}^{}_{\tau 1}$ and ${\cal R}^{}_{\tau 2}$ can be positive, zero, or negative, given the 3$\sigma$ global-fit constraints.
This means that at the electroweak energy scale, both $\triangle^{}_{e}$ and $\triangle^{}_{3}$ are obtuse triangles, while the other four triangles $\triangle^{}_{\mu}$, $\triangle^{}_{\tau}$, $\triangle^{}_{1}$, and $\triangle^{}_{2}$ can be acute, right, or obtuse. 

Hinted by the fact that two of the UTs in the quark sector are nearly right~\cite{Harrison:2009bb,Harrison:2009bz}, the possibility of right leptonic UTs also aroused interesting discussions; see, e.g.,\cite{Xing:2009eg,Antusch:2011sx,Mimura:2018sbc}.
We summarize in Table~\ref{t2} the four possible special conditions where ${\cal R}^{}_{\mu 1}$, ${\cal R}^{}_{\mu 2}$, ${\cal R}^{}_{\tau 1}$, and ${\cal R}^{}_{\tau 2}$ are zero, respectively.
In each scenario, two of the six UTs are right, and we also write down the expressions for CP violating parameters ${\cal J}$ and $\delta$ in each case. 

Moreover, it is clear that if two elements of the ${\cal R}$ matrix in the same row (or column) are equal, that is, ${\cal R}^{}_{\alpha i} = {\cal R}^{}_{\alpha j}$ (or ${\cal R}^{}_{\alpha i} = {\cal R}^{}_{\beta i}$), then the triangle $\triangle^{}_{\alpha}$ (or $\triangle^{}_{\alpha}$) is an isosceles triangle.
However, if two elements of the ${\cal R}$ matrix that are not in the same row or the same column equal to each other, that is, ${\cal R}^{}_{\alpha i} = {\cal R}^{}_{\beta j}$, we can then infer that the triangles $\triangle^{}_{\gamma}$ and $\triangle^{}_{k}$ become congruent with each other.
Here $\alpha$, $\beta$ and $\gamma$ run co-cyclically over $e$, $\mu$ and $\tau$, while $i$, $j$ and $k$ run co-cyclically over $1$, $2$ and $3$.
For example, if we have ${\cal R}^{}_{\mu 1} = {\cal R}^{}_{\tau 2}$, we can then infer that the triangles $\triangle^{}_{e}$ and $\triangle^{}_{3}$ are congruent triangles and equation ${\cal R}^{}_{\mu 2} = {\cal R}^{}_{\tau 1}$ must also hold. 

\begin{table}[]
\footnotesize
\caption{The four possible conditions that some of the UTs are right triangles.}
\label{t2} 
\begin{tabular}{ccccc}
\hline\noalign{\smallskip}
Conditions &~& Right triangles &~& CP violating parameters \\
\noalign{\smallskip}\hline\noalign{\smallskip}
${\cal R}^{}_{\mu 1} = 0$ &&  $\triangle^{}_{\mu}$, $\triangle^{}_{1}$ && ${\cal J}^{2}_{} = {\cal R}^{}_{\mu 2} \; {\cal R}^{}_{\mu 3} = {\cal R}^{}_{\tau 1} \; {\cal R}^{}_{e 1}$, ~~ $\displaystyle \cos\delta = - \frac{s^{}_{12} c^{}_{23} s^{}_{13}}{c^{}_{12} s^{}_{23}}$ \\
\noalign{\smallskip}
${\cal R}^{}_{\mu 2} = 0$ && $\triangle^{}_{\mu}$, $\triangle^{}_{2}$ && ${\cal J}^{2}_{} = {\cal R}^{}_{\mu 1} \; {\cal R}^{}_{\mu 3} = {\cal R}^{}_{\tau 2} \; {\cal R}^{}_{e 2}$, ~~ $\displaystyle \cos\delta = \frac{c^{}_{12} c^{}_{23} s^{}_{13}}{s^{}_{12} s^{}_{23}}$ \\
\noalign{\smallskip}
${\cal R}^{}_{\tau 1} = 0$ && $\triangle^{}_{\tau}$, $\triangle^{}_{1}$ && ${\cal J}^{2}_{} = {\cal R}^{}_{\tau 2} \; {\cal R}^{}_{\tau 3} = {\cal R}^{}_{\mu 1} \; {\cal R}^{}_{e 1}$, ~~ $\displaystyle \cos\delta = \frac{s^{}_{12} s^{}_{23} s^{}_{13}}{c^{}_{12} c^{}_{23}}$ \\
\noalign{\smallskip}
${\cal R}^{}_{\tau 2} = 0$ && $\triangle^{}_{\tau}$, $\triangle^{}_{2}$ && ${\cal J}^{2}_{} = {\cal R}^{}_{\tau 1} \; {\cal R}^{}_{\tau 3} = {\cal R}^{}_{\mu 2} \; {\cal R}^{}_{e 2}$, ~~ $\displaystyle \cos\delta = - \frac{c^{}_{12} s^{}_{23} s^{}_{13}}{s^{}_{12} c^{}_{23}}$ \\
\noalign{\smallskip}\hline
\end{tabular}
\end{table}

In addition, the rescaled UTs intuitively illustrate the relations among these observables.
We find that the magnitude of the CP-violating invariant ${\cal J}$ can be obtained directly from the CP-conserving quantities ${\cal R}^{}_{\alpha i}$ in the following ways~\cite{Harrison:2006bj,Luo:2023xmv}
\begin{eqnarray}
{\cal J}^{2}_{} & = & {\cal R}^{}_{e 1} \; {\cal R}^{}_{e 2} + {\cal R}^{}_{e 2} \; {\cal R}^{}_{e 3} + {\cal R}^{}_{e 3} \; {\cal R}^{}_{e 1} \; , \nonumber\\
& = & {\cal R}^{}_{\mu 1} \; {\cal R}^{}_{\mu 2} + {\cal R}^{}_{\mu 2} \; {\cal R}^{}_{\mu 3} + {\cal R}^{}_{\mu 3} \; {\cal R}^{}_{\mu 1} \; , \nonumber\\
& = & {\cal R}^{}_{\tau 1} \; {\cal R}^{}_{\tau 2} + {\cal R}^{}_{\tau 2} \; {\cal R}^{}_{\tau 3} + {\cal R}^{}_{\tau 3} \; {\cal R}^{}_{\tau 1} \; , \nonumber\\
& = & {\cal R}^{}_{e 1} \; {\cal R}^{}_{\mu 1} + {\cal R}^{}_{\mu 1} \; {\cal R}^{}_{\tau 1} + {\cal R}^{}_{\tau 1} \; {\cal R}^{}_{e 1} \; , \nonumber\\
& = & {\cal R}^{}_{e 2} \; {\cal R}^{}_{\mu 2} + {\cal R}^{}_{\mu 2} \; {\cal R}^{}_{\tau 2} + {\cal R}^{}_{\tau 2} \; {\cal R}^{}_{e 2} \; , \nonumber\\
& = & {\cal R}^{}_{e 3} \; {\cal R}^{}_{\mu 3} + {\cal R}^{}_{\mu 3} \; {\cal R}^{}_{\tau 3} + {\cal R}^{}_{\tau 3} \; {\cal R}^{}_{e 3} \; .
\label{11}
\end{eqnarray}
The above expressions for the Jarlskog invariant ${\cal J}$ are complementary to direct measurements of CP-violation.
Note that with ${\cal J}$ and ${\cal R}^{}_{\alpha i}$ independently measured from various neutrino oscillation experiments, the above relations also give a way to check the unitarity of the PMNS matrix~\cite{Luo:2023xmv} by verifying the consistencies of the different equations in Eq.~(\ref{11}).
Furthermore, if the quartets $|V^{}_{\alpha i}|^{2} |V^{}_{\beta i}|^{2}$ and/or $|V^{}_{\alpha i}|^{2} |V^{}_{\alpha j}|^{2}$ can be independently well determined, Eqs. (\ref{7}) and (\ref{8}) also provide a potential way to verify the unitarity of $V$.
Additionally, the uniformity of the Jarlskog invariant $\cal J$ measured from different oscillation experiments also serves as a straightforward test of the unitarity of the PMNS mixing matrix.

It has been well-known that the matrix of the squared moduli $|V^{}_{\alpha i}|^{2}$ essentially carries the same information as the unitary complex mixing matrix $V$ itself~\footnote{Of course, the information about the Majorana phases are not included in the squared moduli matrix of $V$.}.
Similarly, the matrix ${\cal R}$ contains entirely equivalent information.
Clearly, all these quartets can be expressed in terms of the squared modulus $|V^{}_{\alpha i}|^{2}$ in several ways \cite{Sasaki:1986jv}, for example,
\begin{eqnarray}
2 \; {\cal R}^{}_{\alpha i} & = & 1- |V^{}_{\beta j}|^{2} - |V^{}_{\beta k}|^{2} - |V^{}_{\gamma j}|^{2} - |V^{}_{\gamma k}|^{2} + |V^{}_{\beta j}|^{2} |V^{}_{\gamma k}|^{2} + |V^{}_{\beta k}|^{2} |V^{}_{\gamma j}|^{2} \nonumber\\
& = & |V^{}_{\alpha j}|^{2} |V^{}_{\alpha k}|^{2} - |V^{}_{\beta j}|^{2} |V^{}_{\beta k}|^{2} - |V^{}_{\gamma j}|^{2} |V^{}_{\gamma k}|^{2} \nonumber\\
& = & |V^{}_{\beta i}|^{2} |V^{}_{\gamma i}|^{2} - |V^{}_{\beta j}|^{2} |V^{}_{\gamma j}|^{2} - |V^{}_{\beta k}|^{2} |V^{}_{\gamma k}|^{2} \; , 
\label{12}
\end{eqnarray}
and
\begin{eqnarray}
{\cal J}^{2}_{} & = & |V^{}_{\beta j}|^{2} |V^{}_{\beta k}|^{2} |V^{}_{\gamma j}|^{2} |V^{}_{\gamma k}|^{2} - ( {\cal R}^{}_{\alpha i} )^{2}_{} \; .
\label{13}
\end{eqnarray}
On the contrary, if we have all these quartets available, the nine squared moduli $|V^{}_{\alpha i}|^{2}$ can also be reproduced using, for example, the following relations:
\begin{eqnarray}
|V^{}_{\alpha i}|^{2} ( 1- |V^{}_{\alpha i}|^{2} ) & = & - {\cal R}^{}_{\beta j} - {\cal R}^{}_{\beta k} - {\cal R}^{}_{\gamma j} - {\cal R}^{}_{\gamma k} \; , 
\label{14}
\end{eqnarray}
\begin{eqnarray}
|V^{}_{\alpha i}|^{4} & = & - \frac{({\cal R}^{}_{\beta j} + {\cal R}^{}_{\gamma j})({\cal R}^{}_{\beta k} + {\cal R}^{}_{\gamma k})}{{\cal R}^{}_{\beta i} + {\cal R}^{}_{\gamma i}} 
\; = \; - \frac{({\cal R}^{}_{\beta j} + {\cal R}^{}_{\beta k})({\cal R}^{}_{\gamma j} + {\cal R}^{}_{\gamma k})}{{\cal R}^{}_{\alpha j} + {\cal R}^{}_{\alpha k}} \; .
\label{15}
\end{eqnarray}
Again, in Eqs.~(\ref{12})-(\ref{15}) $\alpha$, $\beta$ and $\gamma$ run co-cyclically over $e$, $\mu$ and $\tau$, while $i$, $j$ and $k$ run co-cyclically over $1$, $2$ and $3$.

We can infer from Eq.~(\ref{11}) that among the ten rephasing invariants (nine ${\cal R}^{}_{\alpha i}$ and $\cal J$), only four are independent. Or in other words, from four independent rephasing invariants, the remains can be reproduced by solving above equations, {\bf however, always with degeneracies}.
This means that to pin down the ${\cal R}$ matrix and the Jarlskog ${\cal J}$ and therefore the unitary PMNS matrix, we need at least {\bf five} independent measurements of the quartets. The set of measurements could be, for example, i) ${\cal R}^{}_{\tau 1}$, ${\cal R}^{}_{\tau 2}$, ${\cal R}^{}_{\tau 3}$, ${\cal R}^{}_{\mu 2}$, ${\cal R}^{}_{\mu 3}$, or ii) ${\cal R}^{}_{\tau 1}$, ${\cal R}^{}_{\tau 2}$, $|V^{}_{e 1}|^{2} |V^{}_{e 2}|^{2}$, $|V^{}_{e 1}|^{2} |V^{}_{e 3}|^{2}$, ${\cal J}$, etc.
Although more quartets are required (compared to the moduli), it provides us with a large variety ways to reproduce the neutrino mixing, which could also be a path to the new physics by checking the consistency of results obtained in different ways.
It is also important to note that the sign of the CP-violating invariant ${\cal J}$ cannot be determined from those CP-conserving measurements such as ${\cal R}^{}_{\alpha i}$ or $|V^{}_{\alpha i}|^{2}$.
In order to pin down the sign of the Jarlskog invariant ${\cal J}$, direct measurement of CP-violating quantities is required.

Before wrapping up this section, we would like to give a brief discussion of the $\mu$-$\tau$ symmetry on the ${\cal R}$ matrix defined in Eq.~(\ref{5}).
It is well known that the observed pattern of lepton flavor mixing has an approximate $\mu$-$\tau$ flavor symmetry $| V^{}_{\mu 1} | \simeq | V^{}_{\tau 1} |$, $| V^{}_{\mu 2} | \simeq | V^{}_{\tau 2} |$, $| V^{}_{\mu 3} | \simeq | V^{}_{\tau 3} |$ (see, e.g. Refs.~\cite{Xing:2014zka, Xing:2015fdg}).
At the same time, there is also a commensurate approximate $\mu$-$\tau$ symmetry in the ${\cal R}$ matrix, which means ${\cal R}^{}_{\mu 1} \simeq {\cal R}^{}_{\tau 1}$, ${\cal R}^{}_{\mu 2} \simeq {\cal R}^{}_{\tau 2}$, ${\cal R}^{}_{\mu 3} \simeq {\cal R}^{}_{\tau 3}$, as can be seen from Table~\ref{t1}.

It has been pointed out in~\cite{Xing:2008fg} that the equality $| V^{}_{\mu i} | = | V^{}_{\tau i} |$ (for $i = 1, 2, 3$) holds exactly if either of the following two sets of conditions can be satisfied: i)  $\displaystyle \theta^{}_{23} = \pi / 4$, $\theta^{}_{13} = 0$; or ii) $\displaystyle \theta^{}_{23} = \pi / 4$, $\displaystyle \delta = \pm \pi / 2$. 
Under the same condition, the equality
\begin{eqnarray}
{\cal R}^{}_{\mu 1} \; = \; {\cal R}^{}_{\tau 1} \; , ~~~ {\cal R}^{}_{\mu 2} \; = \; {\cal R}^{}_{\tau 2} \; , ~~~ {\cal R}^{}_{\mu 3} \; = \; {\cal R}^{}_{\tau 3} \; ,
\label{16}
\end{eqnarray}
exactly holds as well.
Although the possibility of $\theta^{}_{13} = 0$ has been ruled out, $\displaystyle \theta^{}_{23} = \pi / 4$ and $\displaystyle \delta = - \pi / 2$ are both allowed at the $3\sigma$ level, which means that the exact $\mu$-$\tau$ symmetry of $|V|$ at the electroweak energy scale is still allowed by current experimental data. 

The $\mu$-$\tau$ symmetry of the ${\cal R}$ matrix can also be reflected in the (rescaled) unitarity triangles.
From Figs.~\ref{f1} - \ref{f4}, the exact $\mu$-$\tau$ symmetry of the matrix ${\cal R}$ indicates that $\triangle^{}_{\mu} \cong \triangle^{}_{\tau}$, and $\triangle^{}_{1}$, $\triangle^{}_{2}$, $\triangle^{}_{3}$ are all isosceles triangles.

To see the $\mu$-$\tau$ symmetry in ${\cal R}$ and the strength of its breaking more transparently, we can define 
\begin{eqnarray}
{\cal R}^{}_{1} \; \equiv \; \frac{1}{2} \left ( {\cal R}^{}_{\mu 1} + {\cal R}^{}_{\tau 1} \right ) \; , ~~~~~ {\cal R}^{}_{2} \; \equiv \; \frac{1}{2} \left ( {\cal R}^{}_{\mu 2} + {\cal R}^{}_{\tau 2} \right ) \; , ~~~~~ {\cal R}^{}_{3} \; \equiv \; \frac{1}{2} \left ( {\cal R}^{}_{\mu 3} + {\cal R}^{}_{\tau 3} \right ) \; , \nonumber\\[2mm]
\Delta {\cal R}^{}_{1} \; \equiv \; \frac{1}{2} \left ( {\cal R}^{}_{\tau 1} - {\cal R}^{}_{\mu 1} \right ) \; , ~~~ \Delta {\cal R}^{}_{2} \; \equiv \; \frac{1}{2} \left ( {\cal R}^{}_{\tau 2} - {\cal R}^{}_{\mu 2} \right ) \; , ~~~ \Delta {\cal R}^{}_{3} \; \equiv \; \frac{1}{2} \left ( {\cal R}^{}_{\tau 3} - {\cal R}^{}_{\mu 3} \right ) \; ,
\label{18}
\end{eqnarray}
which means
\begin{eqnarray}
{\cal R}^{}_{\mu 1} \; = \; {\cal R}^{}_{1} - \Delta {\cal R}^{}_{1} \; , ~~~ {\cal R}^{}_{\mu 2} \; = \; {\cal R}^{}_{2} - \Delta {\cal R}^{}_{2} \; , ~~~ {\cal R}^{}_{\mu 3} \; = \; {\cal R}^{}_{3} - \Delta {\cal R}^{}_{3} \; , \nonumber\\
{\cal R}^{}_{\tau 1} \; = \; {\cal R}^{}_{1} + \Delta {\cal R}^{}_{1} \; , ~~~ {\cal R}^{}_{\tau 2} \; = \; {\cal R}^{}_{2} + \Delta {\cal R}^{}_{2} \; , ~~~ {\cal R}^{}_{\tau 3} \; = \; {\cal R}^{}_{3} + \Delta {\cal R}^{}_{3} \; . \;
\label{19}
\end{eqnarray}
Apparently, the exact $\mu$-$\tau$ symmetry of the matrix ${\cal R}$ is maintained if $\Delta {\cal R}^{}_{1} = \Delta {\cal R}^{}_{2} = \Delta {\cal R}^{}_{3} = 0$.
We can infer from Eq.~(\ref{11}) that
\begin{eqnarray}
{\cal J}^{2}_{} & = & {\cal R}^{}_{1} \; {\cal R}^{}_{2} + {\cal R}^{}_{2} \; {\cal R}^{}_{3} + {\cal R}^{}_{3} \; {\cal R}^{}_{1} + \Delta {\cal R}^{}_{1} \Delta {\cal R}^{}_{2} + \Delta {\cal R}^{}_{2} \Delta {\cal R}^{}_{3} + \Delta {\cal R}^{}_{3} \Delta {\cal R}^{}_{1} \; , 
\label{20}
\end{eqnarray}
together with
\begin{eqnarray}
\Delta {\cal R}^{}_{1} ( {\cal R}^{}_{2} + {\cal R}^{}_{3} ) + \Delta {\cal R}^{}_{2} ( {\cal R}^{}_{3} + {\cal R}^{}_{1} ) + \Delta {\cal R}^{}_{3} ( {\cal R}^{}_{1} + {\cal R}^{}_{2} ) & = & 0 \; ,
\label{21}
\end{eqnarray}
and
\begin{eqnarray}
{\cal J}^{2}_{} & = & {\cal R}^{}_{e 1} \; {\cal R}^{}_{1} + {\cal R}^{2}_{1} - \Delta {\cal R}^{2}_{1} \; , \nonumber\\
& = & {\cal R}^{}_{e 2} \; {\cal R}^{}_{2} + {\cal R}^{2}_{2} - \Delta {\cal R}^{2}_{2} \; , \nonumber\\
& = & {\cal R}^{}_{e 3} \; {\cal R}^{}_{3} + {\cal R}^{2}_{3} - \Delta {\cal R}^{2}_{3} \; .
\label{22}
\end{eqnarray}

\begin{table}[]
\footnotesize
\caption{The best-fit values, 1$\sigma$ and 3$\sigma$ ranges of ${\cal R}^{}_{i}$ and $\Delta{\cal R}^{}_{i}$ (for $i=1, 2, 3$), where we've adopt the global fit of the unitary PMNS leptonic mixing matrix given in~\cite{Esteban:2024eli,Nufit}.}
\label{t3} 
\begin{tabular}{ccccccccccccc}
\hline\noalign{\smallskip}
&& \multicolumn{5}{c}{Normal Neutrino Mass Ordering} && \multicolumn{5}{c}{Inverted Neutrino Mass Ordering}  \\
&~~~& best-fit &~& 1$\sigma$ range &~& 3$\sigma$ range &~~~& best-fit &~& 1$\sigma$ range &~& 3$\sigma$ range \\
\noalign{\smallskip}\hline\noalign{\smallskip}
${\cal R}^{}_{1}$ && -0.0033 && -0.0035 $\sim$ -0.0031 && -0.0040 $\sim$ -0.0027 && -0.0034 && -0.0036 $\sim$ -0.0032 && -0.0041 $\sim$ -0.0028 \\
${\cal R}^{}_{2}$ && -0.0075 && -0.0078 $\sim$ -0.0072 && -0.0085 $\sim$ -0.0065 && -0.0076 && -0.0079 $\sim$ -0.0072 && -0.0085 $\sim$ -0.0066 \\
${\cal R}^{}_{3}$ && -0.1018 && -0.1040 $\sim$ -0.0996 && -0.1084 $\sim$ -0.0950 && -0.1018 && -0.1040 $\sim$ -0.0995 && -0.1083 $\sim$ -0.0949 \\
\noalign{\smallskip}
$\Delta{\cal R}^{}_{1}$ && -0.0282 && -0.0342 $\sim$ -0.0171 && -0.0360 $\sim$ 0.0360 && 0.0020 && -0.0127 $\sim$ 0.0148 && -0.0338 $\sim$ 0.0329 \\
$\Delta{\cal R}^{}_{2}$ && 0.0289 && 0.0176 $\sim$ 0.0348 && -0.0365 $\sim$ 0.0364 && -0.0031 && -0.0160 $\sim$ 0.0118 && -0.0335 $\sim$ 0.0343 \\
$\Delta{\cal R}^{}_{3}$ && 0.0046 && -0.0031 $\sim$ 0.0111 && -0.0278 $\sim$ 0.0319 && 0.0099 && 0.0013 $\sim$ 0.0195 && -0.0258 $\sim$ 0.0306 \\
\noalign{\smallskip}\hline
\end{tabular}
\end{table}

Again, to give a ballpark figure of the $\mu$-$\tau$ symmetry of ${\cal R}$ and the strength of its breaking, we illustrate in Table~\ref{t3} the best-fit values, 1$\sigma$ and 3$\sigma$ ranges of three ${\cal R}^{}_{i}$ and three corresponding $\Delta {\cal R}^{}_{i}$ (for $i=1, 2, 3$) in both the NMO and IMO scenarios.
The expressions of all ${\cal R}^{}_{i}$ and $\Delta {\cal R}^{}_{i}$ (for $i=1, 2, 3$) using the standard parameterization can be written as
\begin{eqnarray}
{\cal R}^{}_{1} & = & - \frac{1}{2} |V^{}_{e 2}|^{2} |V^{}_{e 3}|^{2} \; = \; - \frac{1}{8} \sin^{2}\theta^{}_{12} \sin^{2}2\theta^{}_{13} \; , \nonumber\\[1mm]
{\cal R}^{}_{2} & = & - \frac{1}{2} |V^{}_{e 1}|^{2} |V^{}_{e 3}|^{2} \; = \; - \frac{1}{8} \cos^{2}\theta^{}_{12} \sin^{2}2\theta^{}_{13} \; , \nonumber\\[1mm]
{\cal R}^{}_{3} & = & - \frac{1}{2} |V^{}_{e 1}|^{2} |V^{}_{e 2}|^{2} \; = \; - \frac{1}{8} \sin^{2}2\theta^{}_{12} \cos^{4}\theta^{}_{13} \; , \nonumber\\[2mm]
\Delta {\cal R}^{}_{1} & = & \frac{1}{8} \sin^{2}\theta^{}_{12} \cos2\theta^{}_{23} \sin^{2}2\theta^{}_{13} + \frac{1}{4} \sin2\theta^{}_{12} \sin2\theta^{}_{23} \cos^{2}\theta^{}_{13} \sin\theta^{}_{13} \cos\delta \; , \nonumber\\[1mm]
\Delta {\cal R}^{}_{2} & = & \frac{1}{8} \cos^{2}\theta^{}_{12} \cos2\theta^{}_{23} \sin^{2}2\theta^{}_{13} - \frac{1}{4} \sin2\theta^{}_{12} \sin2\theta^{}_{23} \cos^{2}\theta^{}_{13} \sin\theta^{}_{13} \cos\delta \; , \nonumber\\[1mm]
\Delta {\cal R}^{}_{3} & = & - \sin2\theta^{}_{12} \cos^{2}\theta^{}_{13} [ \frac{1}{8} \sin2\theta^{}_{12} \cos2\theta^{}_{23} ( 1 + \sin^{2}\theta^{}_{13} ) + \frac{1}{4} \cos2\theta^{}_{12} \sin2\theta^{}_{23} \sin\theta^{}_{13} \cos\delta ] \; . \nonumber\\
\label{23}
\end{eqnarray}
Thanks to the precise measurements of the first row of $V$ (specifically, the values of $\theta^{}_{12}$ and $\theta^{}_{13}$), we can determine ${\cal R}^{}_{i}$ (for $i = 1, 2, 3$) with a reasonably good level of accuracy.
However, most of the uncertainties remain in $\Delta {\cal R}^{}_{i}$ (for $i = 1, 2, 3$).
On the experimental side, it is imperative to measure $\theta^{}_{23}$ and $\delta$ as accurately as possible.
Therefore, measurements of the strength of $\mu$-$\tau$ symmetry breaking of ${\cal R}$, that is, the magnitudes of three $\Delta {\cal R}^{}_{i}$ are crucially important.

\section{Matter Effect}

We now proceed to discuss the matter effect on ${\cal J}$ and ${\cal R}^{}_{\alpha i}$, as these effects are inevitable in many realistic long- and medium-baseline neutrino oscillation experiments.
In the standard three-neutrino framework, the effective Hamiltonian $\widetilde{\cal{H}}$ in the flavor basis responsible for the propagation of neutrinos in matter is an interplay between the vacuum Hamiltonian ${\cal H}$ and the matter term ${\cal H}'$,
\begin{eqnarray}
\widetilde{\cal H} & = & {\cal H} + {\cal H}' \; = \; \frac{1}{2 E} \left [ V \left ( \begin{matrix} m^{2}_{1} & & \cr & m^{2}_{2} & \cr & & m^{2}_{3} \cr \end{matrix} \right ) V^{\dagger}_{} + \left ( \begin{matrix} A^{}_{\rm CC} + A^{}_{\rm NC} & & \cr & A^{}_{\rm NC} & \cr & & A^{}_{\rm NC} \cr \end{matrix} \right ) \right ] \nonumber\\[2mm]
& = & \frac{1}{2 E} \left [ \left ( m^{2}_{1} + A^{}_{\rm NC} \right ) \left ( \begin{matrix} ~ 1 ~ & & \cr & 1 & \cr & & ~ 1 ~ \cr \end{matrix} \right ) + V \left ( \begin{matrix} 0 & & \cr & \Delta^{}_{21} & \cr & & \Delta^{}_{31} \cr \end{matrix} \right ) V^{\dagger}_{} + \left ( \begin{matrix} A^{}_{\rm CC} & & \cr & 0 & \cr & & ~ 0 ~ \cr \end{matrix} \right ) \right ] \; ,
\label{24}
\end{eqnarray}
where $\cal{H}' $ describes the forward coherent scattering of neutrinos with the constituents of the medium via the weak charged-current (CC) and neutral-current (NC) interactions~\cite{Wolfenstein:1977ue, Mikheyev:1985zog}.
Here $A^{}_{\rm CC} = 2 E V^{}_{\rm CC}$, $A^{}_{\rm NC} = 2 E V^{}_{\rm NC}$ (with $V^{}_{\rm CC} = \sqrt{2} G^{}_{\rm F} N^{}_{e}$ and $\displaystyle V^{}_{\rm NC} = - \frac{\sqrt{2}}{2} G^{}_{\rm F} N^{}_{n}$) are parameters that measure the strength of the matter effect, that of the same unit as the mass-squared difference $\Delta^{}_{ji} \equiv m^{2}_{j} - m^{2}_{i}$, and $V$ is just the $3 \times 3$ unitary PMNS leptonic mixing matrix.
For antineutrino oscillations in matter, one can simply replace $V$ by $V^{*}_{}$ and $A^{}_{\rm CC}$ by $-A^{}_{\rm CC}$ in the effective Hamiltonian. 
Note that the diagonal term in $\widetilde{\cal{H}}$ develops just a common phase for all three flavors and does not affect the neutrino oscillation behaviors, which means that the absolute mass scale and the NC interactions of neutrinos are irrelevant to the oscillation behaviors in matter in this standard picture.
For simplicity, we neglect the terms of $A^{}_{\rm NC}$ in the discussion hereafter.

The effective Hamiltonian $\widetilde{\cal{H}}$ can be diagonalized through a unitary transformation
\begin{eqnarray}
\widetilde{\cal{H}} & \equiv & \frac{1}{2 E} \widetilde{V} \left ( \begin{matrix} \widetilde{m}^{2}_{1} & & \cr & \widetilde{m}^{2}_{2} & \cr & & \widetilde{m}^{2}_{3} \cr \end{matrix} \right ) \widetilde{V}^{\dagger}_{} \; ,
\label{25}
\end{eqnarray}
where the effective neutrino masses $\widetilde{m}^{}_{i}$ (for $i = 1, 2, 3$) and flavor mixing matrix $\widetilde{V}$ in matter have been defined. 
The neutrino oscillation probabilities in matter can be written in the same way as those in vacuum (see Eqs.~(\ref{2}) and (\ref{3})) simply by using the effective invariants $\widetilde{\cal R}^{}_{\alpha i}$ and $\widetilde{\cal J}$ together with the effective mass-squared differences $\widetilde{\Delta}^{}_{ji} \equiv \widetilde{m}^{2}_{j} - \widetilde{m}^{2}_{i}$
\begin{eqnarray}
P ( \stackrel{(-)}{\nu}^{}_{\alpha} \rightarrow  \stackrel{(-)}{\nu}^{}_{\alpha} ) & = & 1 + 4 \sum^{}_{j > i} \left ( \widetilde{\cal R}^{}_{\beta k} + \widetilde{\cal R}^{}_{\gamma k} \right ) \sin^2 \frac{\widetilde{\Delta}^{}_{ji} L}{4 E} \; , \\
\label{26}
P ( \stackrel{(-)}{\nu}^{}_{\alpha} \rightarrow  \stackrel{(-)}{\nu}^{}_{\beta} ) & = & - 4 \sum^{}_{j > i} \widetilde{\cal R}^{}_{\gamma k} \sin^2 \frac{\Delta^{}_{ji} L}{4 E} \mp 8 \widetilde{\cal J} \prod^{}_{j > i} \sin \frac{\widetilde{\Delta}^{}_{ji} L}{4 E} \; , 
\label{27}
\end{eqnarray}
where nine effective $\widetilde{\cal R}^{}_{\alpha i}$ and the effective Jarlskog $\widetilde{\cal J}$ in matter are constructed with the effective mixing matrix $\widetilde{V}$ using the same definitions given by Eqs.~(\ref{4}) and (\ref{6}).

The effective Jarlskog $\widetilde{\cal J}$ in matter is related to the vacuum Jarlskog ${\cal J}$ via the well-known Naumov relation~\cite{Naumov:1991ju}
\begin{eqnarray}
\widetilde{\Delta}^{}_{21} \widetilde{\Delta}^{}_{31} \widetilde{\Delta}^{}_{32} \; \widetilde{\cal J} & = & \Delta^{}_{21} \Delta^{}_{31} \Delta^{}_{32} \; {\cal J} \; .
\label{28}
\end{eqnarray}
Similarly, we find that $\widetilde{\cal R}^{}_{\alpha i}$ and ${\cal R}^{}_{\alpha i}$ are related to each other through the following Naumov-like relations
\begin{eqnarray}
\widetilde{\Delta}^{}_{21} \widetilde{\Delta}^{}_{31} \widetilde{\Delta}^{2}_{32} \; \widetilde{\cal R}^{}_{e 1} & = & \Delta^{2}_{21} (m^{2}_{3} - \widetilde{m}^{2}_{2} + A^{}_{\rm CC}) (m^{2}_{3} - \widetilde{m}^{2}_{3} + A^{}_{\rm CC}) \; {\cal R}^{}_{e 3} \nonumber\\
& & + \Delta^{2}_{31} (m^{2}_{2} - \widetilde{m}^{2}_{2} + A^{}_{\rm CC}) (m^{2}_{2} - \widetilde{m}^{2}_{3} + A^{}_{\rm CC}) \; {\cal R}^{}_{e 2} \nonumber\\
& & + \Delta^{2}_{32} (m^{2}_{1} - \widetilde{m}^{2}_{2} + A^{}_{\rm CC}) (m^{2}_{1} - \widetilde{m}^{2}_{3} + A^{}_{\rm CC}) \; {\cal R}^{}_{e 1} \; , \nonumber\\
- \widetilde{\Delta}^{}_{21} \widetilde{\Delta}^{2}_{31} \widetilde{\Delta}^{}_{32} \; \widetilde{\cal R}^{}_{e 2} & = & \Delta^{2}_{21} (m^{2}_{3} - \widetilde{m}^{2}_{1} + A^{}_{\rm CC}) (m^{2}_{3} - \widetilde{m}^{2}_{3} + A^{}_{\rm CC}) \; {\cal R}^{}_{e 3} \nonumber\\
& & + \Delta^{2}_{31} (m^{2}_{2} - \widetilde{m}^{2}_{1} + A^{}_{\rm CC}) (m^{2}_{2} - \widetilde{m}^{2}_{3} + A^{}_{\rm CC}) \; {\cal R}^{}_{e 2} \nonumber\\
& & + \Delta^{2}_{32} (m^{2}_{1} - \widetilde{m}^{2}_{1} + A^{}_{\rm CC}) (m^{2}_{1} - \widetilde{m}^{2}_{3} + A^{}_{\rm CC}) \; {\cal R}^{}_{e 1} \; , \nonumber\\
\widetilde{\Delta}^{2}_{21} \widetilde{\Delta}^{}_{31} \widetilde{\Delta}^{}_{32} \; \widetilde{\cal R}^{}_{e 3} & = & \Delta^{2}_{21} (m^{2}_{3} - \widetilde{m}^{2}_{1} + A^{}_{\rm CC}) (m^{2}_{3} - \widetilde{m}^{2}_{2} + A^{}_{\rm CC}) \; {\cal R}^{}_{e 3} \nonumber\\
& & + \Delta^{2}_{31} (m^{2}_{2} - \widetilde{m}^{2}_{1} + A^{}_{\rm CC}) (m^{2}_{2} - \widetilde{m}^{2}_{2} + A^{}_{\rm CC}) \; {\cal R}^{}_{e 2} \nonumber\\
& & + \Delta^{2}_{32} (m^{2}_{1} - \widetilde{m}^{2}_{1} + A^{}_{\rm CC}) (m^{2}_{1} - \widetilde{m}^{2}_{2} + A^{}_{\rm CC}) \; {\cal R}^{}_{e 1} \; , \nonumber\\
\widetilde{\Delta}^{}_{21} \widetilde{\Delta}^{}_{31} \widetilde{\Delta}^{2}_{32} \; \widetilde{\cal R}^{}_{\mu 1} & = & \Delta^{2}_{21} (m^{2}_{3} - \widetilde{m}^{2}_{2}) (m^{2}_{3} - \widetilde{m}^{2}_{3}) \; {\cal R}^{}_{\mu 3} \nonumber\\
& & + \Delta^{2}_{31} (m^{2}_{2} - \widetilde{m}^{2}_{2}) (m^{2}_{2} - \widetilde{m}^{2}_{3}) \; {\cal R}^{}_{\mu 2} \nonumber\\
& & + \Delta^{2}_{32} (m^{2}_{1} - \widetilde{m}^{2}_{2}) (m^{2}_{1} - \widetilde{m}^{2}_{3}) \; {\cal R}^{}_{\mu 1} \; , \nonumber\\
- \widetilde{\Delta}^{}_{21} \widetilde{\Delta}^{2}_{31} \widetilde{\Delta}^{}_{32} \; \widetilde{\cal R}^{}_{\mu 2} & = & \Delta^{2}_{21} (m^{2}_{3} - \widetilde{m}^{2}_{1}) (m^{2}_{3} - \widetilde{m}^{2}_{3}) \; {\cal R}^{}_{\mu 3} \nonumber\\
& & + \Delta^{2}_{31} (m^{2}_{2} - \widetilde{m}^{2}_{1}) (m^{2}_{2} - \widetilde{m}^{2}_{3}) \; {\cal R}^{}_{\mu 2} \nonumber\\
& & + \Delta^{2}_{32} (m^{2}_{1} - \widetilde{m}^{2}_{1}) (m^{2}_{1} - \widetilde{m}^{2}_{3}) \; {\cal R}^{}_{\mu 1} \; , \nonumber\\
\widetilde{\Delta}^{2}_{21} \widetilde{\Delta}^{}_{31} \widetilde{\Delta}^{}_{32} \; \widetilde{\cal R}^{}_{\mu 3} & = & \Delta^{2}_{21} (m^{2}_{3} - \widetilde{m}^{2}_{1}) (m^{2}_{3} - \widetilde{m}^{2}_{2}) \; {\cal R}^{}_{\mu 3} \nonumber\\
& & + \Delta^{2}_{31} (m^{2}_{2} - \widetilde{m}^{2}_{1}) (m^{2}_{2} - \widetilde{m}^{2}_{2}) \; {\cal R}^{}_{\mu 2} \nonumber\\
& & + \Delta^{2}_{32} (m^{2}_{1} - \widetilde{m}^{2}_{1}) (m^{2}_{1} - \widetilde{m}^{2}_{2}) \; {\cal R}^{}_{\mu 1} \; , \nonumber\\
\widetilde{\Delta}^{}_{21} \widetilde{\Delta}^{}_{31} \widetilde{\Delta}^{2}_{32} \; \widetilde{\cal R}^{}_{\tau 1} & = & \Delta^{2}_{21} (m^{2}_{3} - \widetilde{m}^{2}_{2}) (m^{2}_{3} - \widetilde{m}^{2}_{3}) \; {\cal R}^{}_{\tau 3} \nonumber\\
& & + \Delta^{2}_{31} (m^{2}_{2} - \widetilde{m}^{2}_{2}) (m^{2}_{2} - \widetilde{m}^{2}_{3}) \; {\cal R}^{}_{\tau 2} \nonumber\\
& & + \Delta^{2}_{32} (m^{2}_{1} - \widetilde{m}^{2}_{2}) (m^{2}_{1} - \widetilde{m}^{2}_{3}) \; {\cal R}^{}_{\tau 1} \; , \nonumber\\
- \widetilde{\Delta}^{}_{21} \widetilde{\Delta}^{2}_{31} \widetilde{\Delta}^{}_{32} \; \widetilde{\cal R}^{}_{\tau 2} & = & \Delta^{2}_{21} (m^{2}_{3} - \widetilde{m}^{2}_{1}) (m^{2}_{3} - \widetilde{m}^{2}_{3}) \; {\cal R}^{}_{\tau 3} \nonumber\\
& & + \Delta^{2}_{31} (m^{2}_{2} - \widetilde{m}^{2}_{1}) (m^{2}_{2} - \widetilde{m}^{2}_{3}) \; {\cal R}^{}_{\tau 2} \nonumber\\
& & + \Delta^{2}_{32} (m^{2}_{1} - \widetilde{m}^{2}_{1}) (m^{2}_{1} - \widetilde{m}^{2}_{3}) \; {\cal R}^{}_{\tau 1} \; , \nonumber\\
\widetilde{\Delta}^{2}_{21} \widetilde{\Delta}^{}_{31} \widetilde{\Delta}^{}_{32} \; \widetilde{\cal R}^{}_{\tau 3} & = & \Delta^{2}_{21} (m^{2}_{3} - \widetilde{m}^{2}_{1}) (m^{2}_{3} - \widetilde{m}^{2}_{2}) \; {\cal R}^{}_{\tau 3} \nonumber\\
& & + \Delta^{2}_{31} (m^{2}_{2} - \widetilde{m}^{2}_{1}) (m^{2}_{2} - \widetilde{m}^{2}_{2}) \; {\cal R}^{}_{\tau 2} \nonumber\\
& & + \Delta^{2}_{32} (m^{2}_{1} - \widetilde{m}^{2}_{1}) (m^{2}_{1} - \widetilde{m}^{2}_{2}) \; {\cal R}^{}_{\tau 1} \; .
\label{29}
\end{eqnarray}
The Naumov-like relations for the invariants ${\cal R}^{}_{\alpha i}$ can be derived through a lengthy yet straightforward calculation by comparing the squared moduli of the corresponding matrix elements in Eq.~(\ref{24}) and Eq.~(\ref{25}).
Explicit expressions for $\widetilde{\Delta}^{}_{ji}$ and $m^{2}_{j} - \widetilde{m}^{2}_{i}$ have been given, for example, in Refs.~\cite{Barger:1980tf, Xing:2000gg, Xing:2003ez}.
Note that both $\widetilde{\Delta}^{}_{ji}$ and $m^{2}_{j} - \widetilde{m}^{2}_{i}$ are independent of the absolute neutrino masses and are neatly determined by the three mass-squared differences $\Delta^{}_{ji}$, the structure of the PMNS matrix $V$ in vacuum, and the strength of the CC matter potential $A^{}_{\rm CC}$.
The formulas derived above apply to both the normal and inverted neutrino mass orderings, although the resulting effective neutrino masses may differ between these scenarios.
Also note that these formulas are dealing with parameters related to neutrinos propagating in vacuum and in matter.
They will also apply to antineutrinos if replacement $A^{}_{\rm CC} \rightarrow -A^{}_{\rm CC}$ is made.

We can clearly see that the effective $\widetilde{\cal R}^{}_{\alpha i}$ in matter can be regarded as a linear combination of ${\cal R}^{}_{\alpha 1}$, ${\cal R}^{}_{\alpha 2}$, and ${\cal R}^{}_{\alpha 3}$.
Therefore, we can rearrange Eq.~(\ref{29}) as 
\begin{eqnarray}
\left ( \begin{matrix} \widetilde{\cal R}^{}_{e 1} \cr \widetilde{\cal R}^{}_{e 2} \cr \widetilde{\cal R}^{}_{e 3} \end{matrix} \right ) & = & X^{e}_{} \; \left ( \begin{matrix} {\cal R}^{}_{e 1} \cr {\cal R}^{}_{e 2} \cr {\cal R}^{}_{e 3} \end{matrix} \right ) \;= \; \frac{\Delta^{}_{21} \Delta^{}_{31} \Delta^{}_{32}}{\widetilde{\Delta}^{}_{21} \widetilde{\Delta}^{}_{31} \widetilde{\Delta}^{}_{32}} \; Y^{e}_{} \; \left ( \begin{matrix} {\cal R}^{}_{e 1} \cr {\cal R}^{}_{e 2} \cr {\cal R}^{}_{e 3} \end{matrix} \right ) \; , \nonumber\\[2mm]
\left ( \begin{matrix} \widetilde{\cal R}^{}_{\mu 1} \cr \widetilde{\cal R}^{}_{\mu 2} \cr \widetilde{\cal R}^{}_{\mu 3} \end{matrix} \right ) & = & X^{\mu \tau}_{} \; \left ( \begin{matrix} {\cal R}^{}_{\mu 1} \cr {\cal R}^{}_{\mu 2} \cr {\cal R}^{}_{\mu 3} \end{matrix} \right ) \; = \; \frac{\Delta^{}_{21} \Delta^{}_{31} \Delta^{}_{32}}{\widetilde{\Delta}^{}_{21} \widetilde{\Delta}^{}_{31} \widetilde{\Delta}^{}_{32}} \; Y^{\mu \tau}_{} \; \left ( \begin{matrix} {\cal R}^{}_{\mu 1} \cr {\cal R}^{}_{\mu 2} \cr {\cal R}^{}_{\mu 3} \end{matrix} \right ) \; , \nonumber\\[2mm]
\left ( \begin{matrix} \widetilde{\cal R}^{}_{\tau 1} \cr \widetilde{\cal R}^{}_{\tau 2} \cr \widetilde{\cal R}^{}_{\tau 3} \end{matrix} \right ) & = & X^{\mu \tau}_{} \; \left ( \begin{matrix} {\cal R}^{}_{\tau 1} \cr {\cal R}^{}_{\tau 2} \cr {\cal R}^{}_{\tau 3} \end{matrix} \right ) \; = \; \frac{\Delta^{}_{21} \Delta^{}_{31} \Delta^{}_{32}}{\widetilde{\Delta}^{}_{21} \widetilde{\Delta}^{}_{31} \widetilde{\Delta}^{}_{32}} \; Y^{\mu \tau}_{} \; \left ( \begin{matrix} {\cal R}^{}_{\tau 1} \cr {\cal R}^{}_{\tau 2} \cr {\cal R}^{}_{\tau 3} \end{matrix} \right ) \; ,
\label{30}
\end{eqnarray}
where in the second term of each equation, we have drawn a common coefficient $\displaystyle \frac{\widetilde{\cal J}}{\cal J} = \frac{\Delta^{}_{21} \Delta^{}_{31} \Delta^{}_{32}}{\widetilde{\Delta}^{}_{21} \widetilde{\Delta}^{}_{31} \widetilde{\Delta}^{}_{32}}$.
 The composition matrices $X^{e}_{}$ and $X^{\mu\tau}_{}$ can be easily inferred from Eq.~(\ref{29}), here we just write down the expressions for the composition matrices $Y^{e}_{}$ and $Y^{\mu\tau}_{}$ 
\begin{eqnarray}
Y^{e}_{} & = & \left ( \begin{matrix} \frac{\Delta^{}_{32} (m^{2}_{1} - \widetilde{m}^{2}_{2} + A^{}_{\rm CC}) (m^{2}_{1} - \widetilde{m}^{2}_{3} + A^{}_{\rm CC})}{\widetilde{\Delta}^{}_{32} \Delta^{}_{21} \Delta^{}_{31}} & \frac{\Delta^{}_{31} (m^{2}_{2} - \widetilde{m}^{2}_{2} + A^{}_{\rm CC}) (m^{2}_{2} - \widetilde{m}^{2}_{3} + A^{}_{\rm CC})}{\widetilde{\Delta}^{}_{32} \Delta^{}_{21} \Delta^{}_{32}} & \frac{\Delta^{}_{21} (m^{2}_{3} - \widetilde{m}^{2}_{2} + A^{}_{\rm CC}) (m^{2}_{3} - \widetilde{m}^{2}_{3} + A^{}_{\rm CC})}{\widetilde{\Delta}^{}_{32} \Delta^{}_{31} \Delta^{}_{32}} \cr \frac{\Delta^{}_{32} (m^{2}_{1} - \widetilde{m}^{2}_{1} + A^{}_{\rm CC}) (m^{2}_{1} - \widetilde{m}^{2}_{3} + A^{}_{\rm CC})}{- \widetilde{\Delta}^{}_{31} \Delta^{}_{21} \Delta^{}_{31}} & \frac{\Delta^{}_{31} (m^{2}_{2} - \widetilde{m}^{2}_{1} + A^{}_{\rm CC}) (m^{2}_{2} - \widetilde{m}^{2}_{3} + A^{}_{\rm CC})}{- \widetilde{\Delta}^{}_{31} \Delta^{}_{21} \Delta^{}_{32}} & \frac{\Delta^{}_{21} (m^{2}_{3} - \widetilde{m}^{2}_{1} + A^{}_{\rm CC}) (m^{2}_{3} - \widetilde{m}^{2}_{3} + A^{}_{\rm CC})}{- \widetilde{\Delta}^{}_{31} \Delta^{}_{31} \Delta^{}_{32}}  \cr \frac{\Delta^{}_{32} (m^{2}_{1} - \widetilde{m}^{2}_{1} + A^{}_{\rm CC}) (m^{2}_{1} - \widetilde{m}^{2}_{2} + A^{}_{\rm CC})} {\widetilde{\Delta}^{}_{21} \Delta^{}_{21} \Delta^{}_{31}} & \frac{\Delta^{}_{31} (m^{2}_{2} - \widetilde{m}^{2}_{1} + A^{}_{\rm CC}) (m^{2}_{2} - \widetilde{m}^{2}_{2} + A^{}_{\rm CC})}{\widetilde{\Delta}^{}_{21} \Delta^{}_{21} \Delta^{}_{32}} & \frac{\Delta^{}_{21} (m^{2}_{3} - \widetilde{m}^{2}_{1} + A^{}_{\rm CC}) (m^{2}_{3} - \widetilde{m}^{2}_{2} + A^{}_{\rm CC})}{\widetilde{\Delta}^{}_{21} \Delta^{}_{31} \Delta^{}_{32}} \end{matrix} \right ) \; , \nonumber\\[2mm]
Y^{\mu\tau}_{} & = & \left ( \begin{matrix} \frac{\Delta^{}_{32} (m^{2}_{1} - \widetilde{m}^{2}_{2}) (m^{2}_{1} - \widetilde{m}^{2}_{3})}{\widetilde{\Delta}^{}_{32} \Delta^{}_{21} \Delta^{}_{31}} & \frac{\Delta^{}_{31} (m^{2}_{2} - \widetilde{m}^{2}_{2}) (m^{2}_{2} - \widetilde{m}^{2}_{3})}{\widetilde{\Delta}^{}_{32} \Delta^{}_{21} \Delta^{}_{32}} & \frac{\Delta^{}_{21} (m^{2}_{3} - \widetilde{m}^{2}_{2}) (m^{2}_{3} - \widetilde{m}^{2}_{3})}{\widetilde{\Delta}^{}_{32} \Delta^{}_{31} \Delta^{}_{32}} \cr \frac{\Delta^{}_{32} (m^{2}_{1} - \widetilde{m}^{2}_{1}) (m^{2}_{1} - \widetilde{m}^{2}_{3})}{- \widetilde{\Delta}^{}_{31} \Delta^{}_{21} \Delta^{}_{31}} & \frac{\Delta^{}_{31} (m^{2}_{2} - \widetilde{m}^{2}_{1}) (m^{2}_{2} - \widetilde{m}^{2}_{3})}{- \widetilde{\Delta}^{}_{31} \Delta^{}_{21} \Delta^{}_{32}} & \frac{\Delta^{}_{21} (m^{2}_{3} - \widetilde{m}^{2}_{1}) (m^{2}_{3} - \widetilde{m}^{2}_{3})}{- \widetilde{\Delta}^{}_{31} \Delta^{}_{31} \Delta^{}_{32}}  \cr \frac{\Delta^{}_{32} (m^{2}_{1} - \widetilde{m}^{2}_{1}) (m^{2}_{1} - \widetilde{m}^{2}_{2})} {\widetilde{\Delta}^{}_{21} \Delta^{}_{21} \Delta^{}_{31}} & \frac{\Delta^{}_{31} (m^{2}_{2} - \widetilde{m}^{2}_{1}) (m^{2}_{2} - \widetilde{m}^{2}_{2})}{\widetilde{\Delta}^{}_{21} \Delta^{}_{21} \Delta^{}_{32}} & \frac{\Delta^{}_{21} (m^{2}_{3} - \widetilde{m}^{2}_{1}) (m^{2}_{3} - \widetilde{m}^{2}_{2})}{\widetilde{\Delta}^{}_{21} \Delta^{}_{31} \Delta^{}_{32}} \end{matrix} \right ) \; .
\label{31}
\end{eqnarray}
Although both $X^{e, \; \mu\tau}_{}$ and $Y^{e, \; \mu\tau}_{}$ composition matrices connect ${\cal R}$ in vacuum with the effective $\widetilde{\cal R}$ in matter, the $Y^{e, \; \mu\tau}_{}$ matrices provide more insight into how the shape of these UTs evolves as the matter potential $A^{}_{\rm CC}$ increases.
In this paper, our discussion will focus mainly on the composition matrices $X^{e}_{}$ and $X^{\mu\tau}_{}$.

An interesting observation is that the second and third equations in Eq.~(\ref{30}) share the same composition matrix $X^{\mu\tau}_{}$ (or $Y^{\mu\tau}_{}$). This suggests that the results of medium- or long-baseline $\nu^{}_{\mu} \to \nu^{}_{e}$ and $\nu^{}_{\tau} \to \nu^{}_{e}$ oscillation experiments in which matter effect are included are closely related and can be combined for a more comprehensive analysis.
We will revisit this point later following a general discussion on the matter effect of these quartets.

\begin{figure}[]
\begin{center}
\includegraphics[width=\textwidth]{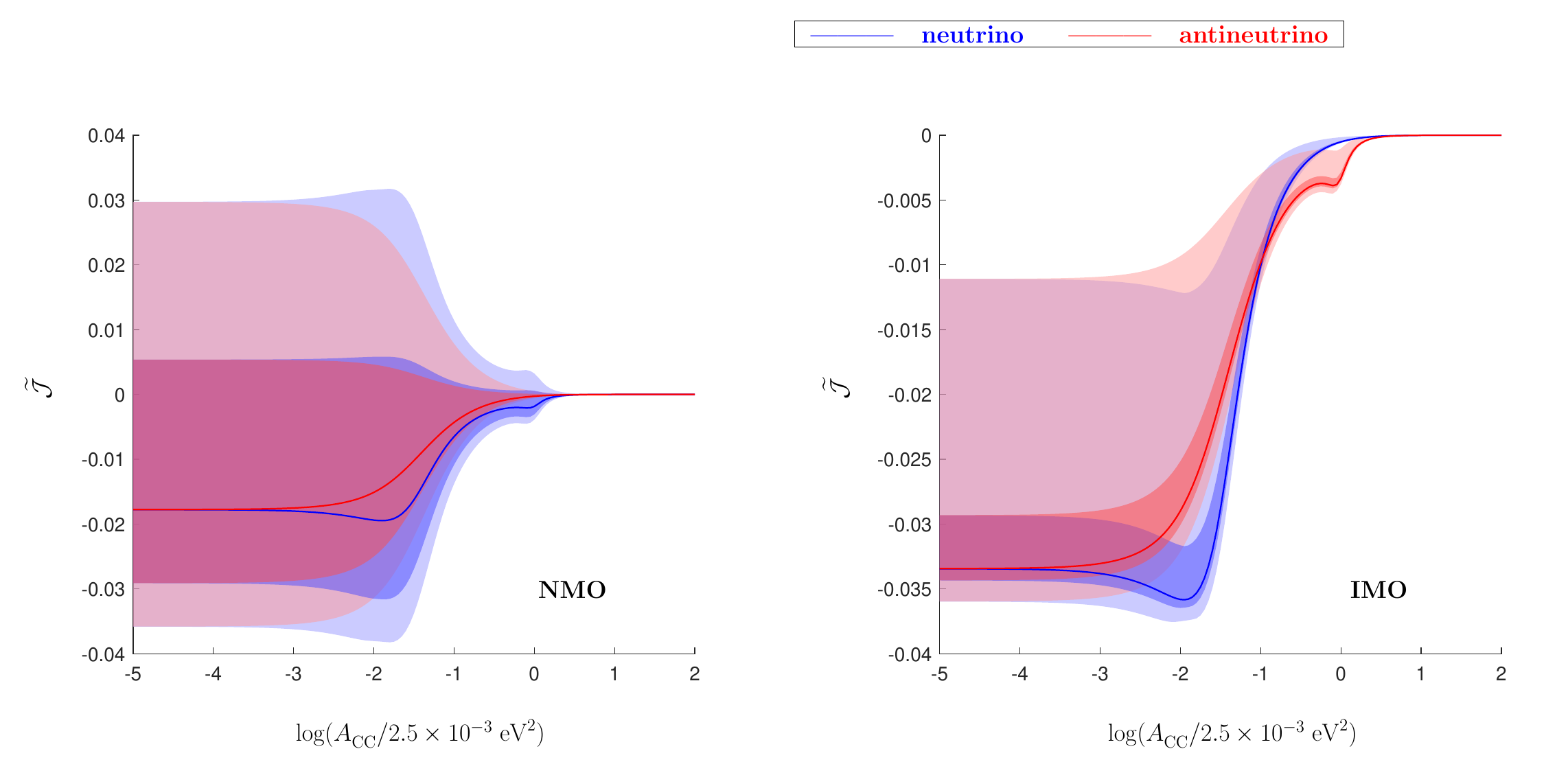}
\caption{The evolution of the effective CP-violating Jarlskog invariant $\widetilde{\cal J}$ in matter
with respect to the dimensionless ratio $A^{}_{\rm CC} / 2.5 \times 10^{-3}_{} {\rm eV}^{2}_{}$ in both the MNO case (left) and the IMO case (right), and for both neutrinos (in blue) and antineutrinos (in red), where the best-fit value, $1\sigma$ and $3\sigma$ ranges of $\widetilde{\cal J}$
are present with solid line, dark and light shadows respectively in the plots.}
\label{f5}
\end{center}
\end{figure}

\begin{figure}[]
\begin{center}
\includegraphics[width=\textwidth]{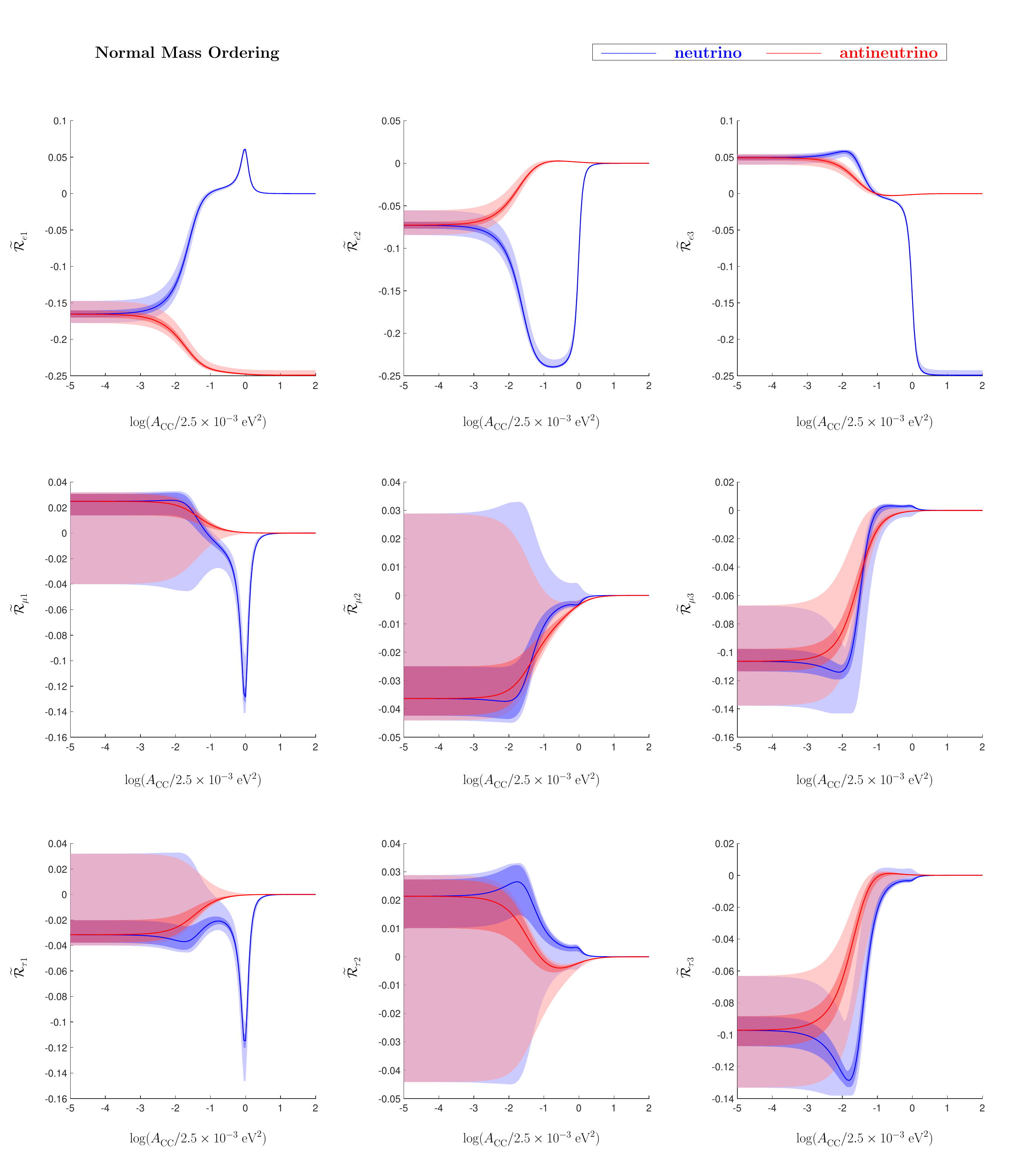}
\caption{The evolution of the effective CP-conserving invariants $\widetilde{\cal R}^{}_{\alpha i}$ in matter with respect to the dimensionless ratio $A^{}_{\rm CC} / 2.5 \times 10^{-3}_{} {\rm eV}^{2}_{}$ in the NMO case for both neutrinos (in blue) and antineutrinos (in red), where the best-fit value, $1\sigma$ and $3\sigma$ ranges of $\widetilde{\cal R}^{}_{\alpha i}$ are present with solid line, dark and light shadows respectively in the plots.}
\label{f6}
\end{center}
\end{figure}

\begin{figure}[]
\begin{center}
\includegraphics[width=\textwidth]{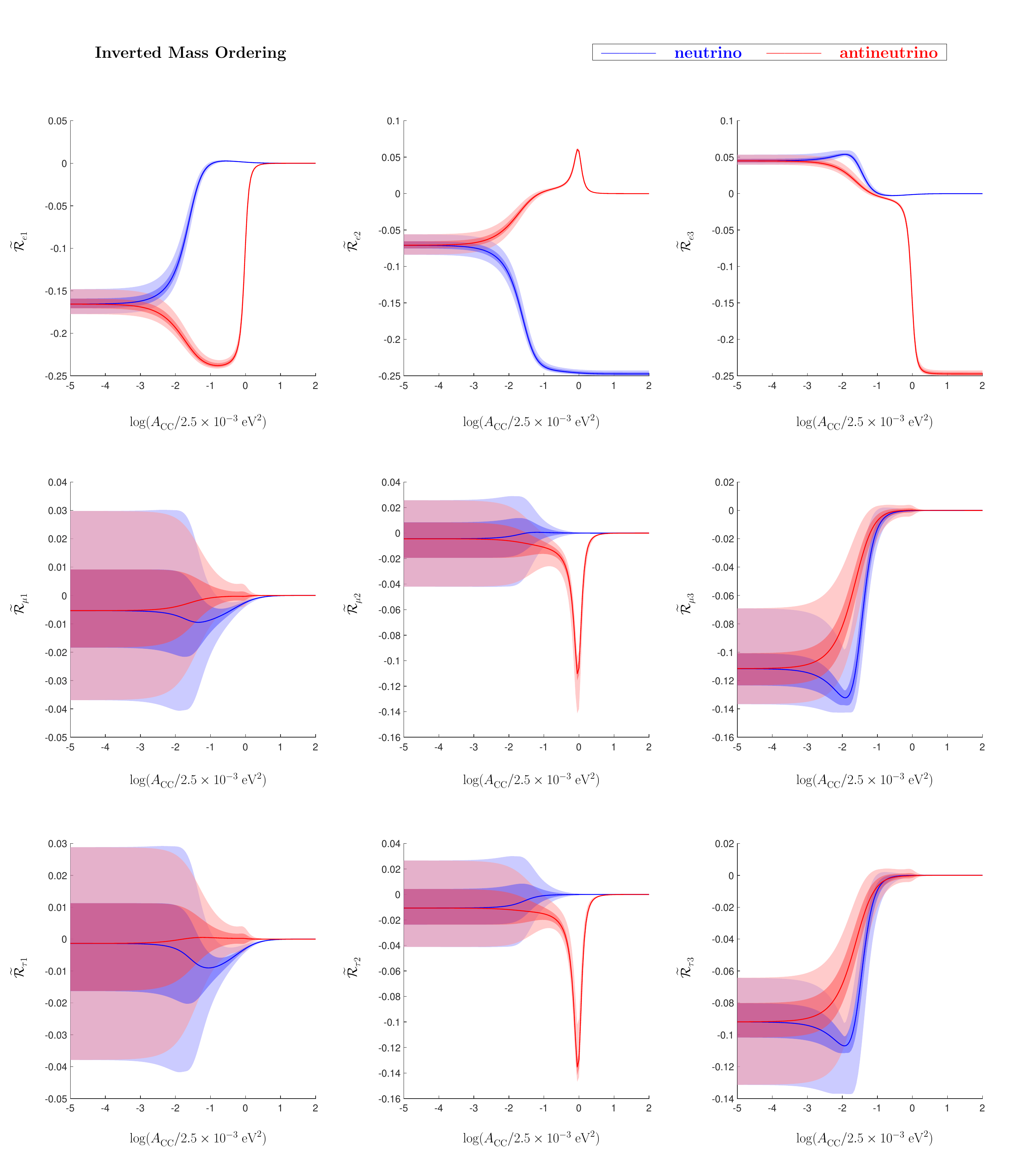}
\caption{The evolution of the effective CP-conserving invariants $\widetilde{\cal R}^{}_{\alpha i}$ in matter with respect to the dimensionless ratio $A^{}_{\rm CC} / 2.5 \times 10^{-3}_{} {\rm eV}^{2}_{}$ in the IMO case for both neutrinos (in blue) and antineutrinos (in red), where the best-fit value, $1\sigma$ and $3\sigma$ ranges of $\widetilde{\cal R}^{}_{\alpha i}$ are present with solid line, dark and light shadows respectively in the plots.}
\label{f7}
\end{center}
\end{figure}

\begin{figure}[]
\begin{center}
\includegraphics[width=\textwidth]{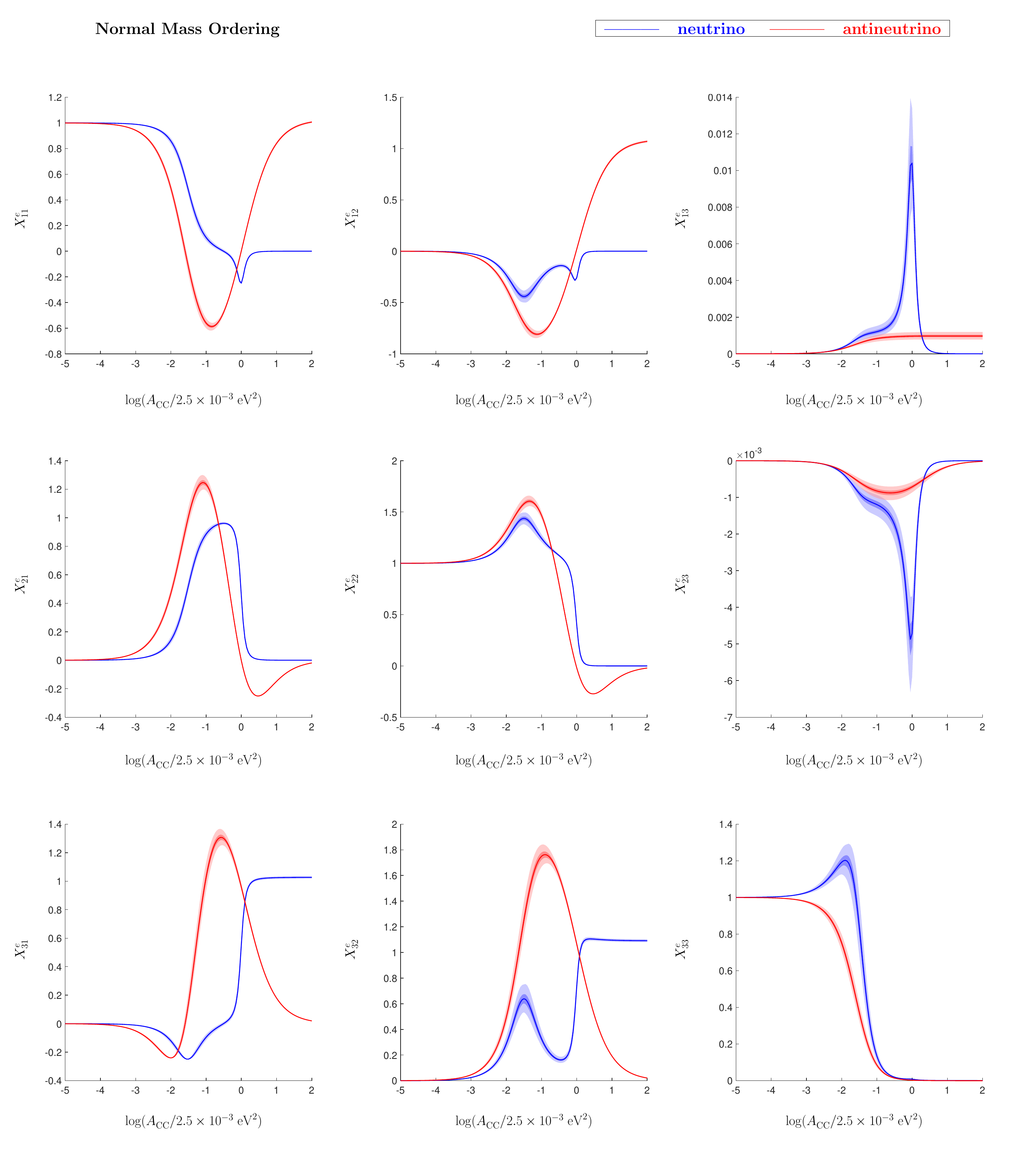}
\caption{The evolution of the composition matrix $X^{e}_{}$ with respect to the dimensionless ratio $A^{}_{\rm CC} / 2.5 \times 10^{-3}_{} {\rm eV}^{2}_{}$ in the NMO case for both neutrinos (in blue) and antineutrinos (in red), where the best-fit value, $1\sigma$ and $3\sigma$ ranges of $X^{e}_{ij}$ are present with solid line, dark and light shadows respectively in the plots.}
\label{f8}
\end{center}
\end{figure}

\begin{figure}[]
\begin{center}
\includegraphics[width=\textwidth]{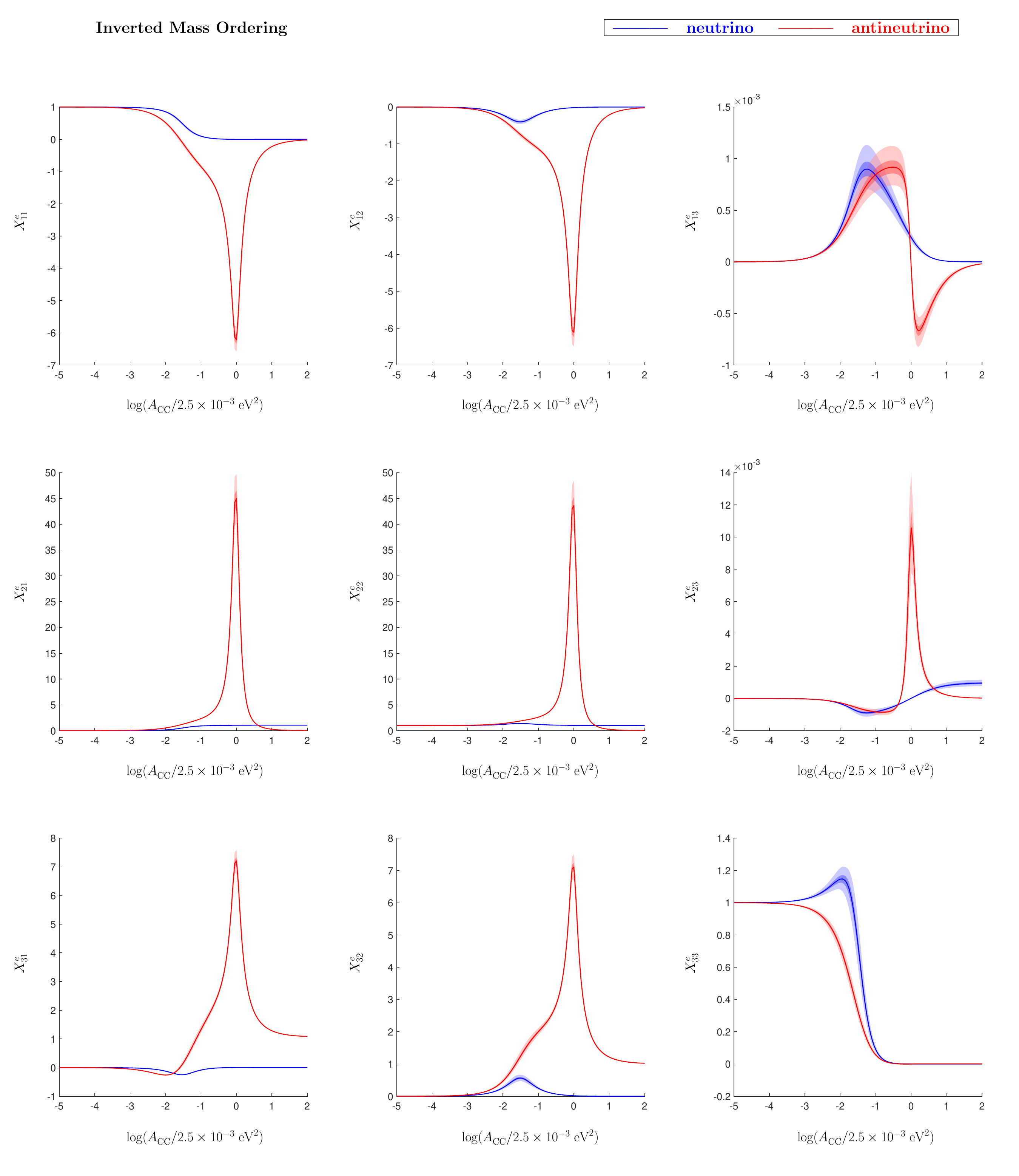}
\caption{The evolution of the  composition matrix $X^{e}_{}$ with respect to the dimensionless ratio $A^{}_{\rm CC} / 2.5 \times 10^{-3}_{} {\rm eV}^{2}_{}$ in the IMO case for both neutrinos (in blue) and antineutrinos (in red), where the best-fit value, $1\sigma$ and $3\sigma$ ranges of $X^{e}_{ij}$ are present with solid line, dark and light shadows respectively in the plots.}
\label{f9}
\end{center}
\end{figure}

\begin{figure}[]
\begin{center}
\includegraphics[width=\textwidth]{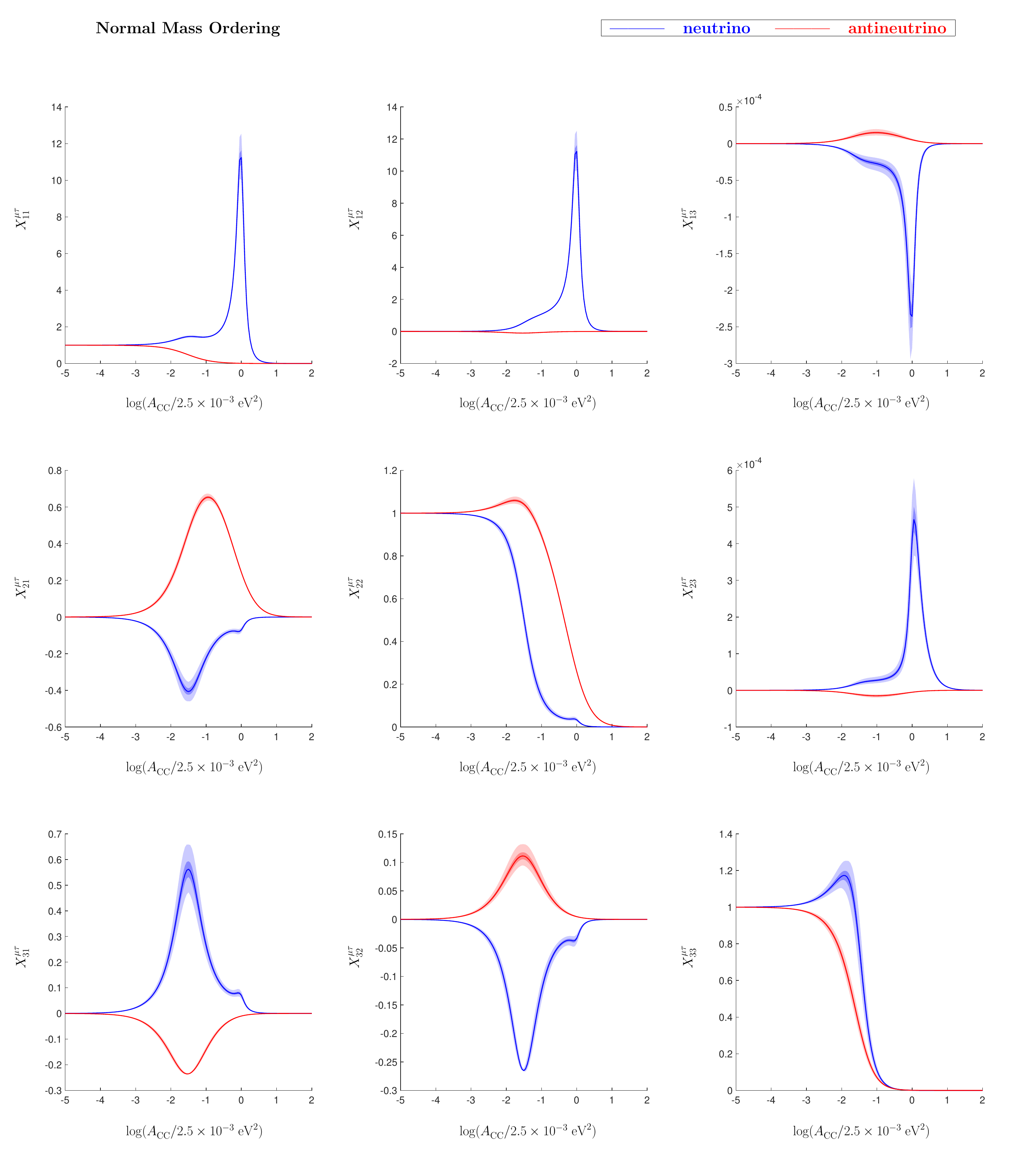}
\caption{The evolution of the composition matrix $X^{\mu\tau}_{}$ with respect to the dimensionless ratio $A^{}_{\rm CC} / 2.5 \times 10^{-3}_{} {\rm eV}^{2}_{}$ in the NMO case for both neutrinos (in blue) and antineutrinos (in red), where the best-fit value, $1\sigma$ and $3\sigma$ ranges of $X^{\mu\tau}_{ij}$ are present with solid line, dark and light shadows respectively in the plots.}
\label{f10}
\end{center}
\end{figure}

\begin{figure}[]
\begin{center}
\includegraphics[width=\textwidth]{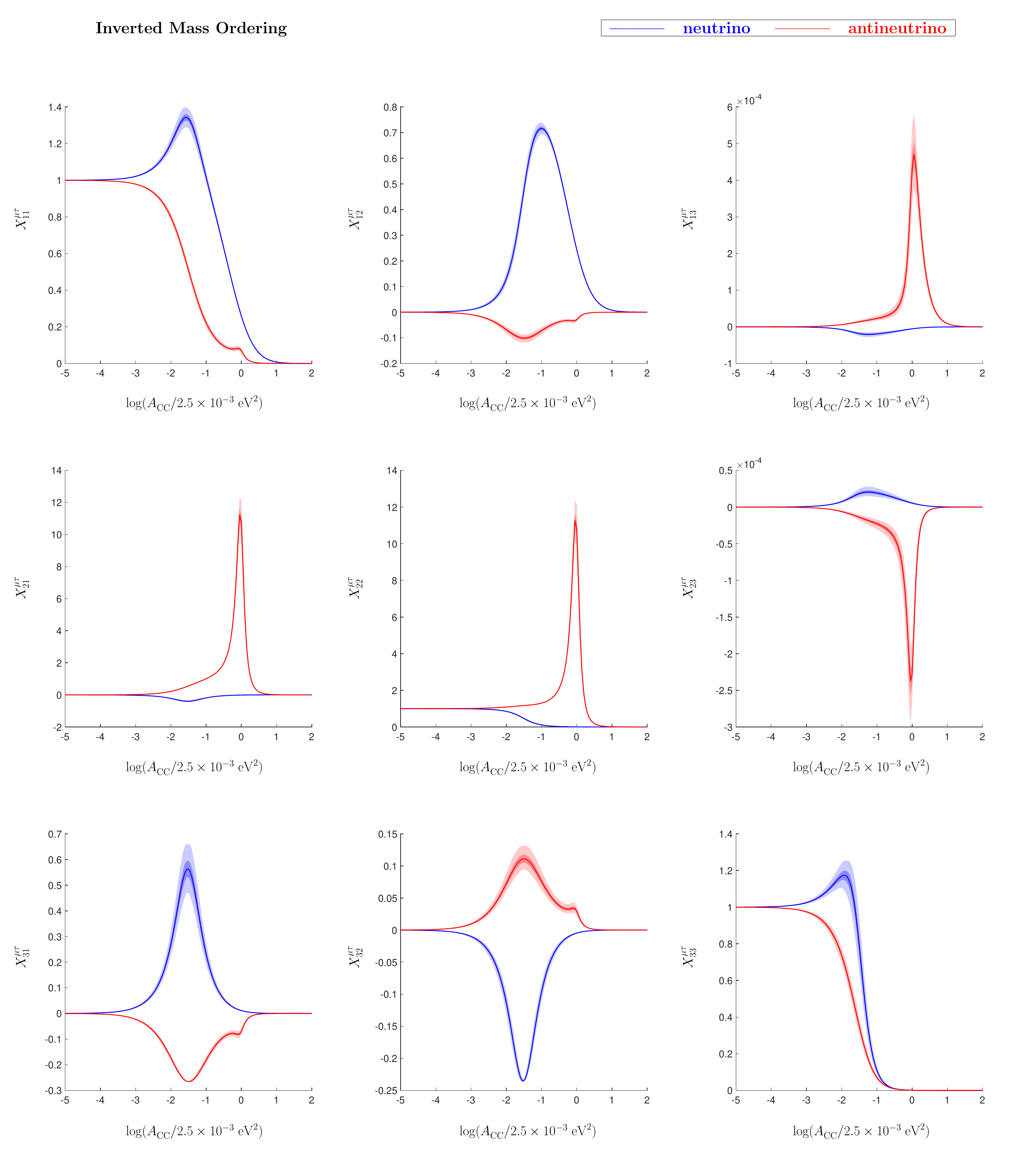}
\caption{The evolution of the  composition matrix $X^{\mu\tau}_{}$ with respect to the dimensionless ratio $A^{}_{\rm CC} / 2.5 \times 10^{-3}_{} {\rm eV}^{2}_{}$ in the IMO case for both neutrinos (in blue) and antineutrinos (in red), where the best-fit value, $1\sigma$ and $3\sigma$ ranges of $X^{\mu\tau}_{ij}$ are present with solid line, dark and light shadows respectively in the plots.}
\label{f11}
\end{center}
\end{figure}

Given the best-fit values, $1\sigma$ and $3\sigma$ ranges of these quartets in vacuum, we illustrated the evolution of the effective $\widetilde{\cal J}$ (in Fig.~\ref{f5}) and the nine effective $\widetilde{\cal R}^{}_{\alpha i}$ (in Figs.~\ref{f6} and \ref{f7}) in matter as the increase of the matter potential.
The evolution of the composition matrices $X^{e}_{}$ and $X^{\mu\tau}_{}$ with respect to $A^{}_{\rm CC}$ is also illustrated in Figs.~\ref{f8}, \ref{f9} (for NMO) and Figs.~\ref{f10}, \ref{f11} (for IMO).
Since only the CC contribution to coherent $\nu^{}_{e} e^{-}_{}$ forward scattering affects the neutrino oscillation behavior in matter, the effective neutrino masses are primarily sensitive to the first row of the PMNS matrix.
Therefore, compared to the broad range of possible values for $\widetilde{\cal R}^{}_{\alpha i}$, the way the composition matrices $X^{e, \; \mu\tau}_{}$ and $Y^{e, \; \mu\tau}_{}$ evolve with $A^{}_{\rm CC}$ is more certain (especially in non-resonance regions), thanks to the significantly reduced uncertainties in $\theta^{}_{12}$, $\theta^{}_{13}$ and the mass-squared differences~\cite{Capozzi:2025wyn}.
With a clear understanding of the evolution behaviors of the effective neutrino masses (see, e.g., discussions in \cite{Barger:1980tf, Xing:2000gg}, Figs. 1 and 2 in \cite{Luo:2019efb} for reference), the evolution of the composition matrices $X^{e}_{}$ and $X^{\mu\tau}_{}$ with respect to $A^{}_{\rm CC}$ as well as the evolution behaviors of the effective invariants $\widetilde{\cal R}^{}_{\alpha i}$ can be easily inferred with the help of Eq.~(\ref{29}).

Depending on the relative magnitude of $\Delta^{}_{21}$, $|\Delta^{}_{31}|$, and $|A^{}_{\rm CC}|$, neutrino oscillation behavior in matter varies significantly.
We will examine the vacuum-dominated, resonances and matter-dominated three scenarios individually.

\begin{itemize}

\item vacuum-dominated region ( $|A^{}_{\rm CC}| \ll \Delta^{}_{21}, |\Delta^{}_{31}|$ )

In the vacuum-dominated scenario, where the vacuum Hamiltonian ${\cal H}$ is the dominant term, useful analytical approximations can be obtained by regarding both $\Delta^{}_{21} / A^{}_{\rm CC}$ and $\Delta^{}_{31} / A^{}_{\rm CC}$ as small parameters and performing the diagonalization of $\widetilde{\cal H}$ using perturbation theory. To the first order of $\Delta^{}_{ji} / A^{}_{\rm CC}$ (for $ji = 21, \; 31, \;32$), we have
\begin{eqnarray}
\widetilde{m}^{2}_{1} & \approx & m^{2}_{1} + A^{}_{\rm CC} |V^{}_{e 1}|^{2} \; , \nonumber\\[1mm]
\widetilde{m}^{2}_{2} & \approx & m^{2}_{2} + A^{}_{\rm CC} |V^{}_{e 2}|^{2} \; , \nonumber\\[1mm]
\widetilde{m}^{2}_{3} & \approx & m^{2}_{3} + A^{}_{\rm CC} |V^{}_{e 3}|^{2} \; .
\label{32}
\end{eqnarray}
With the help of Eq.~(\ref{29}), we can then write down the approximate expressions for the composition matrix $X^{e}_{}$
\begin{eqnarray}
X^{e}_{11} & \approx & 1 - \left ( \frac{A^{}_{\rm CC}}{\Delta^{}_{21}} + \frac{A^{}_{\rm CC}}{\Delta^{}_{31}} \right ) \left ( 1 - |V^{}_{e 1}|^{2} \right ) - \frac{2 A^{}_{\rm CC}}{\Delta^{}_{32}} \left ( |V^{}_{e 3}|^{2} - |V^{}_{e 2}|^{2} \right ) \; , \nonumber\\[1mm]
X^{e}_{12} & \approx & - \left ( \frac{A^{}_{\rm CC}}{\Delta^{}_{21}} + \frac{A^{}_{\rm CC}}{\Delta^{}_{32}} \right ) \left ( 1 - |V^{}_{e 2}|^{2} \right ) \; , \nonumber\\[1mm]
X^{e}_{13} & \approx & \left ( \frac{A^{}_{\rm CC}}{\Delta^{}_{32}} - \frac{A^{}_{\rm CC}}{\Delta^{}_{31}} \right ) \left ( 1 - |V^{}_{e 3}|^{2} \right ) \; , \nonumber\\[1mm]
X^{e}_{21} & \approx & \left ( \frac{A^{}_{\rm CC}}{\Delta^{}_{21}} - \frac{A^{}_{\rm CC}}{\Delta^{}_{31}} \right ) \left ( 1 - |V^{}_{e 1}|^{2} \right ) \; , \nonumber\\[1mm]
X^{e}_{22} & \approx & 1 + \left ( \frac{A^{}_{\rm CC}}{\Delta^{}_{21}} - \frac{A^{}_{\rm CC}}{\Delta^{}_{32}} \right ) \left ( 1 - |V^{}_{e 2}|^{2} \right ) - \frac{2 A^{}_{\rm CC}}{\Delta^{}_{31}} \left ( |V^{}_{e 3}|^{2} - |V^{}_{e 1}|^{2} \right ) \; , \nonumber\\[1mm]
X^{e}_{23} & \approx & \left ( \frac{A^{}_{\rm CC}}{\Delta^{}_{31}} - \frac{A^{}_{\rm CC}}{\Delta^{}_{32}} \right ) \left ( 1 - |V^{}_{e 3}|^{2} \right ) \; , \nonumber\\[1mm]
X^{e}_{31} & \approx & \left ( \frac{A^{}_{\rm CC}}{\Delta^{}_{31}} - \frac{A^{}_{\rm CC}}{\Delta^{}_{21}} \right ) \left ( 1 - |V^{}_{e 1}|^{2} \right ) \; , \nonumber\\[1mm]
X^{e}_{32} & \approx & \left ( \frac{A^{}_{\rm CC}}{\Delta^{}_{21}} + \frac{A^{}_{\rm CC}}{\Delta^{}_{32}} \right ) \left ( 1 - |V^{}_{e 2}|^{2} \right ) \; , \nonumber\\[1mm]
X^{e}_{33} & \approx & 1 + \left ( \frac{A^{}_{\rm CC}}{\Delta^{}_{31}} + \frac{A^{}_{\rm CC}}{\Delta^{}_{32}} \right ) \left ( 1 - |V^{}_{e 3}|^{2} \right ) - \frac{2 A^{}_{\rm CC}}{\Delta^{}_{21}} \left ( |V^{}_{e 2}|^{2} - |V^{}_{e 1}|^{2} \right ) \; ,
\label{33}
\end{eqnarray}
together with the approximations for $X^{\mu\tau}_{}$
\begin{eqnarray}
X^{\mu\tau}_{11} & \approx & 1 + \left ( \frac{A^{}_{\rm CC}}{\Delta^{}_{21}} + \frac{A^{}_{\rm CC}}{\Delta^{}_{31}} \right ) |V^{}_{e 1}|^{2} - \frac{2 A^{}_{\rm CC}}{\Delta^{}_{32}} \left ( |V^{}_{e 3}|^{2} - |V^{}_{e 2}|^{2} \right ) \; , \nonumber\\[1mm]
X^{\mu\tau}_{12} & \approx & \left ( \frac{A^{}_{\rm CC}}{\Delta^{}_{21}} + \frac{A^{}_{\rm CC}}{\Delta^{}_{32}} \right ) |V^{}_{e 2}|^{2} \; , \nonumber\\[1mm]
X^{\mu\tau}_{13} & \approx & \left ( \frac{A^{}_{\rm CC}}{\Delta^{}_{31}} - \frac{A^{}_{\rm CC}}{\Delta^{}_{32}} \right ) |V^{}_{e 3}|^{2} \; , \nonumber\\[1mm]
X^{\mu\tau}_{21} & \approx & \left ( \frac{A^{}_{\rm CC}}{\Delta^{}_{31}} - \frac{A^{}_{\rm CC}}{\Delta^{}_{21}} \right ) |V^{}_{e 1}|^{2} \; , \nonumber\\[1mm]
X^{\mu\tau}_{22} & \approx & 1 + \left ( \frac{A^{}_{\rm CC}}{\Delta^{}_{32}} - \frac{A^{}_{\rm CC}}{\Delta^{}_{21}} \right ) |V^{}_{e 2}|^{2} - \frac{2 A^{}_{\rm CC}}{\Delta^{}_{31}} \left ( |V^{}_{e 3}|^{2} - |V^{}_{e 1}|^{2} \right ) \; , \nonumber\\[1mm]
X^{\mu\tau}_{23} & \approx & \left ( \frac{A^{}_{\rm CC}}{\Delta^{}_{32}} - \frac{A^{}_{\rm CC}}{\Delta^{}_{31}} \right ) |V^{}_{e 3}|^{2} \; , \nonumber\\[1mm]
X^{\mu\tau}_{31} & \approx & \left ( \frac{A^{}_{\rm CC}}{\Delta^{}_{21}} - \frac{A^{}_{\rm CC}}{\Delta^{}_{31}} \right ) |V^{}_{e 1}|^{2} \; , \nonumber\\[1mm]
X^{\mu\tau}_{32} & \approx & - \left ( \frac{A^{}_{\rm CC}}{\Delta^{}_{21}} + \frac{A^{}_{\rm CC}}{\Delta^{}_{32}} \right ) |V^{}_{e 2}|^{2} \; , \nonumber\\[1mm]
X^{\mu\tau}_{33} & \approx & 1 - \left ( \frac{A^{}_{\rm CC}}{\Delta^{}_{31}} + \frac{A^{}_{\rm CC}}{\Delta^{}_{32}} \right ) |V^{}_{e 3}|^{2} - \frac{2 A^{}_{\rm CC}}{\Delta^{}_{21}} \left ( |V^{}_{e 2}|^{2} - |V^{}_{e 1}|^{2} \right ) \; .
\label{34}
\end{eqnarray}
In the vacuum-dominated region, the composition matrices $X^{e}_{}$ and $X^{\mu\tau}_{}$ can both be approximated by the identity matrix plus a small perturbation of the order $\Delta^{}_{ji} / A^{}_{\rm CC}$.
It is evident that both $X^{e}_{}$ and $X^{\mu\tau}_{}$ can be accurately determined for a given matter potential $A^{}_{\rm CC}$ and neutrino mass ordering, owing to the precise measurements of the mass-squared differences and the first row of the PMNS matrix \cite{Capozzi:2025wyn}.
We can conclude that neutrino oscillations receive only predictable small corrections from the matter effect in the vacuum-dominated region which is the case for many of ongoing and upcoming long-baseline oscillation experiments aimed at the measuring of $\theta^{}_{23}$ and the CP violation in the lepton sector, such as T2K~\cite{T2K:2023smv} and NO$\nu$A~\cite{NOvA:2021nfi}.
With the help of equations above, the effective quantities $\widetilde{\cal R}^{}_{\alpha i}$ and $\widetilde{\cal J}$ obtained in different experiments can all be converted to the ${\cal R}^{}_{\alpha i}$ and ${\cal J}$, allowing for a combined analysis.

\item the resonance region ($A^{}_{\rm CC} \sim \Delta^{}_{21}, \Delta^{}_{31}$)

During resonance, one of the effective mass-squared differences approaches zero, causing significant changes in $\widetilde{\cal R}^{}_{\alpha i}$ and $\widetilde{\cal J}$.
Within the standard three-neutrino framework, it is well known that there are two possible resonance regions when studying neutrino oscillations in matter: the solar resonance (around $A^{}_{\rm CC} \sim \Delta^{}_{21}$) and the atmospheric resonance (around $A^{}_{\rm CC} \sim \Delta^{}_{31}$).
However, since the sign of $A^{}_{\rm CC}$ are different for neutrino or antineutrino oscillation and the sign of $\Delta^{}_{31}$ varies between the NMO and IMO cases, above two resonance conditions are not always fulfilled even if the magnitude of $A^{}_{\rm CC}$ is carefully chosen.
Taking neutrino oscillations in the NMO case as an example, as the matter potential increases, neutrino oscillations experience both the solar and the atmospheric resonances sequentially. During the solar resonance, $\widetilde{\Delta}^{}_{21}$ approaches zero, causing noticeable changes in all nine effective $\widetilde{\cal R}^{}_{\alpha i}$ and the effective Jarlskog ${\cal J}$. Among these, $\widetilde{\cal R}^{}_{e 3}$, $\widetilde{\cal R}^{}_{\mu 3}$ and $\widetilde{\cal R}^{}_{\tau 3}$ exhibit significant resonance for they involve terms proportional to $\widetilde{\Delta}^{-2}_{21}$, while the others contain only terms proportional to $\widetilde{\Delta}^{-1}_{21}$. Likely, during the atmospheric resonance, it is the effective $\widetilde{\Delta}^{}_{32}$ that approaches zero and while all these CP-conserving and CP-violating invariants change noticeably, $\widetilde{\cal R}^{}_{e 3}$, $\widetilde{\cal R}^{}_{\mu 3}$ and $\widetilde{\cal R}^{}_{\tau 3}$ undergo an even more significant resonance, as clearly shown in Eq.~(\ref{27}).
We summarize in Table~\ref{t4} the various resonances neutrinos or antineutrinos with different mass orderings may experience along with the invariants that undergo significant resonance in each case.
Exploring the oscillation behavior of any channel in the atmospheric resonance region can help distinguish the NMO from the IMO.
In our numerical analysis, the three neutrino mass eigenvalues $\widetilde{\lambda}^{}_{i}$ are ordered in such a way that in all four scenarios the same correct order $\{ \widetilde{\lambda}^{}_{1}, \widetilde{\lambda}^{}_{2}, \widetilde{\lambda}^{}_{3} \} = \{ m^{2}_{1}, m^{2}_{2}, m^{2}_{3} \} / 2E$ can be obtained in the limit $A^{}_{\rm CC} \to 0$. In this case, after the resonance, related two mass eigenstates and two eigenvectors are ``swapped'', so as the related two $\widetilde{\cal R}^{}_{\alpha i}$.

\begin{table}
\caption{A summary of the evolution behavior of the effective CP-conserving invariants $\widetilde{\cal R}^{}_{\alpha}$ in the resonance and matter-dominated regions in both the NMO and IMO cases and for both neutrino and antineutrino oscillations. Note that the three eigenvalues $\widetilde{\lambda}^{}_{i}$ are ordered in such a way that in all four scenarios the same correct order $\{ \widetilde{\lambda}^{}_{1}, \widetilde{\lambda}^{}_{2}, \widetilde{\lambda}^{}_{3} \} = \{ m^{2}_{1}, m^{2}_{2}, m^{2}_{3} \} / 2E$ can be obtained in the limit $A^{}_{\rm CC}$ through continuous evolution as $|A^{}_{\rm CC}|$ decreasing.}
\label{t4}
\begin{tabular}{c|cc|cc}
\toprule[1pt]
&&&&\\[-6mm]
& \multicolumn{2}{c|}{ ~ Normal Mass Ordering \; ($\Delta^{}_{31} > 0$) ~ } & \multicolumn{2}{c}{ ~ Inverted Mass Ordering \; ($\Delta^{}_{31} < 0$) ~ } \\[3mm]
& \begin{tabular}{c} neutrinos \\[-1mm] ($A^{}_{\rm CC} > 0$) \\[1mm] \end{tabular} &\begin{tabular}{c} antineutrinos \\[-1mm] ($A^{}_{\rm CC} < 0$) \\[1mm] \end{tabular} & \begin{tabular}{c} neutrinos \\[-1mm] ($A^{}_{\rm CC} > 0$) \\[1mm] \end{tabular} & \begin{tabular}{c} antineutrinos \\[-1mm] ($A^{}_{\rm CC} < 0$) \\[1mm] \end{tabular} \\
\midrule[1pt]
\multicolumn{5}{l}{~ Resonances} \\
\hline
\begin{tabular}{c} solar \\[-1mm] ($A ^{}_{\rm CC} \sim \Delta^{}_{21}$) \\[1mm] \end{tabular} & \begin{tabular}{c} $\widetilde{\Delta}^{}_{21} \simeq 0$ \\[2mm] $\widetilde{\cal R}^{}_{e 3}$, $\widetilde{\cal R}^{}_{\mu 3}$, $\widetilde{\cal R}^{}_{\tau 3}$ \\[-2mm] resonance \\[-2mm] significantly \end{tabular} & None & \begin{tabular}{c} $\widetilde{\Delta}^{}_{21} \simeq 0$ \\[2mm] $\widetilde{\cal R}^{}_{e 3}$, $\widetilde{\cal R}^{}_{\mu 3}$, $\widetilde{\cal R}^{}_{\tau 3}$ \\[-2mm] resonance \\[-2mm] significantly \end{tabular} & None \\
\hline
\begin{tabular}{c} atmospheric \\[-1mm] ($A ^{}_{\rm CC} \sim \Delta^{}_{31}, \Delta^{}_{32}$) \\[1mm] \end{tabular} & \begin{tabular}{c} $\widetilde{\Delta}^{}_{32} \simeq 0$ \\[2mm] $\widetilde{\cal R}^{}_{e 1}$, $\widetilde{\cal R}^{}_{\mu 1}$, $\widetilde{\cal R}^{}_{\tau 1}$ \\[-2mm] resonance \\[-2mm] significantly \end{tabular} & None & None & \begin{tabular}{c} $\widetilde{\Delta}^{}_{31} \simeq 0$ \\[2mm] $\widetilde{\cal R}^{}_{e 2}$, $\widetilde{\cal R}^{}_{\mu 2}$, $\widetilde{\cal R}^{}_{\tau 2}$ \\[-2mm] resonance \\[-2mm] significantly \end{tabular} \\
\midrule[1pt]
\multicolumn{5}{l}{~ Matter-Dominated ~ (in the limit $|A^{}_{\rm CC}| \rightarrow \infty$)} \\
\hline
&&&&\\[-5mm]
\begin{tabular}{c} eigenvalues \\[-1mm] of $\widetilde{\cal H}$ \\[1mm] \end{tabular} & $\left ( \begin{matrix} ~ \widetilde{\lambda}^{f}_{2} ~ \cr ~ \widetilde{\lambda}^{f}_{3} ~ \cr ~ \widetilde{\lambda}^{f}_{1} ~ \end{matrix} \right )$ & $\left ( \begin{matrix} ~ \widetilde{\lambda}^{f}_{1} ~ \cr ~ \widetilde{\lambda}^{f}_{2} ~ \cr ~ \widetilde{\lambda}^{f}_{3} ~ \end{matrix} \right )$ & $\left ( \begin{matrix} ~ \widetilde{\lambda}^{f}_{2} ~ \cr ~ \widetilde{\lambda}^{f}_{1} ~ \cr ~ \widetilde{\lambda}^{f}_{3} ~ \end{matrix} \right )$ & $\left ( \begin{matrix} ~ \widetilde{\lambda}^{f}_{3} ~ \cr ~ \widetilde{\lambda}^{f}_{2} ~ \cr ~ \widetilde{\lambda}^{f}_{1} ~ \end{matrix} \right )$ \\
&&&&\\[-4mm]
$\widetilde{\cal R}$ & $\left ( \begin{matrix} ~ 0 ~ & ~ 0 ~ & \widetilde{\cal R}^{f}_{} \cr 0 & 0 & 0 \cr 0 & 0 & 0 \cr \end{matrix} \right )$ & $\left ( \begin{matrix} \widetilde{\cal R}^{f}_{} & ~ 0 ~ & ~ 0 ~ \cr 0 & 0 & 0 \cr 0 & 0 & 0 \cr \end{matrix} \right )$ & $\left ( \begin{matrix} ~ 0 ~ & \widetilde{\cal R}^{f}_{} & ~ 0 ~ \cr 0 & 0 & 0 \cr 0 & 0 & 0 \cr \end{matrix} \right )$ & $\left ( \begin{matrix} ~ 0 ~ & ~ 0 ~ & \widetilde{\cal R}^{f}_{} \cr 0 & 0 & 0 \cr 0 & 0 & 0 \cr \end{matrix} \right )$ \\
&&&&\\[-4mm]
$X^{e}_{}$ & $\left ( \begin{matrix} 0 & 0 & 0 \cr 0 & 0 & 0 \cr X^{f}_{1} & X^{f}_{2} & X^{f}_{3} \cr \end{matrix} \right )$ & $\left ( \begin{matrix} X^{f}_{1} & X^{f}_{2} & X^{f}_{3} \cr 0 & 0 & 0 \cr 0 & 0 & 0 \cr \end{matrix} \right )$ & $\left ( \begin{matrix} 0 & 0 & 0 \cr X^{f}_{1} & X^{f}_{2} & X^{f}_{3} \cr 0 & 0 & 0 \cr \end{matrix} \right )$ & $\left ( \begin{matrix} 0 & 0 & 0 \cr 0 & 0 & 0 \cr X^{f}_{1} & X^{f}_{2} & X^{f}_{3} \cr \end{matrix} \right )$ \\[9mm]
\bottomrule[1pt]
\end{tabular}
\end{table}

\item the matter-dominated region ($|A^{}_{\rm CC}| \gg \Delta^{}_{21}, |\Delta^{}_{31}|$)

As the magnitude of $|A^{}_{\rm CC}|$ increases, following the atmospheric resonance, the matter term ${\cal H}^{\prime}_{}$ begins to dominate over the vacuum term ${\cal H}$.
Consequently, electron neutrino decouples due to its intense CC interaction with electrons in the medium~\cite{Xing:2018lob, Parke:2018brr, Wang:2019yfp, Xing:2019owb, Luo:2019efb}.
Here, we refer to the results of our previous studies~\cite{Luo:2019efb} to briefly describe the evolution behavior of these rephasing invariants in the matter-dominated region.
As the increase of $A^{}_{\rm CC}$, the matter term becomes dominant over the vacuum term, three eigenvalues of $\widetilde{\cal H}$ approach a set of fixed values
\begin{eqnarray}
\widetilde{\lambda}^{f}_{1} & = & \frac{1}{2 E} \left ( m^{2}_{1} + A^{}_{\rm NC} + A^{}_{\rm CC} + \Omega^{}_{11} \right ) \; , \nonumber\\[1mm]
\widetilde{\lambda}^{f}_{2} & = & \frac{1}{2 E} \left ( m^{2}_{1} + A^{}_{\rm NC} + \Omega^{}_{22} \cos^2\widetilde{\theta} +\Omega^{}_{33} \sin^2\widetilde{\theta} - | \Omega^{}_{23} | \sin2\widetilde{\theta} \right ) \; , \nonumber\\[1mm]
\widetilde{\lambda}^{f}_{3} & = & \frac{1}{2 E} \left ( m^{2}_{1} + A^{}_{\rm NC} + \Omega^{}_{33} \cos^2\widetilde{\theta} +\Omega^{}_{22} \sin^2\widetilde{\theta} + | \Omega^{}_{23} | \sin2\widetilde{\theta} \right ) \; .
\label{35}
\end{eqnarray}
At the same time, the effective mixing matrix in matter $\widetilde{V}$ evolves toward a fixed $3 \times 3$ real matrix
\begin{eqnarray}
\widetilde{V}^{fixed}_{} & = & \left ( \begin{matrix} ~ 1 ~ & ~ 0 ~ & ~ 0 ~ \cr ~ 0 ~ & \cos \widetilde{\theta} & \sin \widetilde{\theta} \cr ~ 0 ~ & - \sin \widetilde{\theta} & \cos \widetilde{\theta} \cr \end{matrix} \right )  \; .
\label{36}
\end{eqnarray}
Here the $3 \times 3$ Hermitian matrix $\Omega$ is defined as $\Omega \; \equiv \; V \; {\rm diag}\{ 0, \Delta^{}_{21}, \Delta^{}_{31} \} \; V^{\dagger}_{}$, and the complete expressions for its nine elements $\Omega^{}_{ij}$ can be found in \cite{Luo:2019efb}.
Clearly, in this matter-dominated case, $\widetilde{\lambda}^{f}_{2}$ and $\widetilde{\lambda}^{f}_{3}$ are nearly degenerate, both exhibiting strong hierarchies compared to $\widetilde{\lambda}^{f}_{1}$. And the effective mixing matrix presents a two-flavor-mixing structure, with the only mixing angle $\widetilde{\theta}$ defined by
\begin{eqnarray}
\tan 2 \widetilde{\theta} & = & \frac{2 \left | \Delta^{}_{21} V^{}_{\mu 2} V^{*}_{\tau 2}  + \Delta^{}_{31} V^{}_{\mu 3} V^{*}_{\tau 3} \right |}{\Delta^{}_{21} \left ( | V^{}_{\tau 2} |^{2}_{} - | V^{}_{\mu 2} |^{2}_{} \right ) + \Delta^{}_{31} \left ( | V^{}_{\tau 3} |^{2}_{} - | V^{}_{\mu 3} |^{2}_{} \right )} \; .
\label{37}
\end{eqnarray}
Considering the strong hierarchy $\Delta^{}_{21} \ll |\Delta^{}_{31}|$ and the smallness of $s^{}_{13}$, it follows that $\widetilde{\theta} \approx \theta^{}_{23}$.
Moreover, the mixing angle $\widetilde{\theta}$ defined in Eq.~(\ref{37}) serves as an indicator of the $\mu$-$\tau$ symmetry breaking. If the neutrino mass matrix in vacuum possess the exact $\mu$-$\tau$ symmetry, we then have $\widetilde{\theta} = \pi/4$. 

Correspondingly, there is only one non-vanishing $\widetilde{\cal R}^{}_{e i}$ (in the first row) that approaches a fixed value
\begin{eqnarray}
\widetilde{\cal R}^{f}_{} & = & - \frac{1}{4} \sin^{2}_{} 2 \widetilde{\theta} \; ,
\label{38}
\end{eqnarray}
corresponding to a fixed mass-squared difference $\widetilde{\Delta}^{f}_{} \; \equiv \; \widetilde{\lambda}^{f}_{3} - \widetilde{\lambda}^{f}_{2}$.
All the other eight CP-conserving invariants, as well as the effective CP-violating invariant ${\cal J}$, vanish in the matter-dominated region. Which means all the UTs collapse in this case~\cite{Zhang:2004hf}.
In this matter-dominated limit, the composition matrix $X^{\mu\tau}_{}$ becomes a zero matrix, and only one row in $X^{e}_{}$ remains non-vanishing, with its three elements
\begin{eqnarray}
X^{f}_{1} & = & - \frac{\Delta^{2}_{32}}{4 |\Omega^{}_{23}| \sin^2 2\widetilde{\theta}} \; , \nonumber\\
X^{f}_{2} & = & - \frac{\Delta^{2}_{31}}{4 |\Omega^{}_{23}| \sin^2 2\widetilde{\theta}} \; , \nonumber\\
X^{f}_{3} & = & - \frac{\Delta^{2}_{21}}{4 |\Omega^{}_{23}| \sin^2 2\widetilde{\theta}} \; .
\label{39}
\end{eqnarray}
Note that when passing through the resonances region, the related two eigenvalues ``exchange'' their evolution behaviors. The patterns of the fixed points can differ depending on the resonances they experienced.
We list in Table~\ref{t4} also the different patterns of ${\cal R}$ matrix and the composition matrix $X^{e}_{}$ in each scenario.
\begin{table}[]
\footnotesize
\caption{The best-fit values, 1$\sigma$ and 3$\sigma$ ranges of the fixed values of $\widetilde{\cal R}^{f}_{}$, $\widetilde{\Delta}^{f}_{}$ and $X^{f}_{i}$ (for $i=1, 2, 3$) in the matter-dominated limit, where we've adopt the global fit of the unitary PMNS leptonic mixing matrix given in~\cite{Esteban:2024eli,Nufit}.}
\label{t5} 
\begin{tabular}{ccccccccccccc}
\hline\noalign{\smallskip}
&& \multicolumn{5}{c}{Normal Neutrino Mass Ordering} && \multicolumn{5}{c}{Inverted Neutrino Mass Ordering}  \\
&~~~& best-fit &~& 1$\sigma$ range &~& 3$\sigma$ range &~~~& best-fit &~& 1$\sigma$ range &~& 3$\sigma$ range \\
\noalign{\smallskip}\hline\noalign{\smallskip}
$\widetilde{\cal R}^{f}_{}$ && -0.2492 && -0.2499 $\sim$ -0.2482 && -0.25 $\sim$ -0.2423 && -0.2474 && -0.2488 $\sim$ -0.2450 && -0.25 $\sim$ -0.2427 \\
\noalign{\smallskip}
$\displaystyle \frac{|\widetilde{\Delta}^{f}_{}|}{10^{-3}_{} {\rm eV}^{2}_{}}$ && 2.4060 && 2.3838 $\sim$ 2.4301 && 2.3347 $\sim$ 2.4808 && 2.4067 && 2.3843 $\sim$ 2.4291 && 2.3360$\sim$ 2.4766\\
\noalign{\smallskip}
$X^{f}_{1}$ && 1.0268 && 1.0241 $\sim$ 1.0293 && 1.0182 $\sim$ 1.0343 && 1.0653 && 1.0628 $\sim$ 1.0681 && 1.0578 $\sim$ 1.0740 \\
$X^{f}_{2}$ && 1.0908 && 1.0872 $\sim$ 1.0945 && 1.0798 $\sim$ 1.1020 && 1.0021 && 0.9987 $\sim$ 1.0054 && 0.9918 $\sim$ 1.0122 \\
$X^{f}_{3} / 10^{-3}_{}$ && 0.9692 && 0.9025 $\sim$ 1.0380 && 0.7781 $\sim$ 1.1890 && 0.9684 && 0.9031 $\sim$ 1.0374 && 0.7807 $\sim$ 1.1872 \\
\noalign{\smallskip}\hline
\end{tabular}
\end{table}

However, whether it is neutrino or antineutrino oscillation, and whether it is NMO or IMO, in the matter-dominated region, $\nu^{}_{e}$ decouples, and oscillations occur only between $\nu^{}_{\mu}$ and $\nu^{}_{\tau}$ with a fixed magnitude $\widetilde{\cal R}^{f}_{}$ and a fixed frequency $\widetilde{\Delta}^{f}_{} L / 4 E$.
In other word, the remaining non-vanishing neutrino/antineutrino oscillation probabilities in the matter-dominated region are
\begin{eqnarray}
P ( \stackrel{(-)}{\nu}^{}_{\mu} \rightarrow \stackrel{(-)}{\nu}^{}_{\mu} ) \; = \; P ( \stackrel{(-)}{\nu}^{}_{\tau} \rightarrow \stackrel{(-)}{\nu}^{}_{\tau} ) & = & 1 + 4 \; \widetilde{\cal R}^{f}_{} \sin^2 \frac{\widetilde{\Delta}^{f}_{} L}{4 E} \; , \nonumber\\
P ( \stackrel{(-)}{\nu}^{}_{\mu} \rightarrow \stackrel{(-)}{\nu}^{}_{\tau} ) \; = \; P ( \stackrel{(-)}{\nu}^{}_{\tau} \rightarrow \stackrel{(-)}{\nu}^{}_{\mu} ) & = & - 4 \; \widetilde{\cal R}^{f}_{} \sin^2 \frac{\widetilde{\Delta}^{f}_{} L}{4 E} \; ,
\label{40}
\end{eqnarray}
This is in agreement with the vanishing of CP violation in the matter-dominated limit.
Although CP violation vanishes and neutrino and antineutrino oscillation probabilities equals each other in this limit, the resulting $\widetilde{\cal R}^{f}_{}$ and $\widetilde{\Delta}^{f}_{}$ differ between the NMO and IMO case (as shown in Table~\ref{t5}). Therefore it is possible to determine the neutrino mass ordering by examining the oscillation between $\nu^{}_{\mu}$ and $\nu^{}_{\tau}$ in the matter-dominated region.

For completeness, we illustrate in Table~\ref{t5} the best-fit, 1$\sigma$ and 3$\sigma$ ranges of the fixed values of $\widetilde{\cal R}^{f}_{}$, $\widetilde{\Delta}^{f}_{}$, and $X^{f}_{i}$ (for $i=1, 2, 3$) in the matter-dominated limit for both the NMO and IMO scenarios.
We observe that, in this limit, $X^{f}_{1} \sim X^{f}_{2} \sim 1$, $X^{f}_{3} \sim 10^{-3}_{}$, and $\widetilde{\cal R}^{f}_{}$ is negative and close to -0.25.
If the $\mu$-$\tau$ symmetry hols, we then have $\widetilde{\theta} = \pi/4$, with $\nu^{}_{\mu}$ and $\mu^{}_{\tau}$ maximally mixed, and the oscillation magnitude $\widetilde{\cal R}^{f}_{}$ reaches it maximal negative value of $- 1/4$, a possibility still allowed within the $3\sigma$ constraints.

\end{itemize}

In the end, to give a more intuitive and visible illustration on the matter effects, we choose some benchmark values of $A^{}_{\rm CC}$ and present in the appendix the evolution of these rephasing invariants and the rescaled UTs with our numerical results.

Before wrapping up this section, we come back to the $\mu$-$\tau$ symmetry of ${\cal R}$ when the matter effect is taken into account.
From the last six equations in Eq.~(\ref{29}), we can straightforwardly write down the relationships between the effective quantities $\widetilde{\cal R}^{}_{i}$, $\Delta \widetilde{\cal R}^{}_{i}$ in matter and their counterpart ${\cal R}^{}_{i}$, $\Delta {\cal R}^{}_{i}$ (defined by Eq.~(\ref{18})) in vacuum
\begin{eqnarray}
\widetilde{\Delta}^{}_{21} \widetilde{\Delta}^{}_{31} \widetilde{\Delta}^{2}_{32} \; \widetilde{\cal R}^{}_{1} & = & \Delta^{2}_{21} (m^{2}_{3} - \widetilde{m}^{2}_{2}) (m^{2}_{3} - \widetilde{m}^{2}_{3}) \; {\cal R}^{}_{3} \nonumber\\
& & + \Delta^{2}_{31} (m^{2}_{2} - \widetilde{m}^{2}_{2}) (m^{2}_{2} - \widetilde{m}^{2}_{3}) \; {\cal R}^{}_{2} \nonumber\\
& & + \Delta^{2}_{32} (m^{2}_{1} - \widetilde{m}^{2}_{2}) (m^{2}_{1} - \widetilde{m}^{2}_{3}) \; {\cal R}^{}_{1} \; , \nonumber\\
- \widetilde{\Delta}^{}_{21} \widetilde{\Delta}^{2}_{31} \widetilde{\Delta}^{}_{32} \; \widetilde{\cal R}^{}_{2} & = & \Delta^{2}_{21} (m^{2}_{3} - \widetilde{m}^{2}_{1}) (m^{2}_{3} - \widetilde{m}^{2}_{3}) \; {\cal R}^{}_{3} \nonumber\\
& & + \Delta^{2}_{31} (m^{2}_{2} - \widetilde{m}^{2}_{1}) (m^{2}_{2} - \widetilde{m}^{2}_{3}) \; {\cal R}^{}_{2} \nonumber\\
& & + \Delta^{2}_{32} (m^{2}_{1} - \widetilde{m}^{2}_{1}) (m^{2}_{1} - \widetilde{m}^{2}_{3}) \; {\cal R}^{}_{1} \; , \nonumber\\
\widetilde{\Delta}^{2}_{21} \widetilde{\Delta}^{}_{31} \widetilde{\Delta}^{}_{32} \; \widetilde{\cal R}^{}_{3} & = & \Delta^{2}_{21} (m^{2}_{3} - \widetilde{m}^{2}_{1}) (m^{2}_{3} - \widetilde{m}^{2}_{2}) \; {\cal R}^{}_{3} \nonumber\\
& & + \Delta^{2}_{31} (m^{2}_{2} - \widetilde{m}^{2}_{1}) (m^{2}_{2} - \widetilde{m}^{2}_{2}) \; {\cal R}^{}_{2} \nonumber\\
& & + \Delta^{2}_{32} (m^{2}_{1} - \widetilde{m}^{2}_{1}) (m^{2}_{1} - \widetilde{m}^{2}_{2}) \; {\cal R}^{}_{1} \; , \nonumber\\[1mm]
\widetilde{\Delta}^{}_{21} \widetilde{\Delta}^{}_{31} \widetilde{\Delta}^{2}_{32} \; \Delta \widetilde{\cal R}^{}_{1} & = & \Delta^{2}_{21} (m^{2}_{3} - \widetilde{m}^{2}_{2}) (m^{2}_{3} - \widetilde{m}^{2}_{3}) \; \Delta {\cal R}^{}_{3} \nonumber\\
& & + \Delta^{2}_{31} (m^{2}_{2} - \widetilde{m}^{2}_{2}) (m^{2}_{2} - \widetilde{m}^{2}_{3}) \; \Delta {\cal R}^{}_{2} \nonumber\\
& & + \Delta^{2}_{32} (m^{2}_{1} - \widetilde{m}^{2}_{2}) (m^{2}_{1} - \widetilde{m}^{2}_{3}) \; \Delta {\cal R}^{}_{1} \; , \nonumber\\
- \widetilde{\Delta}^{}_{21} \widetilde{\Delta}^{2}_{31} \widetilde{\Delta}^{}_{32} \; \Delta \widetilde{\cal R}^{}_{2} & = & \Delta^{2}_{21} (m^{2}_{3} - \widetilde{m}^{2}_{1}) (m^{2}_{3} - \widetilde{m}^{2}_{3}) \; \Delta {\cal R}^{}_{3} \nonumber\\
& & + \Delta^{2}_{31} (m^{2}_{2} - \widetilde{m}^{2}_{1}) (m^{2}_{2} - \widetilde{m}^{2}_{3}) \; \Delta {\cal R}^{}_{2} \nonumber\\
& & + \Delta^{2}_{32} (m^{2}_{1} - \widetilde{m}^{2}_{1}) (m^{2}_{1} - \widetilde{m}^{2}_{3}) \; \Delta {\cal R}^{}_{1} \; , \nonumber\\
\widetilde{\Delta}^{2}_{21} \widetilde{\Delta}^{}_{31} \widetilde{\Delta}^{}_{32} \; \Delta \widetilde{\cal R}^{}_{3} & = & \Delta^{2}_{21} (m^{2}_{3} - \widetilde{m}^{2}_{1}) (m^{2}_{3} - \widetilde{m}^{2}_{2}) \; \Delta {\cal R}^{}_{3} \nonumber\\
& & + \Delta^{2}_{31} (m^{2}_{2} - \widetilde{m}^{2}_{1}) (m^{2}_{2} - \widetilde{m}^{2}_{2}) \; \Delta {\cal R}^{}_{2} \nonumber\\
& & + \Delta^{2}_{32} (m^{2}_{1} - \widetilde{m}^{2}_{1}) (m^{2}_{1} - \widetilde{m}^{2}_{2}) \; \Delta {\cal R}^{}_{1} \; .
\label{41}
\end{eqnarray}
In other words, the effective $\widetilde{\cal R}^{}_{i}$ in matter can be regarded as a linear combination of ${\cal R}^{}_{1}$, ${\cal R}^{}_{2}$, and ${\cal R}^{}_{3}$, while $\Delta \widetilde{\cal R}^{}_{i}$ is a linear combination of $\Delta {\cal R}^{}_{1}$, $\Delta {\cal R}^{}_{2}$, and $\Delta {\cal R}^{}_{3}$ both using the same composition matrix $X^{\mu\tau}_{}$ 
\begin{eqnarray}
\left ( \begin{matrix} \widetilde{\cal R}^{}_{1} \cr \widetilde{\cal R}^{}_{2} \cr \widetilde{\cal R}^{}_{3} \end{matrix} \right ) & = & X^{\mu \tau}_{} \; \left ( \begin{matrix} {\cal R}^{}_{1} \cr {\cal R}^{}_{2} \cr {\cal R}^{}_{3} \end{matrix} \right )  \; = \; \frac{\Delta^{}_{21} \Delta^{}_{31} \Delta^{}_{32}}{\widetilde{\Delta}^{}_{21} \widetilde{\Delta}^{}_{31} \widetilde{\Delta}^{}_{32}} \; Y^{\mu \tau}_{} \; \left ( \begin{matrix} {\cal R}^{}_{1} \cr {\cal R}^{}_{2} \cr {\cal R}^{}_{3} \end{matrix} \right ) \; , \nonumber\\[2mm]
\left ( \begin{matrix} \Delta \widetilde{\cal R}^{}_{1} \cr \Delta \widetilde{\cal R}^{}_{2} \cr \Delta \widetilde{\cal R}^{}_{3} \end{matrix} \right ) & = &  X^{\mu \tau}_{} \; \left ( \begin{matrix} \Delta {\cal R}^{}_{1} \cr \Delta {\cal R}^{}_{2} \cr \Delta {\cal R}^{}_{3} \end{matrix} \right ) \; = \; \frac{\Delta^{}_{21} \Delta^{}_{31} \Delta^{}_{32}}{\widetilde{\Delta}^{}_{21} \widetilde{\Delta}^{}_{31} \widetilde{\Delta}^{}_{32}} \; Y^{\mu \tau}_{} \; \left ( \begin{matrix} \Delta {\cal R}^{}_{1} \cr \Delta {\cal R}^{}_{2} \cr \Delta {\cal R}^{}_{3} \end{matrix} \right ) \; .
\label{42}
\end{eqnarray}

\begin{figure}[]
\begin{center}
\includegraphics[width=\textwidth]{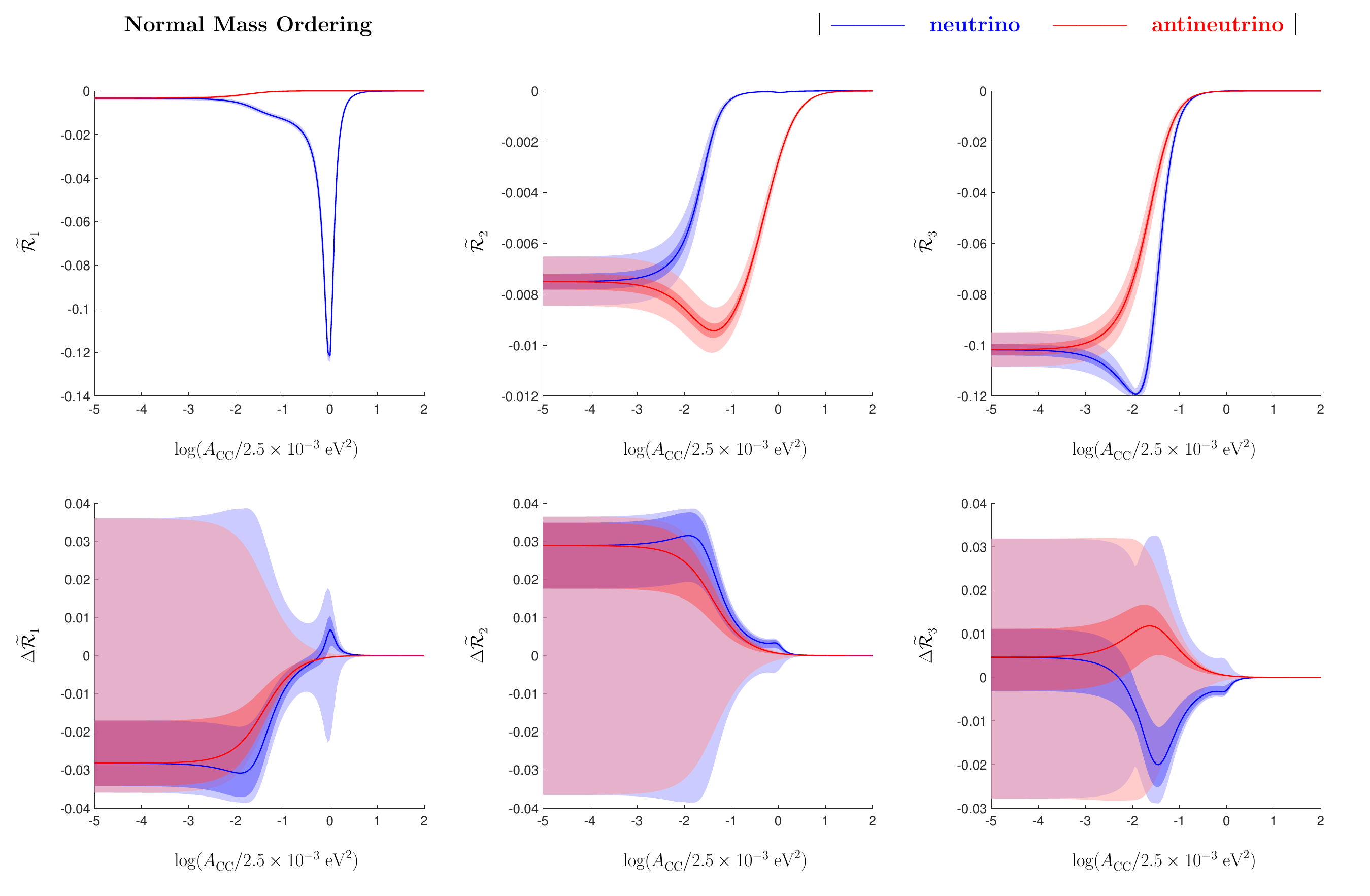}
\caption{The evolution of the effective invariants $\widetilde{\cal R}^{}_{i}$ and $\Delta \widetilde{\cal R}^{}_{i}$ (defined in Eqs.~(\ref{18})) in matter with respect to the dimensionless ratio $A^{}_{\rm CC} / 2.5 \times 10^{-3}_{} {\rm eV}^{2}_{}$ in the normal mass ordering case for both neutrinos (in blue) and antineutrinos (in red), where the best-fit value, $1\sigma$ and $3\sigma$ ranges of $\widetilde{\cal R}^{}_{i}$ and $\Delta \widetilde{\cal R}^{}_{i}$ are present with solid line, dark and light shadows respectively in the plots.}
\label{f12}
\end{center}
\end{figure}

\begin{figure}[]
\begin{center}
\includegraphics[width=\textwidth]{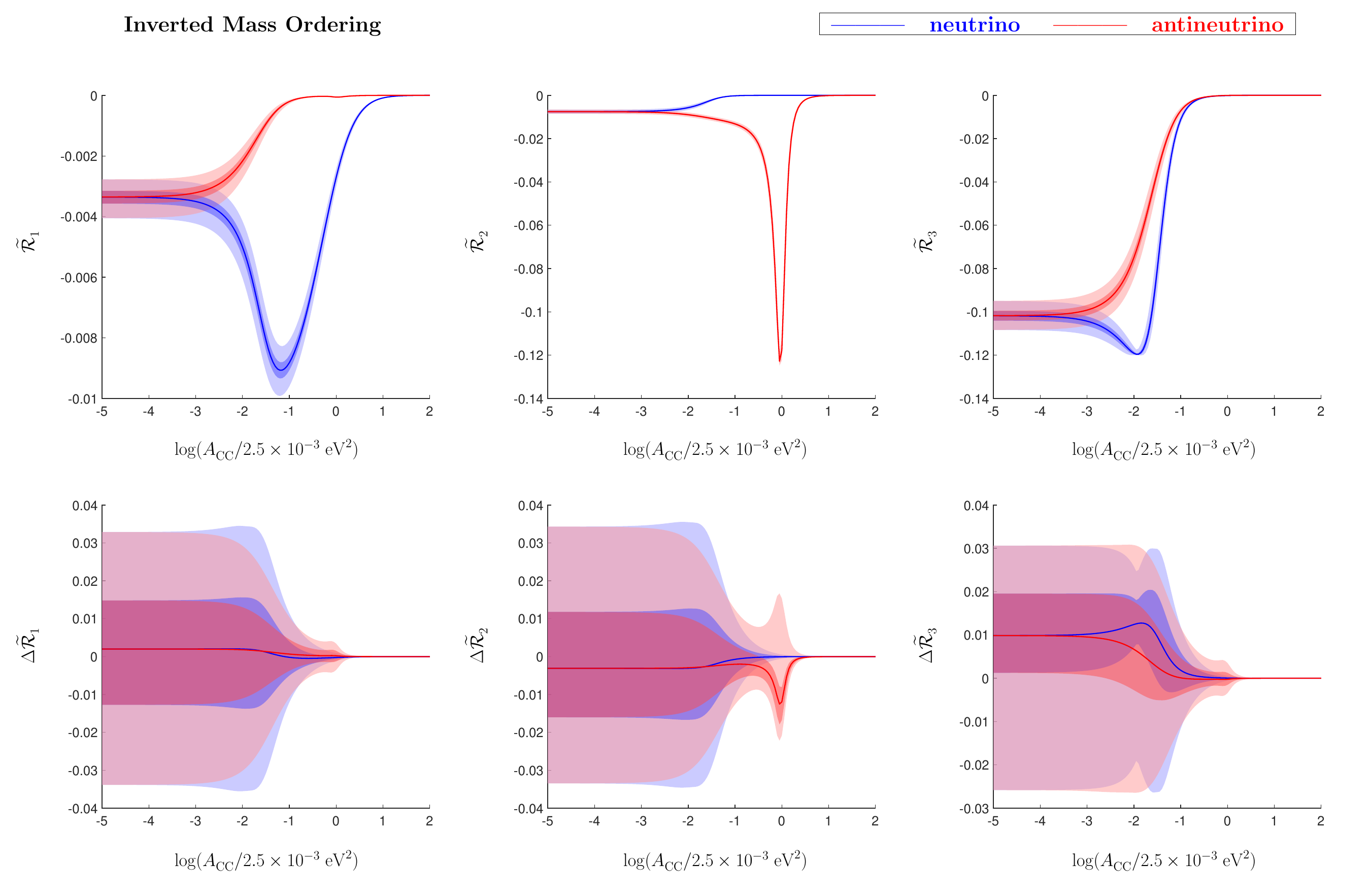}
\caption{The evolution of the effective invariants $\widetilde{\cal R}^{}_{i}$ and $\Delta \widetilde{\cal R}^{}_{i}$ (defined in Eqs.~(\ref{18})) in matter with respect to the dimensionless ratio $A^{}_{\rm CC} / 2.5 \times 10^{-3}_{} {\rm eV}^{2}_{}$ in the inverted mass ordering case for both neutrinos (in blue) and antineutrinos (in red), where the best-fit value, $1\sigma$ and $3\sigma$ ranges of $\widetilde{\cal R}^{}_{i}$ and $\Delta \widetilde{\cal R}^{}_{i}$ are present with solid line, dark and light shadows respectively in the plots.}
\label{f13}
\end{center}
\end{figure}

Given the best-fit values, $1\sigma$ and $3\sigma$ ranges of these invariants in vacuum, we illustrated in Fig.~\ref{f12} (for NMO) and Fig.~\ref{f13} (for IMO) the evolution of the effective $\widetilde{\cal R}^{}_{i}$ and $\Delta \widetilde{\cal R}^{}_{i}$ in matter as the increase of the matter potential.
It is clear that the evolution of $\widetilde{\cal R}^{}_{i}$ is quite certain while most of the uncertainties lie in the evolution of $\Delta \widetilde{\cal R}^{}_{i}$, which is crucial for the determination of $\theta^{}_{23}$ and $\delta$.

We can observe that the pattern of the $\mu$-$\tau$ symmetry is somehow preserved when the matter effect is considered. If the ${\cal R}$ matrix in vacuum exhibits the $\mu$-$\tau$ symmetry, so is the effective $\widetilde{\cal R}$ in matter. The matter effect does not introduce additional $\mu$-$\tau$ symmetry breaking.
Moreover, we can observe that $\widetilde{\cal R}^{}_{\mu i}$ (magnitudes of $\nu^{}_{\tau} \rightarrow \nu^{}_{e}$ oscillations), $\widetilde{\cal R}^{}_{\tau i}$ (magnitudes of $\nu^{}_{\mu} \rightarrow \nu^{}_{e}$ oscillations), $\widetilde{\cal R}^{}_{i}$ (amplitudes of $\nu^{}_{e} \rightarrow \nu^{}_{e}$ oscillations) and $\Delta \widetilde{\cal R}^{}_{i}$ (measures of the $\mu$-$\tau$ symmetry, sensitive to the undetermined parameter $\theta^{}_{23}$ and $\delta$), which may be obtained from various neutrino oscillations including matter effect, can all be converted to ${\cal R}^{}_{\mu i}$, ${\cal R}^{}_{\tau i}$, ${\cal R}^{}_{i}$ and $\Delta {\cal R}^{}_{i}$ using the same composition matrix $X^{\mu\tau}_{}$. This strongly suggests that, to accurately determine $\theta^{}_{12}$ and $\delta$, data from these different experiments could be combined for a comprehensive analysis.

Of course, the above discussions are based on the assumption that the NC interaction is the same for all three flavors, and only the CC interaction contributes to the matter effect. If loop contributions are considered~\cite{Botella:1986wy, Huang:2025apv, Huang:2025iww}, or new physics such as non-standard interaction (NSI) is taken into account~\cite{Farzan:2017xzy, Ge:2018uhz}, the additional $\mu$-$\tau$ symmetry introduced by the matter effect and the corrections to Eqs.~(\ref{30}) and (\ref{42}) must be carefully examined.

\section{Summary}

The UTs provide us with a geometrical language to describe the flavor issues of massive neutrinos, making underlying physics such as leptonic CP violation more transparent and intuitive.
In the quark sector, one unitarity triangle $\triangle^{}_{s}$ has been established at the KEK-B and SLAC-B factories, which is complementary to direct measurements of CP-asymmetry and is considered a valuable method to test the unitarity of the CKM matrix. Further studies of the other five triangles will be implemented at LHC-B in the near future. 
We are convinced that similar steps will be adopted for the experimental reconstruction of the UTs in the lepton sector and the determination of leptonic CP violation in the era of precision neutrino physics.
In the lepton sector, varieties of neutrino oscillation experiments play a crucial role in clarifying the pattern of neutrino masses and mixing. 
The challenge lies in establishing a connection between leptonic UTs and realistic oscillation experiments.
To address this, we proposed the rescaled UTs characterized entirely by the rephasing invariants ${\cal R}^{}_{\gamma k} \equiv {\rm Re} \left [ V^{}_{\alpha i} V^{}_{\beta j} V^{*}_{\alpha j} V^{*}_{\beta i}\right ]$ and ${\cal J} \equiv \sum_\gamma \epsilon^{}_{\alpha\beta\gamma} \sum_k \epsilon^{}_{ijk} \; {\rm Im} \left [ V^{}_{\alpha i} V^{}_{\beta j} V^{*}_{\alpha j} V^{*}_{\beta i} \right ]$, which correspond precisely to the oscillation amplitudes observable in various neutrino oscillation experiments.
In this way, we bridge the gap between the UTs and realistic oscillation experiments.
We can now readily determine how many measurements and of what kind (from appearance/disappearance neutrino oscillation experiments or astrophysical measurements) are necessary to reconstruct a chosen UT.

In this paper, the matter effect is discussed in detail.
Instead of the classical approaches making use of the effective PMNS matrix and the effective neutrino mass-squared differences, we employ the well-known Naumov relation of the Jarlskog invariant ${\cal J}$ and introduce Naumov-like relations for ${\cal R}^{}_{\alpha i}$. These relations concisely connect these rephasing invariants with their effective counterparts $\widetilde{\cal J}$ and $\widetilde{\cal R}^{}_{\alpha i}$ in matter.
With these direct and concise relations, we can not only easily describe how the effective UTs in matter evolve as the matter potential increases but also transparently and intuitively explore matter effect on the behaviors of neutrino oscillations in the vacuum-dominated, resonances, or matter-dominated regions.

We find that the effective CP-conserving invariants $\widetilde{\cal R}^{}_{\alpha i}$ in matter can be regarded as  a linear combination of their vacuum counterparts ${\cal R}^{}_{\alpha 1}$, ${\cal R}^{}_{\alpha 2}$, and ${\cal R}^{}_{\alpha 3}$. The corresponding composition matrices $X^{e}_{}$ and $X^{\mu\tau}_{}$ are mostly determined by the matter potential $A^{}_{CC}$, neutrino mass-squared differences, and the first row of the PMNS matrix, due to the fact that only the CC contribution to the coherent $\nu^{}_{e} e^{-}_{}$ forward scattering affects the neutrino oscillation behaviors in matter.
Thanks to appreciable progress in the last few years, the measurements of the first row of the PMNS matrix and the neutrino mass-squared differences have entered the subpercent precision era. We observe that the evolution of the composition matrices $X^{e}_{}$ and $X^{\mu\tau}_{}$ with respect to the matter potential is well understood once the neutrino mass ordering is established.
The JUNO reactor experiment recently measured $\theta^{}_{12}$ and $\Delta^{}_{21}$ with improved accuracy~\cite{JUNO:2025gmd, JUNO:2022mxj}, and it is expected to achieve increasing sensitivity to the mass ordering options~\cite{JUNO:2015zny, JUNO:2024jaw}. This kind of improvement increases the accuracy we predict the composition matrices.
With the language of composition matrices, data from accelerator, reactor, and atmospheric neutrino experiments, which probe different energy ranges, targets, and baselines, can be effectively converted to ${\cal R}^{}_{\alpha i}$ and ${\cal J}$ in vacuum using these relations. The synergy among the different experiments, combined in a global analysis, will certainly help reduce uncertainties in those unknown mixing parameters ($\theta^{}_{23}$ octant, CP phase $\delta$, neutrino mass ordering). 

Finally, it is worth mentioning that although the discussions in this paper are carried out within the standard three-neutrino framework, the relations expressed using these rephasing invariants can be very helpful for testing the unitarity of the PMNS matrix (by checking the consistencies of the relations from different experimental measurements) and  for identifying possible deviations caused by new physics such as light or heavy sterile neutrinos, or non-standard neutrino interactions.
The above discussion also offers a convenient foundation or starting point that can be extended to examine potential corrections induced by the new physics.
A further study along this line of thought is beyond the scope of this paper and will be presented elsewhere.

\begin{acknowledgements}
This work is supported in part by the National Natural Science Foundation of China under grant No. 11775183.
\end{acknowledgements}

\appendix

\section{Numerical results on the evolution of the effective rephasing invariants and the rescaled UTs in matter}

\begin{figure}[h]
\begin{center}
\vspace{-1cm}
\includegraphics[width=\textwidth]{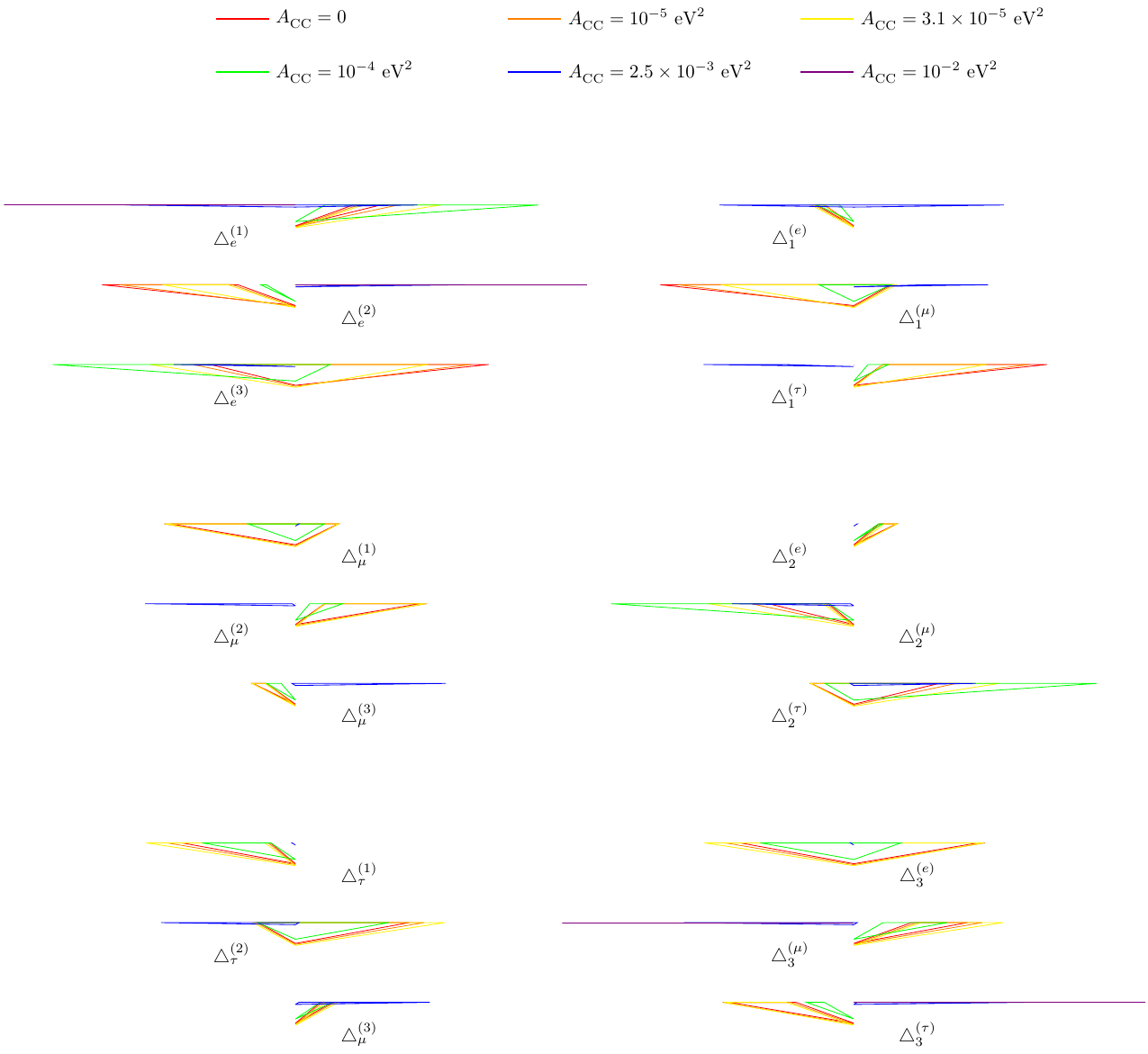}
\vspace{-4cm}
\caption{The evolution of 18 rescaled UTs as the increasing of the matter potential $A^{}_{\rm CC}$ for {\bf neutrinos} in the {\bf NMO} case, where the explicit values of these rephasing invariants can be found in Table~\ref{t6}}
\label{f14}
\end{center}
\end{figure}

\begin{figure}[h]
\begin{center}
\vspace{-1cm}
\includegraphics[width=\textwidth]{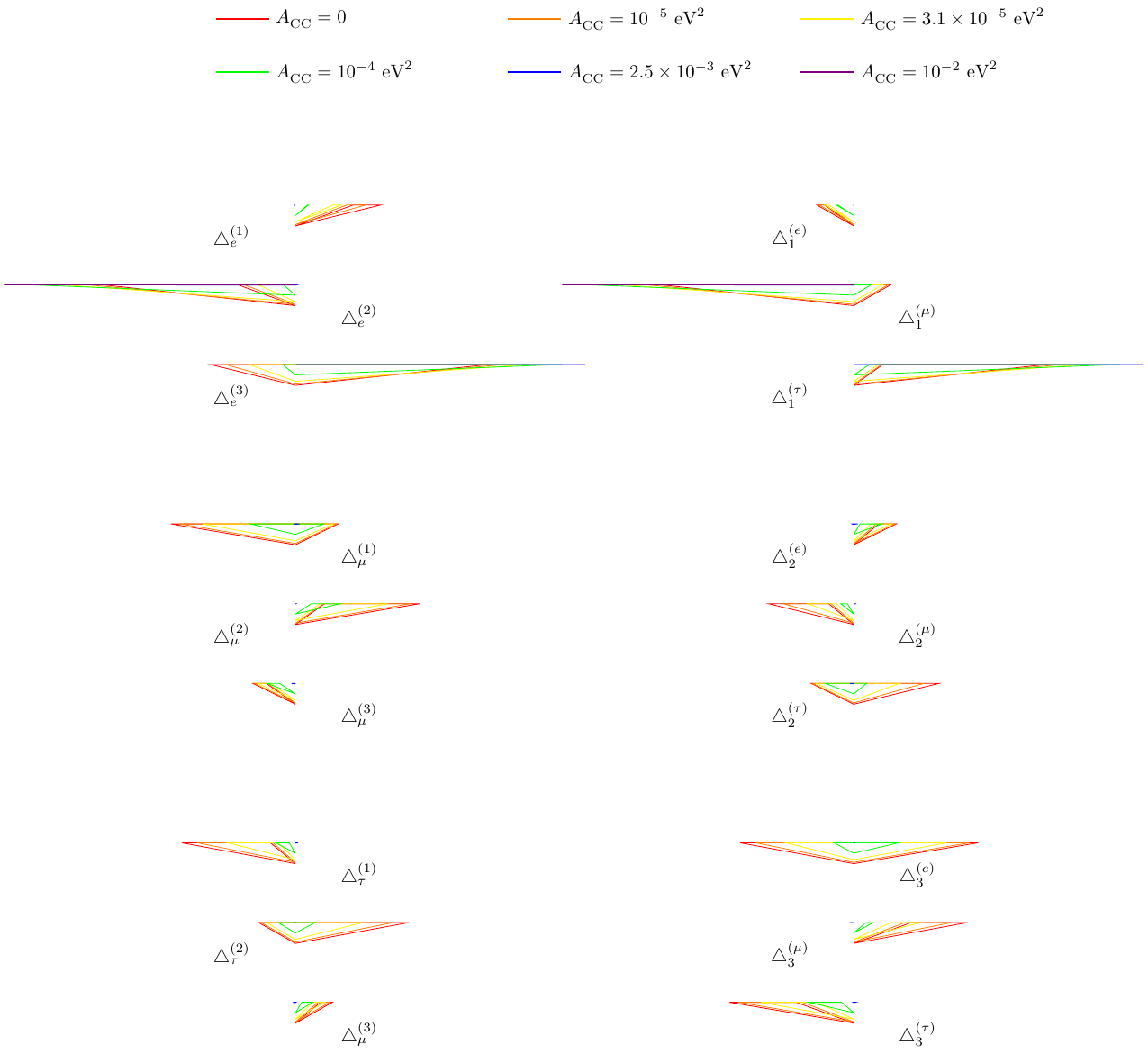}
\vspace{-4cm}
\caption{The evolution of 18 rescaled UTs as the increasing of the matter potential $A^{}_{\rm CC}$ for {\bf antineutrinos} in the {\bf NMO} case, where the explicit values of these rephasing invariants can be found in Table~\ref{t7}}
\label{f15}
\end{center}
\end{figure}

\begin{figure}[h]
\begin{center}
\vspace{-1cm}
\includegraphics[width=\textwidth]{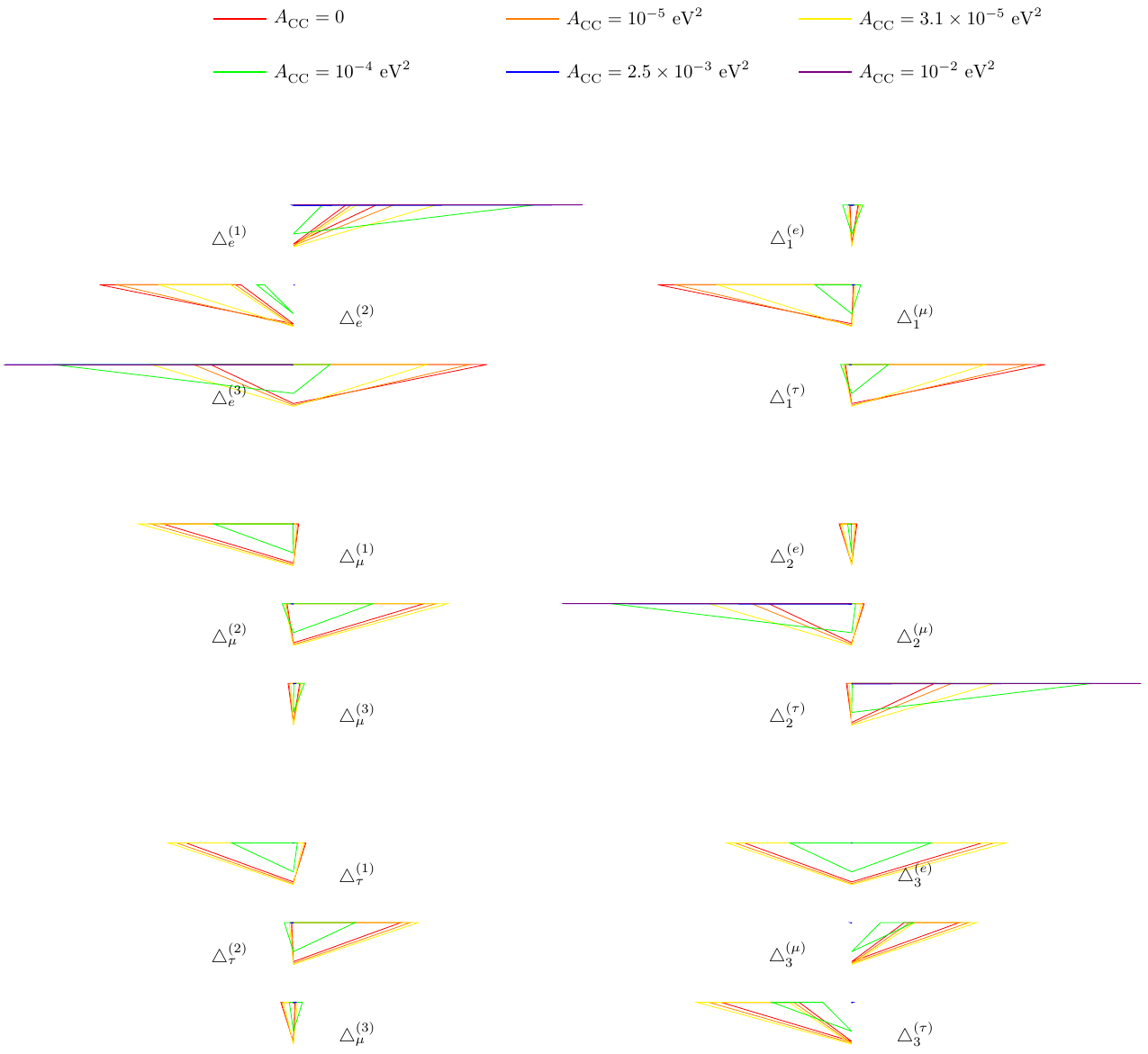}
\vspace{-4cm}
\caption{The evolution of 18 rescaled UTs as the increasing of the matter potential $A^{}_{\rm CC}$ for {\bf neutrinos} in the {\bf IMO} case, where the explicit values of these rephasing invariants can be found in Table~\ref{t8}}
\label{f16}
\end{center}
\end{figure}

\begin{figure}[h]
\begin{center}
\vspace{-1cm}
\includegraphics[width=\textwidth]{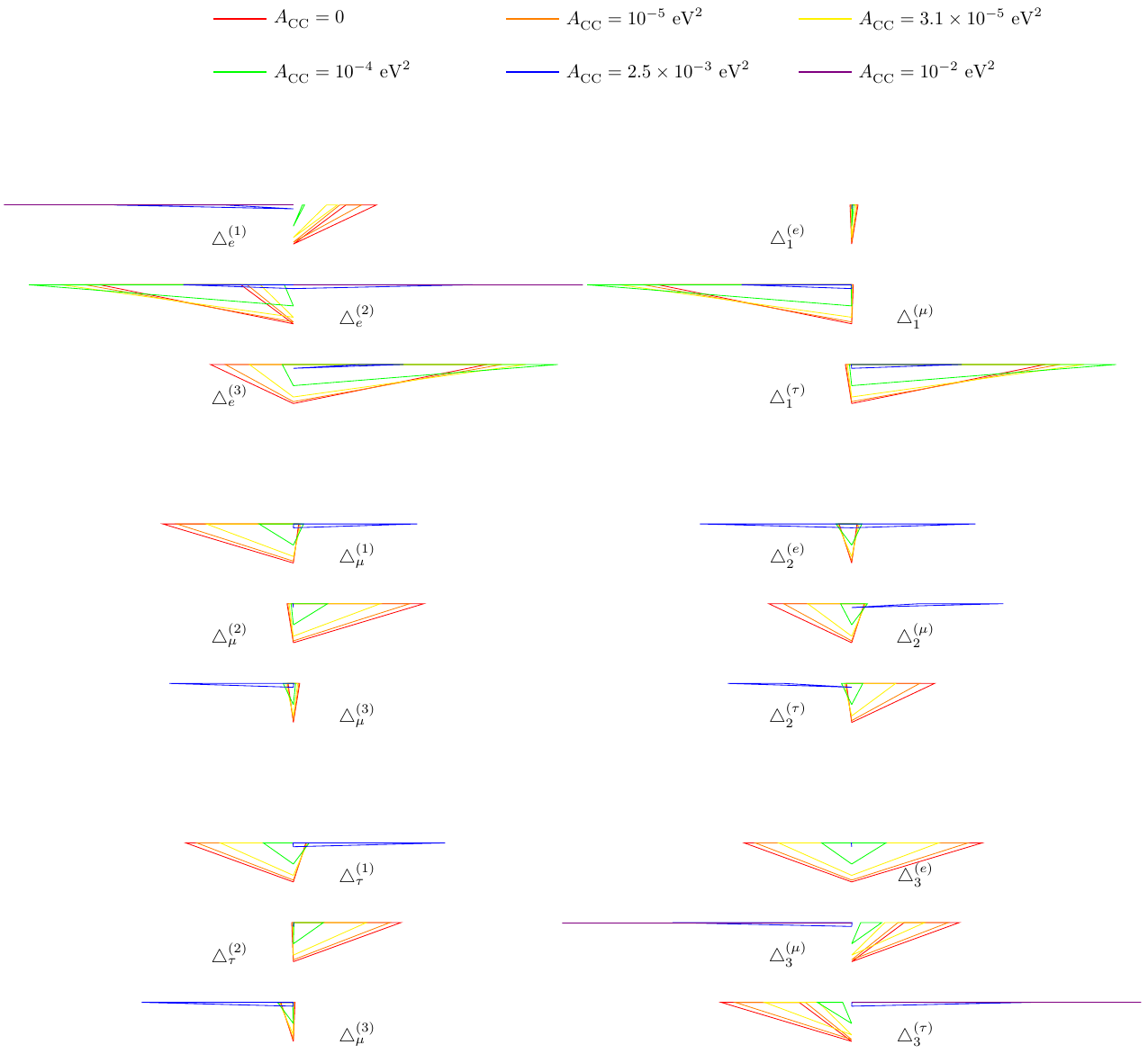}
\vspace{-4cm}
\caption{The evolution of 18 rescaled UTs as the increasing of the matter potential $A^{}_{\rm CC}$ for {\bf antineutrinos} in the {\bf IMO} case, where the explicit values of these rephasing invariants can be found in Table~\ref{t9}}
\label{f17}
\end{center}
\end{figure}

In this appendix, we present the evolution of effective rescaled UTs as the increasing of the matter potential $A^{}_{\rm CC}$ with our numerical results. Six benchmark values of $A^{}_{\rm CC}$ are chosen to cover all the regions from ``vacuum-dominated'' through ``resonances'' to the ``matter-dominated'', and the corresponding values of these effective rephasing invariants are given in Tables~\ref{t6}-\ref{t9}, where the best-fit values of the unitary PMNS leptonic mixing matrix together with the two mass-squared differences from the latest global fit~\cite{Esteban:2024eli,Nufit} are adopted as inputs for the vacuum case. The evolution of these 18 rescaled UTs in each scenario is also visually illustrated in Figures~\ref{f14}-\ref{f17}. Note that in the matter-dominated region, the effective $\widetilde{\cal J}$ and eight out of the nine $\widetilde{\cal R}^{}_{\alpha i}$ are all approaching zero, all the UTs either collapse or shrink to a point and therefore are not clearly shown in the figures.

\begin{table}[h]
\footnotesize
\caption{The evolution of the Jarlskog CP-violating invariant ${\cal J}$ and the nine CP-conserving invariants ${\cal R}^{}_{\alpha i}$ as the increasing of the matter potential $A^{}_{\rm CC}$ for {\bf neutrinos} in the {\bf NMO} case.}
\label{t6} 
\begin{tabular}{ccccccccccccc}
\hline\noalign{\smallskip}
&& \multicolumn{3}{c}{Vacuum-Dominated} && \multicolumn{5}{c}{Resonances} && Matter-Dominated \\
\noalign{\smallskip}
$A^{}_{\rm CC}$ && $0$ && $10^{-5} ~{\rm eV}^{2}$ && $3.1 \times 10^{-5}_{} ~{\rm eV}^{2}_{}$ && $10^{-4}_{} ~{\rm eV}^{2}_{}$ && $2.5 \times 10^{-3}_{} ~{\rm eV}^{2}_{}$ && $10^{-2}_{} ~{\rm eV}^{2}_{}$ \\[-2mm]
&& (vacuum) &&  && (solar) &&  && (atmospheric) &&  \\[-1mm]
\noalign{\smallskip}\hline\noalign{\smallskip}
$\widetilde{\cal J}$ ~ & ~~~ & -0.0178 & ~ & -0.0186 & ~ & -0.0195 & ~ & -0.0142 & ~ & -0.0018 & ~ & $-3.5 \times 10^{-7}$ \\
\noalign{\smallskip}
$\widetilde{\cal R}^{}_{e1}$ && -0.1653 && -0.1504 && -0.1136 && -0.0297 && 0.0606 && $3.5 \times 10^{-6}$ \\
$\widetilde{\cal R}^{}_{e2}$ && -0.0729 && -0.0877 && -0.1243 && -0.2076 && -0.1042 && $-3.5 \times 10^{-6}$ \\
$\widetilde{\cal R}^{}_{e3}$ && 0.0493 && 0.0539 && 0.0578 && 0.0252 && -0.1451 && -0.2492 \\
$\widetilde{\cal R}^{}_{\mu1}$ && 0.0249 && 0.0256 && 0.0251 && 0.0122 && -0.1285 && $-8.1 \times 10^{-6}$ \\
$\widetilde{\cal R}^{}_{\mu2}$ && -0.0364 && -0.0371 && -0.0369 && -0.0246 && -0.0030 && $-6.2 \times 10^{-7}$ \\
$\widetilde{\cal R}^{}_{\mu3}$ && -0.1064 && -0.1121 && -0.1110 && -0.0404 && 0.0029 && $5.6 \times 10^{-7}$ \\
$\widetilde{\cal R}^{}_{\tau1}$ && -0.0316 && -0.0336 && -0.0365 && -0.0321 && -0.1148 && $-6.1 \times 10^{-6}$ \\
$\widetilde{\cal R}^{}_{\tau2}$ && 0.0214 && 0.0233 && 0.0261 && 0.0211 && 0.0029 && $5.0 \times 10^{-7}$ \\
$\widetilde{\cal R}^{}_{\tau3}$ && -0.0972 && -0.1100 && -0.1276 && -0.0799 && -0.0030 && $-5.6 \times 10^{-7}$ \\
\noalign{\smallskip}
\hline
\end{tabular}
\end{table}

\begin{table}[h]
\footnotesize
\caption{The evolution of the Jarlskog CP-violating invariant ${\cal J}$ and the nine CP-conserving invariants ${\cal R}^{}_{\alpha i}$ as the increasing of the matter potential $A^{}_{\rm CC}$ for {\bf antineutrinos} in the {\bf NMO} case.}
\label{t7} 
\begin{tabular}{ccccccccccccc}
\hline\noalign{\smallskip}
&& \multicolumn{3}{c}{Vacuum-Dominated} && \multicolumn{5}{c}{Resonances} && Matter-Dominated \\
\noalign{\smallskip}
$A^{}_{\rm CC}$ && $0$ && $10^{-5} ~{\rm eV}^{2}$ && $3.1 \times 10^{-5}_{} ~{\rm eV}^{2}_{}$ && $10^{-4}_{} ~{\rm eV}^{2}_{}$ && $2.5 \times 10^{-3}_{} ~{\rm eV}^{2}_{}$ && $10^{-2}_{} ~{\rm eV}^{2}_{}$ \\[-2mm]
&& (vacuum) &&  && (solar) &&  && (atmospheric) &&  \\[-1mm]
\noalign{\smallskip}\hline\noalign{\smallskip}
$\widetilde{\cal J}$ ~ & ~~~ & -0.0178 & ~ & -0.0167 & ~ & -0.0144 & ~ & -0.0089 & ~ & -0.0003 & ~ & $-3.3 \times 10^{-7}$ \\
\noalign{\smallskip}
$\widetilde{\cal R}^{}_{e1}$ && -0.1653 && -0.178 && -0.1992 && -0.2279 && -0.2477 && -0.2492 \\
$\widetilde{\cal R}^{}_{e2}$ && -0.0729 && -0.0601 && -0.0393 && -0.0111 && 0.0013 && $3.2 \times 10^{-6}$ \\
$\widetilde{\cal R}^{}_{e3}$ && 0.0493 && 0.0438 && 0.0319 && 0.0103 && -0.0013 && $-3.2 \times 10^{-6}$ \\
$\widetilde{\cal R}^{}_{\mu1}$ && 0.0249 && 0.0239 && 0.0211 && 0.0136 && 0.0004 && $5.3 \times 10^{-7}$ \\
$\widetilde{\cal R}^{}_{\mu2}$ && -0.0364 && -0.0353 && -0.0323 && -0.0243 && -0.0034 && $-7.4 \times 10^{-6}$ \\
$\widetilde{\cal R}^{}_{\mu3}$ && -0.1064 && -0.0986 && -0.0794 && -0.0385 && -0.0005 && $-5.9 \times 10^{-7}$ \\
$\widetilde{\cal R}^{}_{\tau1}$ && -0.0316 && -0.0294 && -0.0248 && -0.0148 && -0.0004 && $-5.3 \times 10^{-7}$ \\
$\widetilde{\cal R}^{}_{\tau2}$ && 0.0214 && 0.0193 && 0.0148 && 0.0054 && -0.0022 && $-5.6 \times 10^{-6}$ \\
$\widetilde{\cal R}^{}_{\tau3}$ && -0.0972 && -0.0838 && -0.0578 && -0.0169 && 0.0003 && $4.7 \times 10^{-7}$ \\
\noalign{\smallskip}
\hline
\end{tabular}
\end{table}

\begin{table}[h]
\footnotesize
\caption{The evolution of the Jarlskog CP-violating invariant ${\cal J}$ and the nine CP-conserving invariants ${\cal R}^{}_{\alpha i}$ as the increasing of the matter potential $A^{}_{\rm CC}$ for {\bf neutrinos} in the {\bf IMO} case.}
\label{t8} 
\begin{tabular}{ccccccccccccc}
\hline\noalign{\smallskip}
&& \multicolumn{3}{c}{Vacuum-Dominated} && \multicolumn{5}{c}{Resonances} && Matter-Dominated \\
\noalign{\smallskip}
$A^{}_{\rm CC}$ && $0$ && $10^{-5} ~{\rm eV}^{2}$ && $3.1 \times 10^{-5}_{} ~{\rm eV}^{2}_{}$ && $10^{-4}_{} ~{\rm eV}^{2}_{}$ && $2.5 \times 10^{-3}_{} ~{\rm eV}^{2}_{}$ && $10^{-2}_{} ~{\rm eV}^{2}_{}$ \\[-2mm]
&& (vacuum) &&  && (solar) &&  && (atmospheric) &&  \\[-1mm]
\noalign{\smallskip}\hline\noalign{\smallskip}
$\widetilde{\cal J}$ ~ & ~~~ & -0.0334 & ~ & -0.0348 & ~ & -0.0358 & ~ & -0.0248 & ~ & -0.0005 & ~ & $-6.1 \times 10^{-7}$ \\
\noalign{\smallskip}
$\widetilde{\cal R}^{}_{e1}$ && -0.1656 && -0.1510 && -0.1147 && -0.0314 && 0.0013 && $3.0 \times 10^{-6}$ \\
$\widetilde{\cal R}^{}_{e2}$ && -0.0709 && -0.0856 && -0.1221 && -0.2059 && -0.2460 && -0.2474 \\
$\widetilde{\cal R}^{}_{e3}$ && 0.0449 && 0.0495 && 0.0537 && 0.0247 && -0.0013 && $-3.0 \times 10^{-6}$ \\
$\widetilde{\cal R}^{}_{\mu1}$ && -0.0054 && -0.0060 && -0.0074 && -0.0095 && -0.0025 && $-5.6 \times 10^{-6}$ \\
$\widetilde{\cal R}^{}_{\mu2}$ && -0.0045 && -0.0037 && -0.0021 && 0.0004 && $3.5 \times 10^{-5}$ && $4.3 \times 10^{-8}$ \\
$\widetilde{\cal R}^{}_{\mu3}$ && -0.1116 && -0.1221 && -0.1322 && -0.0680 && -0.0001 && $-1.1 \times 10^{-7}$ \\
$\widetilde{\cal R}^{}_{\tau1}$ && -0.0014 && -0.0019 && -0.0035 && -0.0077 && -0.0029 && $-6.8 \times 10^{-6}$ \\
$\widetilde{\cal R}^{}_{\tau2}$ && -0.0107 && -0.0100 && -0.0082 && -0.0034 && $-3.7 \times 10^{-5}$ && $-4.3 \times 10^{-8}$ \\
$\widetilde{\cal R}^{}_{\tau3}$ && -0.0919 && -0.1000 && -0.1068 && -0.0529 && $-5.3 \times 10^{-5}$ && $-1.3 \times 10^{-8}$ \\
\noalign{\smallskip}
\hline
\end{tabular}
\end{table}

\begin{table}[h]
\footnotesize
\caption{The evolution of the Jarlskog CP-violating invariant ${\cal J}$ and the nine CP-conserving invariants ${\cal R}^{}_{\alpha i}$ as the increasing of the matter potential $A^{}_{\rm CC}$ for {\bf antineutrinos} in the {\bf IMO} case.}
\label{t9} 
\begin{tabular}{ccccccccccccc}
\hline\noalign{\smallskip}
&& \multicolumn{3}{c}{Vacuum-Dominated} && \multicolumn{5}{c}{Resonances} && Matter-Dominated \\
\noalign{\smallskip}
$A^{}_{\rm CC}$ && $0$ && $10^{-5} ~{\rm eV}^{2}$ && $3.1 \times 10^{-5}_{} ~{\rm eV}^{2}_{}$ && $10^{-4}_{} ~{\rm eV}^{2}_{}$ && $2.5 \times 10^{-3}_{} ~{\rm eV}^{2}_{}$ && $10^{-2}_{} ~{\rm eV}^{2}_{}$ \\[-2mm]
&& (vacuum) &&  && (solar) &&  && (atmospheric) &&  \\[-1mm]
\noalign{\smallskip}\hline\noalign{\smallskip}
$\widetilde{\cal J}$ ~ & ~~~ & -0.0334 & ~ & -0.0318 & ~ & -0.0278 & ~ & -0.0181 & ~ & -0.0033 & ~ & $-6.4 \times 10^{-7}$ \\
\noalign{\smallskip}
$\widetilde{\cal R}^{}_{e1}$ && -0.1656 && -0.1782 && -0.1987 && -0.2261 && -0.0941 && $-3.4 \times 10^{-6}$ \\
$\widetilde{\cal R}^{}_{e2}$ && -0.0709 && -0.0582 && -0.0376 && -0.0095 && 0.0583 && $3.4 \times 10^{-6}$ \\
$\widetilde{\cal R}^{}_{e3}$ && 0.0449 && 0.0396 && 0.0283 && 0.0077 && -0.1532 && -0.2474 \\
$\widetilde{\cal R}^{}_{\mu1}$ && -0.0054 && -0.0048 && -0.0037 && -0.0019 && -0.0003 && $-1.1 \times 10^{-7}$ \\
$\widetilde{\cal R}^{}_{\mu2}$ && -0.0045 && -0.0051 && -0.0064 && -0.0087 && -0.1059 && $-6.2 \times 10^{-6}$ \\
$\widetilde{\cal R}^{}_{\mu3}$ && -0.1116 && -0.0996 && -0.0744 && -0.0293 && 0.0002 && $4.4 \times 10^{-8}$ \\
$\widetilde{\cal R}^{}_{\tau1}$ && -0.0014 && -0.0009 && -0.0002 && 0.0005 && 0.0002 && $-1.1 \times 10^{-8}$ \\
$\widetilde{\cal R}^{}_{\tau2}$ && -0.0107 && -0.0112 && -0.0121 && -0.0135 && -0.1297 && $-7.5 \times 10^{-6}$ \\
$\widetilde{\cal R}^{}_{\tau3}$ && -0.0919 && -0.0825 && -0.0625 && -0.0257 && -0.0003 && $-4.5 \times 10^{-8}$ \\
\noalign{\smallskip}
\hline
\end{tabular}
\end{table}

\begin{thebibliography}{99}

\bibitem{ParticleDataGroup:2022pth}
R.~L.~Workman \textit{et al.} [Particle Data Group],
``Review of Particle Physics,''
PTEP \textbf{2022}, 083C01 (2022).

\bibitem{Maki:1962mu}
Z.~Maki, M.~Nakagawa and S.~Sakata,
``Remarks on the unified model of elementary particles,''
Prog. Theor. Phys. \textbf{28}, 870-880 (1962).

\bibitem{Pontecorvo:1967fh}
B.~Pontecorvo,
``Neutrino Experiments and the Problem of Conservation of Leptonic Charge,''
Zh. Eksp. Teor. Fiz. \textbf{53}, 1717-1725 (1967).

\bibitem{Harrison:2006bj}
P.~F.~Harrison, W.~G.~Scott and T.~J.~Weiler,
``Real Invariant Matrices and Flavour-Symmetric Mixing Variables with Emphasis on Neutrino Oscillations,''
Phys. Lett. B \textbf{641}, 372-380 (2006)
[arXiv:hep-ph/0607335 [hep-ph]].

\bibitem{Jarlskog:1985ht}
C.~Jarlskog,
``Commutator of the Quark Mass Matrices in the Standard Electroweak Model and a Measure of Maximal CP Nonconservation,''
Phys. Rev. Lett. \textbf{55}, 1039 (1985).

\bibitem{Jarlskog:1985cw}
C.~Jarlskog,
``A Basis Independent Formulation of the Connection Between Quark Mass Matrices, CP Violation and Experiment,''
Z. Phys. C \textbf{29}, 491-497 (1985).

\bibitem{T2K:2019bcf}
K.~Abe \textit{et al.} [T2K],
``Constraint on the matter\textendash{}antimatter symmetry-violating phase in neutrino oscillations,''
Nature \textbf{580}, no.7803, 339-344 (2020)
[erratum: Nature \textbf{583}, no.7814, E16 (2020)]
[arXiv:1910.03887 [hep-ex]].

\bibitem{T2K:2023smv}
K.~Abe \textit{et al.} [T2K],
``Measurements of neutrino oscillation parameters from the T2K experiment using $3.6\times 10^{21}$ protons on target,''
Eur. Phys. J. C \textbf{83}, no.9, 782 (2023)
[arXiv:2303.03222 [hep-ex]].

\bibitem{Fritzsch:1999ee}
H.~Fritzsch and Z.~Z.~Xing,
``Mass and flavor mixing schemes of quarks and leptons,''
Prog. Part. Nucl. Phys. \textbf{45}, 1-81 (2000)
[arXiv:hep-ph/9912358 [hep-ph]].

\bibitem{Aguilar-Saavedra:2000jom}
J.~A.~Aguilar-Saavedra and G.~C.~Branco,
``Unitarity triangles and geometrical description of CP violation with Majorana neutrinos,''
Phys. Rev. D \textbf{62}, 096009 (2000)
[arXiv:hep-ph/0007025 [hep-ph]].

\bibitem{Farzan:2002ct}
Y.~Farzan and A.~Y.~Smirnov,
``Leptonic unitarity triangle and CP violation,''
Phys. Rev. D \textbf{65}, 113001 (2002)
[arXiv:hep-ph/0201105 [hep-ph]].

\bibitem{Ahuja:2007cu}
G.~Ahuja and M.~Gupta,
``Constructing the Leptonic Unitarity Triangle,''
Phys. Rev. D \textbf{77}, 057301 (2008)
[arXiv:hep-ph/0702129 [hep-ph]].

\bibitem{Dueck:2010fa}
A.~Dueck, S.~Petcov and W.~Rodejohann,
``On Leptonic Unitary Triangles and Boomerangs,''
Phys. Rev. D \textbf{82}, 013005 (2010)
[arXiv:1006.0227 [hep-ph]].

\bibitem{He:2013rba}
H.~J.~He and X.~J.~Xu,
``Connecting Leptonic Unitarity Triangle to Neutrino Oscillation,''
Phys. Rev. D \textbf{89}, no.7, 073002 (2014)
[arXiv:1311.4496 [hep-ph]].

\bibitem{He:2016dco}
H.~J.~He and X.~J.~Xu,
``Connecting the leptonic unitarity triangle to neutrino oscillation with CP violation in the vacuum and in matter,''
Phys. Rev. D \textbf{95}, no.3, 033002 (2017)
[arXiv:1606.04054 [hep-ph]].

\bibitem{Xing:2012zv}
Z.~Z.~Xing,
``Model-independent access to the structure of quark flavor mixing,''
Phys. Rev. D \textbf{86}, 113006 (2012)
[arXiv:1211.3890 [hep-ph]].

\bibitem{Luo:2023xmv}
S.~Luo and Z.~Z.~Xing,
``A Pythagoras-like theorem for CP violation in neutrino oscillations,''
Phys. Lett. B \textbf{845}, 138142 (2023)
[arXiv:2306.16231 [hep-ph]].

\bibitem{Esteban:2024eli}
I.~Esteban, M.~C.~Gonzalez-Garcia, M.~Maltoni, I.~Martinez-Soler, J.~P.~Pinheiro and T.~Schwetz,
``NuFit-6.0: Updated global analysis of three-flavor neutrino oscillations,''
[arXiv:2410.05380 [hep-ph]].

\bibitem{Nufit}
NuFIT 6.0 (2024), www.nu-fit.org.

\bibitem{Harrison:2009bb}
P.~F.~Harrison, D.~R.~J.~Roythorne and W.~G.~Scott,
``Is the Unitarity Triangle Right?,''
[arXiv:0904.3014 [hep-ph]].

\bibitem{Harrison:2009bz}
P.~F.~Harrison, S.~Dallison and W.~G.~Scott,
``The Matrix of Unitarity Triangle Angles for Quarks,''
Phys. Lett. B \textbf{680}, 328-333 (2009)
[arXiv:0904.3077 [hep-ph]].

\bibitem{Xing:2009eg}
Z.~Z.~Xing,
``Right Unitarity Triangles, Stable CP-violating Phases and Approximate Quark-Lepton Complementarity,''
Phys. Lett. B \textbf{679}, 111-117 (2009)
[arXiv:0904.3172 [hep-ph]].

\bibitem{Antusch:2011sx}
S.~Antusch, S.~F.~King, C.~Luhn and M.~Spinrath,
``Right Unitarity Triangles and Tri-Bimaximal Mixing from Discrete Symmetries and Unification,''
Nucl. Phys. B \textbf{850}, 477-504 (2011)
[arXiv:1103.5930 [hep-ph]].

\bibitem{Mimura:2018sbc}
Y.~Mimura,
``Nearly right unitarity triangle and CP phase in quark and lepton flavor mixings,''
[arXiv:1805.07773 [hep-ph]].

\bibitem{Sasaki:1986jv}
K.~Sasaki,
``Renormalization Group Equations for the {Kobayashi-Maskawa} Matrix,''
Z. Phys. C \textbf{32}, 149-152 (1986)

\bibitem{Xing:2014zka}
Z.~Z.~Xing and S.~Zhou,
``A partial \ensuremath{\mu} \textendash{} \ensuremath{\tau} symmetry and its prediction for leptonic CP violation,''
Phys. Lett. B \textbf{737}, 196-200 (2014)
[arXiv:1404.7021 [hep-ph]].

\bibitem{Xing:2015fdg}
Z.~Z.~Xing and Z.~H.~Zhao,
``A review of \ensuremath{\mu}-\ensuremath{\tau} flavor symmetry in neutrino physics,''
Rept. Prog. Phys. \textbf{79}, no.7, 076201 (2016)
[arXiv:1512.04207 [hep-ph]].

\bibitem{Xing:2008fg}
Z.~Z.~Xing and S.~Zhou,
``Implications of Leptonic Unitarity Violation at Neutrino Telescopes,''
Phys. Lett. B \textbf{666}, 166-172 (2008)
[arXiv:0804.3512 [hep-ph]].

\bibitem{Wolfenstein:1977ue}
L.~Wolfenstein,
``Neutrino Oscillations in Matter,''
Phys. Rev. D \textbf{17}, 2369-2374 (1978).

\bibitem{Mikheyev:1985zog}
S.~P.~Mikheyev and A.~Y.~Smirnov,
``Resonance Amplification of Oscillations in Matter and Spectroscopy of Solar Neutrinos,''
Sov. J. Nucl. Phys. \textbf{42}, 913-917 (1985).

\bibitem{Naumov:1991ju}
V.~A.~Naumov,
``Three neutrino oscillations in matter, CP violation and topological phases,''
Int. J. Mod. Phys. D \textbf{1}, 379-399 (1992).

\bibitem{Barger:1980tf}
V.~D.~Barger, K.~Whisnant, S.~Pakvasa and R.~J.~N.~Phillips,
``Matter Effects on Three-Neutrino Oscillations,''
Phys. Rev. D \textbf{22}, 2718 (1980).

\bibitem{Xing:2000gg}
Z.~Z.~Xing,
``New formulation of matter effects on neutrino mixing and CP violation,''
Phys. Lett. B \textbf{487}, 327-333 (2000)
[arXiv:hep-ph/0002246 [hep-ph]].

\bibitem{Xing:2003ez}
Z.~Z.~Xing,
``Flavor mixing and CP violation of massive neutrinos,''
Int. J. Mod. Phys. A \textbf{19}, 1-80 (2004)
[arXiv:hep-ph/0307359 [hep-ph]].

\bibitem{Capozzi:2025wyn}
F.~Capozzi, W.~Giar{\`e}, E.~Lisi, A.~Marrone, A.~Melchiorri and A.~Palazzo,
``Neutrino masses and mixing: Entering the era of subpercent precision,''
Phys. Rev. D \textbf{111}, no.9, 093006 (2025)
[arXiv:2503.07752 [hep-ph]].

\bibitem{Luo:2019efb}
S.~Luo,
``Neutrino Oscillation in Dense Matter,''
Phys. Rev. D \textbf{101}, no.3, 033005 (2020)
[arXiv:1911.06301 [hep-ph]].

\bibitem{NOvA:2021nfi}
M.~A.~Acero \textit{et al.} [NOvA],
``Improved measurement of neutrino oscillation parameters by the NOvA experiment,''
Phys. Rev. D \textbf{106}, no.3, 032004 (2022)
[arXiv:2108.08219 [hep-ex]].

\bibitem{Xing:2018lob}
Z.~Z.~Xing, S.~Zhou and Y.~L.~Zhou,
``Renormalization-Group Equations of Neutrino Masses and Flavor Mixing Parameters in Matter,''
JHEP \textbf{05}, 015 (2018)
[arXiv:1802.00990 [hep-ph]].

\bibitem{Parke:2018brr}
S.~J.~Parke, P.~B.~Denton and H.~Minakata,
``Analytic Neutrino Oscillation Probabilities in Matter: Revisited,''
PoS \textbf{NuFact2017}, 055 (2018)
[arXiv:1801.00752 [hep-ph]].

\bibitem{Wang:2019yfp}
X.~Wang and S.~Zhou,
``Analytical solutions to renormalization-group equations of effective neutrino masses and mixing parameters in matter,''
JHEP \textbf{05}, 035 (2019)
[arXiv:1901.10882 [hep-ph]].

\bibitem{Xing:2019owb}
Z.~Z.~Xing and J.~Y.~Zhu,
``Sum rules and asymptotic behaviors of neutrino mixing in dense matter,''
Nucl. Phys. B \textbf{949}, 114803 (2019)
[arXiv:1905.08644 [hep-ph]].

\bibitem{Zhang:2004hf}
H.~Zhang and Z.~Z.~Xing,
``Leptonic unitarity triangles in matter,''
Eur. Phys. J. C \textbf{41}, 143-152 (2005)
[arXiv:hep-ph/0411183 [hep-ph]].

\bibitem{Botella:1986wy}
F.~J.~Botella, C.~S.~Lim and W.~J.~Marciano,
``Radiative Corrections to Neutrino Indices of Refraction,''
Phys. Rev. D \textbf{35}, 896 (1987)

\bibitem{Huang:2025apv}
J.~Huang, T.~Ohlsson, S.~Vihonen and S.~Zhou,
``Effects of the matter potential at one-loop level on neutrino oscillations in long-baseline experiments,''
Phys. Rev. D \textbf{111}, no.11, 116024 (2025)
[arXiv:2504.15998 [hep-ph]].

\bibitem{Huang:2025iww}
J.~Huang, T.~Ohlsson, S.~Vihonen and S.~Zhou,
``One-Loop Effects in the Neutrino Matter Potential and Implications for Non-Standard Interactions,''
[arXiv:2510.04841 [hep-ph]].

\bibitem{Farzan:2017xzy}
Y.~Farzan and M.~Tortola,
``Neutrino oscillations and Non-Standard Interactions,''
Front. in Phys. \textbf{6}, 10 (2018)
[arXiv:1710.09360 [hep-ph]].

\bibitem{Ge:2018uhz}
S.~F.~Ge and S.~J.~Parke,
``Scalar Nonstandard Interactions in Neutrino Oscillation,''
Phys. Rev. Lett. \textbf{122}, no.21, 211801 (2019)
[arXiv:1812.08376 [hep-ph]].

\bibitem{JUNO:2025gmd}
A.~Abusleme \textit{et al.} [JUNO],
``First measurement of reactor neutrino oscillations at JUNO,''
[arXiv:2511.14593 [hep-ex]].

\bibitem{JUNO:2022mxj}
A.~Abusleme \textit{et al.} [JUNO],
``Sub-percent precision measurement of neutrino oscillation parameters with JUNO,''
Chin. Phys. C \textbf{46}, no.12, 123001 (2022)
[arXiv:2204.13249 [hep-ex]].

\bibitem{JUNO:2015zny}
F.~An \textit{et al.} [JUNO],
``Neutrino Physics with JUNO,''
J. Phys. G \textbf{43}, no.3, 030401 (2016)
[arXiv:1507.05613 [physics.ins-det]].

\bibitem{JUNO:2024jaw}
A.~Abusleme \textit{et al.} [JUNO],
``Potential to identify neutrino mass ordering with reactor antineutrinos at JUNO,''
Chin. Phys. C \textbf{49}, no.3, 033104 (2025)
[arXiv:2405.18008 [hep-ex]].


\end{thebibliography}
\end{document}